\documentclass[10pt]{article}

\usepackage{amsmath}
\usepackage{amssymb}

\usepackage[proposal]{gmu_project}

\titleformat{\section}
    {\large\bfseries\color{GMUGreen}}
    {\thesection}{0.6em}{}
    [\nobreak\vspace{1.5pt}\nobreak{\color{GMUGold}\nobreak\titlerule[1.1pt]}\nobreak]
\titleformat{\subsection}
    {\normalsize\bfseries\color{GMUGreen}}
    {\thesubsection}{0.6em}{}

\usepackage{booktabs}
\usepackage{array}
\usepackage{longtable}
\usepackage{float}
\usepackage{siunitx}

\newcommand{\nChecks}{eighteen}
\newcommand{\bezpyVersion}{0.5.0}
\newcommand{\bezpyCommit}{39ed8ee}
\newcommand{\verifyPeriodCount}{121}
\newcommand{\verifyPeriodMin}{1}
\newcommand{\verifyPeriodMax}{10000}
\newcommand{\floorPeriodCount}{26}

\newcommand{\floorSynthetic}{0.0063}
\newcommand{\floorStorm}{0.2}

\newcommand{\nSites}{1616}
\newcommand{\nStormSites}{1614}
\newcommand{\nLateStormSites}{1611}
\newcommand{\nHighAsymmetry}{1196}
\newcommand{\nLowAsymmetry}{420}
\newcommand{\medianAsymmetry}{0.39}
\newcommand{\upperAsymmetry}{0.73}
\newcommand{\nFitTarget}{1150}

\newcommand{\nFitAboveTwo}{185}
\newcommand{\fitWorstRms}{162}
\newcommand{\minRetainedPeriods}{12}
\newcommand{\maxRetainedPeriods}{28}
\newcommand{\syntheticPeakMedian}{20}

\newcommand{\syntheticMagnitudeMedian}{28}
\newcommand{\syntheticVectorMedian}{43}
\newcommand{\syntheticVectorNinety}{94}
\newcommand{\syntheticReductionMedian}{42}
\newcommand{\syntheticMisfitMedian}{2.1}
\newcommand{\syntheticProcessingMedian}{1.3}
\newcommand{\syntheticProcessingNinety}{3.3}

\newcommand{\stormPeakMedian}{20}

\newcommand{\stormMagnitudeMedian}{30}
\newcommand{\stormVectorMedian}{46}

\newcommand{\stormReductionMedian}{46}
\newcommand{\stormMisfitMedian}{1.9}
\newcommand{\stormProcessingMedian}{1.3}
\newcommand{\stormProcessingNinety}{2.8}

\newcommand{\lateStormPeakMedian}{20}

\newcommand{\lateStormVectorMedian}{46}

\newcommand{\nObservatoriesTested}{20}
\newcommand{\withheldCode}{BRW}
\newcommand{\withheldPeak}{4954}
\newcommand{\withheldPredictedPeak}{2221}
\newcommand{\closeKm}{305}
\newcommand{\closeVectorWorst}{63}
\newcommand{\farCode}{HON}
\newcommand{\farKm}{3790}
\newcommand{\farVector}{43}

\newcommand{\sourceVectorMedian}{48}
\newcommand{\sourceBandVectorMedian}{80}
\newcommand{\observatoryKm}{920}
\newcommand{\siteKm}{785}
\newcommand{\isolationRatio}{1.2}
\newcommand{\budgetTensors}{1614}
\newcommand{\budgetPairs}{32280}
\newcommand{\budgetObservatories}{20}
\newcommand{\budgetExceedPercent}{96}
\newcommand{\budgetMagneticMedian}{93}
\newcommand{\budgetMagneticPeakMedian}{35}
\newcommand{\budgetGroundMedian}{46}
\newcommand{\budgetGroundPeakMedian}{20}
\newcommand{\budgetChainMedian}{96}
\newcommand{\budgetChainPeakMedian}{43}
\newcommand{\secsLatitudes}{36}
\newcommand{\secsLongitudes}{76}
\newcommand{\secsSystems}{2736}
\newcommand{\secsHeightKm}{110}
\newcommand{\secsLatMin}{15}
\newcommand{\secsLatMax}{85}
\newcommand{\secsLonWest}{175}
\newcommand{\secsLonEast}{25}

\newcommand{\secsEpsilon}{0.05}
\newcommand{\pysecsVersion}{0.4.0}
\newcommand{\gapSamples}{5}
\newcommand{\gapPercent}{2}
\newcommand{\taperEachPercent}{2.5}
\newcommand{\padWord}{twice}
\newcommand{\syntheticExponent}{14}
\newcommand{\syntheticIntervalS}{10}
\newcommand{\bandEdgeDecades}{0.15}
\newcommand{\nArchiveFiles}{2470}

\newcommand{\nArchiveBundles}{four}
\newcommand{\emtfBundleDate}{8 October 2025}

\newcommand{\nPeriodsRetained}{41901}
\newcommand{\minimumPeriods}{8}
\newcommand{\errorScreenPercent}{25}
\newcommand{\covVarianceMismatch}{1e-06}
\newcommand{\covCrossMedian}{5}
\newcommand{\covCrossNinety}{17}
\newcommand{\covCrossMax}{100}
\newcommand{\covCrossAboveTen}{9952}
\newcommand{\covCrossSitesAboveTen}{1196}
\newcommand{\covInvalidPeriods}{20}
\newcommand{\covInvalidSites}{ARU42, AZW15, and FL007}
\newcommand{\covInvalidSiteCount}{three}
\newcommand{\covKrigingFloor}{99}
\newcommand{\covGroundRhoFloor}{95}
\newcommand{\covGroundPhaseFloor}{79}
\newcommand{\procAntisymMedian}{0.5}
\newcommand{\procAntisymMax}{10}
\newcommand{\procLeakMedian}{3.8}
\newcommand{\procLeakMax}{18}
\newcommand{\sensSettings}{12}
\newcommand{\sensReferenceScalar}{57.3}
\newcommand{\sensReferenceCoverage}{82.5}
\newcommand{\sensNearestScalar}{63.5}
\newcommand{\sensMeanScalar}{59.1}
\newcommand{\sensReferenceTensor}{75.7}
\newcommand{\sensNearestTensor}{76.8}
\newcommand{\sensMeanTensor}{72.7}
\newcommand{\sensReferenceHundred}{11}
\newcommand{\sensNearestHundred}{22}
\newcommand{\sensMeanHundred}{7}
\newcommand{\sensErrorLow}{53.0}
\newcommand{\sensErrorHigh}{58.2}
\newcommand{\sensCoverageLow}{65.2}
\newcommand{\sensCoverageHigh}{91.1}
\newcommand{\sensDeconvolvedScalar}{53.0}

\newcommand{\extremeSites}{11}
\newcommand{\extremeBandSites}{10}
\newcommand{\extremeIsolatedSites}{1}
\newcommand{\extremeOver}{84}
\newcommand{\extremeUnder}{101}
\newcommand{\extremeLocalAlso}{18}

\newcommand{\extremeLocalSites}{MEB61 and VAS55}

\newcommand{\extremeMeasurementMax}{4.0}
\newcommand{\priorRatioLow}{0.015}
\newcommand{\priorRatioHigh}{0.21}
\newcommand{\priorTopSites}{284}
\newcommand{\priorBottomSites}{301}
\newcommand{\priorShareLow}{5}
\newcommand{\priorShareHigh}{51}
\newcommand{\nObservatories}{21}

\newcommand{\obsDefinitive}{four}
\newcommand{\obsQuasiDefinitive}{five}
\newcommand{\obsProvisional}{one}
\newcommand{\obsVariation}{11}
\newcommand{\refusalClause}{two sites, CAY08 and REK58, are excluded because their longest retained periods, \SI{273}{\second} and \SI{529}{\second}, are shorter than \SI{600}{\second}}

\newcommand{\krigeSites}{1616}
\newcommand{\krigeTargets}{149}
\newcommand{\krigeRegions}{6}
\newcommand{\krigeRadius}{200}
\newcommand{\krigeLayers}{10}
\newcommand{\krigeNeighbours}{24}
\newcommand{\krigeDraws}{1000}
\newcommand{\krigeSingleScalar}{34.9}
\newcommand{\krigeSingleTensor}{53.2}
\newcommand{\krigeSingleDistance}{59.7}
\newcommand{\krigeGapScalar}{46.9}
\newcommand{\krigeGapTensor}{62.9}
\newcommand{\krigeGapDistance}{152.8}
\newcommand{\krigeRegionalScalar}{57.3}
\newcommand{\krigeRegionalTensor}{75.7}
\newcommand{\krigeRegionalDistance}{87.0}
\newcommand{\krigeMatchedScalar}{44.4}
\newcommand{\krigeMatchedTensor}{66.6}
\newcommand{\krigeMatchedDistance}{57.9}
\newcommand{\krigeHundredSites}{11}
\newcommand{\krigeHundredPeriods}{185}
\newcommand{\krigePeriods}{3869}

\newcommand{\krigeCoverageLow}{62.1}
\newcommand{\krigeCoverageHigh}{98.6}
\newcommand{\krigeCoverageSiteMean}{82.2}
\newcommand{\krigeCoverageCovered}{6379}
\newcommand{\krigeCoverageComponents}{7738}
\newcommand{\krigeCoveragePooled}{82.4}
\newcommand{\krigeCoverageFullSites}{98}
\newcommand{\krigeCoverageEmptySites}{13}
\newcommand{\krigeMidAtlanticSites}{28}
\newcommand{\krigeCentralPlainsSites}{28}
\newcommand{\krigePacificNorthwestSites}{23}
\newcommand{\krigeSouthernRockiesSites}{23}
\newcommand{\krigeCentralTexasSites}{26}
\newcommand{\krigeNorthernNewEnglandSites}{21}
\newcommand{\fitLayerCount}{16}
\newcommand{\fitShallowInterfaceM}{300}
\newcommand{\fitDeepInterfaceKm}{247.6}
\newcommand{\fitLowSite}{WYL18}
\newcommand{\fitLowName}{Farson, WY, USA}
\newcommand{\fitLowLatitude}{42.146650}
\newcommand{\fitLowLongitude}{-109.667500}
\newcommand{\fitHighSite}{WYI21}
\newcommand{\fitHighName}{Ten Sleep, WY, USA}
\newcommand{\fitHighLatitude}{43.970600}
\newcommand{\fitHighLongitude}{-107.251000}
\newcommand{\halfSpaceGain}{2.236}
\newcommand{\halfSpaceField}{223.6}
\newcommand{\exampleSkinDepthKm}{50.329}
\newcommand{\examplePhaseDelay}{15.92}
\newcommand{\exampleGroupDelay}{7.96}
\newcommand{\measuredPeriodCount}{26}
\newcommand{\measuredPeriodMin}{11.636}
\newcommand{\measuredPeriodMax}{7281.778}
\newcommand{\measuredReferencePeriod}{102.4}

\setstudent{Dennies Bor}
\setproposalunit{George Mason University}
\setproposaldate{September 2026}
\settitle{Revisiting magnetotelluric theory: layered induction and geoelectric fields}
\setgmulogo{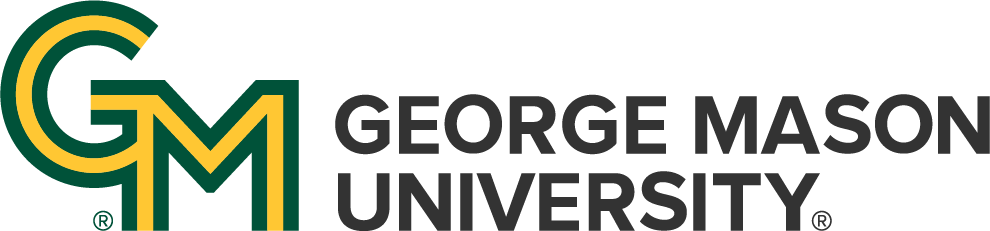}

\graphicspath{{./}}

\hypersetup{
    pdftitle={Revisiting magnetotelluric theory: layered induction and geoelectric fields},
    pdfauthor={Dennies Bor},
    pdfsubject={A review of layered magnetotelluric induction, the
                transmission line analogy, and geoelectric applications}
}

\newcommand{\ee}{\mathrm{e}}
\newcommand{\ii}{\mathrm{i}}
\newcommand{\dd}{\mathrm{d}}

\begin{document}

\makeproposalheader

\begin{proposalabstract}
We review the magnetotelluric (MT) theory of a horizontally layered Earth under plane wave forcing and its use in calculating geoelectric fields. Starting from Maxwell's equations, we derive the diffusion approximation and the impedance recursion by matching electric and magnetic fields at layer boundaries. The equivalent transmission line connects the surface impedance to attenuation with depth, apparent resistivity, phase, and the electric response to a magnetic time series. Three evaluations accompany the derivation, each against its own reference. First, under the storm of 10 and 11 May 2024, electric fields calculated from layered models fitted to measured impedance tensors differ from those calculated from the tensors by a median vector error of \SI{\stormVectorMedian}{\percent} across \num{\nStormSites} EarthScope MT sites with sufficient period coverage. Reducing each tensor to its antisymmetric projection alone gives a median of \SI{\stormReductionMedian}{\percent}, while the layered fit and numerical processing give medians of \SI{\stormMisfitMedian}{\percent} and \SI{\stormProcessingMedian}{\percent}. Second, the magnetic field reconstructed at each of \num{\nObservatoriesTested} withheld observatories from the remaining network has a median vector error of \SI{\sourceVectorMedian}{\percent}. In a common electric field comparison, the magnetic reconstruction gives the larger electric vector error in \SI{\budgetExceedPercent}{\percent} of \num{\budgetPairs} observatory--tensor pairs, with medians of \SI{\budgetMagneticMedian}{\percent} against \SI{\budgetGroundMedian}{\percent} for the layered representation. Third, when conductivity inferred at MT sites is interpolated by kriging into withheld regions, the median relative error of the predicted impedance is \SI{\krigeRegionalScalar}{\percent}, and \num{\krigeHundredSites} of \num{\krigeTargets} withheld sites have an apparent resistivity discrepancy of a factor of 100 or more at one or more periods. The vector comparison retains differences in the amplitude and direction of the electric field that are relevant to calculating induced voltages along transmission line routes.
\end{proposalabstract}

\begin{proposalcolumns}

\section{Introduction}

Temporal variations in the geomagnetic field, described by $\partial\mathbf{B}/\partial t$, induce electric fields in the conducting Earth. These geoelectric fields drive geomagnetically induced currents (GIC) through grounded electricity transmission networks \citep{pirjola2002}. Magnetotelluric (MT) impedance relates the horizontal electric and magnetic fields in the frequency domain. Its amplitude and phase depend on the electrical conductivity of the crust and upper mantle and on the frequency of the disturbance.

The relation between geomagnetic variations and electric currents in the Earth was examined in the early work of \citet{rikitake1948}, \citet{tikhonov1950}, and \citet{cagniard1953}. For a horizontally layered Earth, the recursion of \citet{wait1954} relates the impedance at the surface to the resistivity and thickness of each layer. Representing the induction equations as an equivalent transmission line gives a common formulation for the surface response, attenuation with depth, and the geoelectric field in the time domain.

The United States Magnetotelluric Array (USMTArray) provides measured MT transfer functions at survey sites across the contiguous United States, forming the National Impedance Map \citep{kelbert2026usmtarray}. The transfer functions are available through the EarthScope electromagnetic transfer function (EMTF) data product \citep{kelbert_em_2011}. Conductivity models are inferred by inversion of these responses and can be used to calculate modelled impedances. \citet{bedrosian2015} demonstrated that a spatially uniform magnetic disturbance can produce electric fields that differ in amplitude, direction, and phase between sites because their measured impedances differ. \citet{lucas2020} extended the use of empirical responses to statistical estimates of extreme geoelectric fields and transmission line voltages.

Applications in the literature use geoelectric fields to estimate voltages along transmission lines \citep{kelbertlucas2020} and GICs in power networks represented through their substation components \citep{bor2026components}. The Coupled Space Weather Impact Model (C-SWIM) combines magnetic interpolation and measured MT tensors with calculations of GICs, transformer vulnerability, and economic consequences \citep{oughton2026coupled}. These applications connect geomagnetic observations and the Earth's electrical response to infrastructure risk assessment.

\subsection{Scope}

The theory covers a horizontally layered Earth under plane wave forcing. From Maxwell's equations we derive the coupled equations for the horizontal fields in each layer with displacement current neglected, impose continuity of the tangential fields at each boundary and decay in the deepest half space, and obtain the surface impedance. The equivalent transmission line provides an equivalent circuit interpretation of the same equations \citep{ward1988,chave2012}.

We verify the implementation against analytic solutions and an independent implementation. Three worked evaluations follow, summarised in Table~\ref{tab:roadmap}. The first compares conductivity representations: it applies one magnetic input to a measured MT tensor and to a layered model fitted to it, and measures the discrepancy between the two electric fields. The second evaluates magnetic reconstruction against observation: each observatory is withheld during the storm of 10 and 11 May 2024 and its record is reconstructed from the others. The third predicts the impedance at withheld MT sites from conductivity inferred at other sites and interpolated by kriging. The contribution is the derivation in one notation together with these evaluations, which state for each step of a geoelectric estimate the size of the discrepancy it introduces, the reference against which it is measured, and the number and distribution of evaluation locations.

\end{proposalcolumns}

\begin{table}[H]
    \centering
    \caption{The three worked evaluations. Each compares a prediction with a reference of a different kind, so their errors are reported separately. Section~\ref{sec:source} also passes the magnetic reconstruction through the measured tensors, which places the first two on one scale.}
    \label{tab:roadmap}
    \small
    \renewcommand{\arraystretch}{1.18}
    \begin{tabular}{@{}>{\raggedright\arraybackslash}p{0.11\textwidth}*{3}{>{\raggedright\arraybackslash}p{0.27\textwidth}}@{}}
    \toprule
    & Conductivity representation & Magnetic reconstruction & Impedance at withheld sites \\
    & Section~\ref{sec:ground} & Section~\ref{sec:source} & Section~\ref{sec:kriging} \\
    \midrule
    Input & one magnetic record at each MT site, synthetic or reconstructed & records of the remaining observatories & layered profiles inferred at retained MT sites \\
    Reference & electric field from the measured tensor & recorded magnetic field at the withheld observatory & measured impedance at the withheld site \\
    Prediction & electric field from the fitted layered model & reconstructed magnetic field & impedance of the kriged profile \\
    Metric & peak, horizontal magnitude, and vector errors & peak and vector errors & relative error of the impedance and of the full tensor, apparent resistivity factor, interval coverage \\
    Implication & discrepancy from representing the ground by layers & error in the magnetic input where no observatory exists & error in the ground response where no MT site exists \\
    \bottomrule
    \end{tabular}
\end{table}

\begin{figure}[H]
    \centering
    \includegraphics[width=\textwidth]{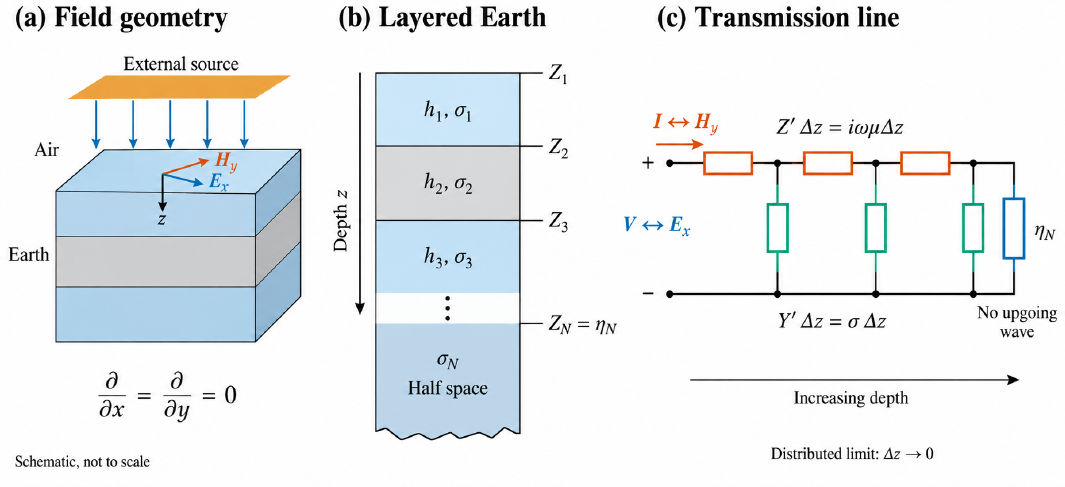}
    \caption{The magnetotelluric problem in one dimension and its circuit equivalent. (a)~An external source above the surface, drawn as a horizontal sheet, is represented locally by a plane wave: the fields are taken as uniform in both horizontal directions, so horizontal derivatives vanish and the fields reduce to the orthogonal pair $E_x$ and $H_y$, with $\hat{x}\times\hat{y}$ pointing down along $z$. The drawing shows neither the source currents nor the attenuation within the ground, which Figure~\ref{fig:skin} shows. (b)~The Earth is a stack of uniform layers terminating in a half space, and $Z_n$ is the input impedance at the top of layer $n$. (c)~The same system as a ladder network in the limit of vanishing segment thickness, with series impedance $\ii\omega\mu$ and shunt admittance $\sigma$ per unit depth, terminated by the intrinsic impedance of the half space. Because no upgoing wave returns from the termination, the line is matched at its far end. The geometry is schematic and is not drawn to scale.}
    \label{fig:geometry}
\end{figure}

\begin{proposalcolumns}

\section{Assumptions}
\label{sec:assumptions}

The derivation uses a planar, layered Earth with the following material properties, source geometry, and boundary conditions \citep{ward1988,chave2012}. Section~\ref{sec:notation} collects the symbols, units, and sign conventions used throughout the text and figures.

\textbf{Uniform, isotropic layers.} Within each layer, the fields satisfy the linear constitutive relations
\begin{equation}
    \mathbf{J}=\sigma\mathbf{E}, \qquad
    \mathbf{D}=\epsilon\mathbf{E}, \qquad
    \mathbf{B}=\mu\mathbf{H},
    \label{eq:constitutive}
\end{equation}
where $\mathbf{E}$ is the electric field (\si{\volt\per\metre}), $\mathbf{J}$ the conduction current density (\si{\ampere\per\metre\squared}), $\mathbf{D}$ the electric displacement field (\si{\coulomb\per\metre\squared}), $\mathbf{B}$ the magnetic flux density (\si{\tesla}), and $\mathbf{H}$ the magnetic field strength (\si{\ampere\per\metre}). The conductivity $\sigma>0$ (\si{\siemens\per\metre}) and permittivity $\epsilon$ (\si{\farad\per\metre}) are constant scalars within each layer, independent of time and frequency. The resistivity is $\rho=1/\sigma$ (\si{\ohm\metre}). The ground is nonmagnetic, with permeability $\mu=\mu_0$ (\si{\henry\per\metre}), where $\mu_0$ is the permeability of free space. The interfaces are horizontal and the surface is flat.

\textbf{Horizontally uniform source.} The imposed disturbance originates above the Earth. Its horizontal variation is assumed small over the distances relevant to induction, so $\partial/\partial x=\partial/\partial y=0$. This assumption concerns the source and is separate from horizontal layering. For a uniform ground, the surface magnetic field should vary little over the length $1/|k|$, with $k$ defined in Equation~\ref{eq:helmholtz}. A source with appreciable spatial variation can require horizontal derivatives even above a layered Earth \citep{wait1954}.

\textbf{Quasi static induction in the ground.} At angular frequency $\omega$, conduction current dominates displacement current when $\omega\epsilon/\sigma\ll1$. The derivation below identifies the term omitted under this approximation. The time variation responsible for electromagnetic induction is retained.

\textbf{Interface continuity.} The tangential electric and magnetic fields are continuous across each interface. The model contains no imposed surface current sheet at a layer boundary.

\textbf{Terminating half space.} The deepest layer extends to infinite depth. At nonzero frequency the field remains bounded and decays as $z\to\infty$. This condition selects the decaying solution in that layer.

\section{From Maxwell's equations to the telegrapher's equations}
\label{sec:telegrapher}

\subsection{Current balance and diffusion}

The Cartesian coordinate $z$ increases downward from the surface, as in Figure~\ref{fig:geometry}a. Within a uniform layer, with all imposed source currents outside the conducting Earth, Faraday's law and the Ampere--Maxwell law give
\begin{align}
    \nabla\times\mathbf{E}
        &= -\mu\frac{\partial\mathbf{H}}{\partial t},
        \label{eq:faraday-time}\\
    \nabla\times\mathbf{H}
        &= \sigma\mathbf{E}
         + \epsilon\frac{\partial\mathbf{E}}{\partial t}.
        \label{eq:ampere-full}
\end{align}
The terms on the right of Equation~\ref{eq:ampere-full} are current densities. The conduction term is $\mathbf{J}_{\mathrm{conduction}}=\mathbf{J}=\sigma\mathbf{E}$. The displacement term is $\mathbf{J}_{\mathrm{displacement}}=\partial\mathbf{D}/\partial t =\epsilon\,\partial\mathbf{E}/\partial t$, where $t$ denotes time. The bulk of each uniform layer is charge free for the transverse fields considered here, so $\nabla\cdot\mathbf{E}=0$. Taking the curl of Equation~\ref{eq:faraday-time}, using $\nabla\times(\nabla\times\mathbf{E})=-\nabla^2\mathbf{E}$ in that region, and substituting Equation~\ref{eq:ampere-full} gives
\begin{equation}
    \nabla^2\mathbf{E}
      = \mu\sigma\frac{\partial\mathbf{E}}{\partial t}
      + \mu\epsilon\frac{\partial^2\mathbf{E}}{\partial t^2}.
    \label{eq:full-field}
\end{equation}
For harmonic fields with time dependence $\ee^{\ii\omega t}$ and $\omega=2\pi/T$, the ratio of displacement to conduction current is
\begin{equation}
    \frac{|\mathbf{J}_{\mathrm{displacement}}|}
         {|\mathbf{J}_{\mathrm{conduction}}|}
    = \frac{\omega\epsilon}{\sigma}.
    \label{eq:current-ratio}
\end{equation}
When this ratio is small, neglecting displacement current in the ground reduces Equation~\ref{eq:full-field} to
\begin{equation}
    \frac{\partial\mathbf{E}}{\partial t}
       = \frac{1}{\mu\sigma}\nabla^2\mathbf{E}.
    \label{eq:diffusion}
\end{equation}
The result is a diffusion equation with diffusivity $1/(\mu\sigma)$. The magnetic field satisfies the same equation within a uniform layer. For fixed material properties, the displacement current becomes less important as the period increases \citep{chave2012}.

The approximation applies within the conducting ground. In the limit of negligible conductivity, as in air away from the sources, retaining displacement current in Equation~\ref{eq:full-field} gives the electromagnetic wave equation. The boundary value problem below uses the diffusion limit in the ground, with the time dependence of Faraday's law retained.

\subsection{Horizontal field components}

In the diffusion limit, the curl equations for the complex amplitudes are
\begin{align}
    \nabla\times\mathbf{E} &= -\ii\omega\mu\mathbf{H}, \label{eq:faraday}\\
    \nabla\times\mathbf{H} &= \sigma\mathbf{E}. \label{eq:ampere}
\end{align}
Horizontal uniformity sets the horizontal derivatives to zero. The relevant curl components then become
\begin{equation}
    (\nabla\times\mathbf{E})_y=\frac{\dd E_x}{\dd z},
    \qquad
    (\nabla\times\mathbf{H})_x=-\frac{\dd H_y}{\dd z}.
    \label{eq:curl-components}
\end{equation}
Substitution in Equations~\ref{eq:faraday} and~\ref{eq:ampere} gives
\begin{equation}
    \frac{\dd E_x}{\dd z} = -\ii\omega\mu H_y,
    \qquad
    \frac{\dd H_y}{\dd z} = -\sigma E_x.
    \label{eq:coupled}
\end{equation}
These equations involve only $E_x$ and $H_y$. The other horizontal pair obeys
\begin{equation}
    \frac{\dd E_y}{\dd z} = \ii\omega\mu H_x,
    \qquad
    \frac{\dd H_x}{\dd z} = \sigma E_y.
    \label{eq:orthogonal-pair}
\end{equation}
The sign changes follow from the orientation of the coordinate axes. Isotropy makes the conductivity response identical for the two horizontal orientations. For $\omega>0$, the vertical components of the curl equations also give $E_z=H_z=0$. Thus either horizontal pair is sufficient to derive the impedance of this layered model.

\subsection{Equivalent transmission line}

The equations for voltage $V$ and current $I$ on a uniform transmission line have the form
\begin{equation}
    \frac{\dd V}{\dd z} = -Z' I, \qquad \frac{\dd I}{\dd z} = -Y' V,
    \label{eq:telegrapher}
\end{equation}
where $Z'$ is the series impedance per unit length and $Y'$ is the shunt admittance per unit length. The series impedance relates the voltage gradient to the current along the line. The shunt admittance relates the current leaving the line per unit length to the local voltage. Equation~\ref{eq:coupled} has the same differential form under the correspondence
\begin{equation}
    E_x \leftrightarrow V, \quad H_y \leftrightarrow I, \quad
    Z' = \ii\omega\mu, \quad Y' = \sigma .
    \label{eq:analogy}
\end{equation}
Depth plays the role of distance along the equivalent line. The correspondence is between equations, not between physical quantities. $E_x$ and $H_y$ are field amplitudes in \si{\volt\per\metre} and \si{\ampere\per\metre}, whereas $V$ and $I$ are a voltage in \si{\volt} and a current in \si{\ampere}. Each field equation therefore carries one more factor of inverse length than its circuit counterpart, while $Z'$ and $Y'$ keep their units of \si{\ohm\per\metre} and \si{\siemens\per\metre} and the ratio $E_x/H_y$ has the unit of $V/I$, the ohm. The equivalent line is a mathematical device and does not represent a conductor in the ground. The correspondence represents magnetic induction by series inductance and conduction by shunt conductance. Retaining displacement current would give $Y'=\sigma+\ii\omega\epsilon$, adding shunt capacitance to the analogy. The quasi static limit omits that capacitive term. Figure~\ref{fig:geometry}c shows a discretisation that approaches the distributed equations as the segment thickness tends to zero.

\end{proposalcolumns}

\begin{figure}[H]
    \centering
    \includegraphics[width=\textwidth]{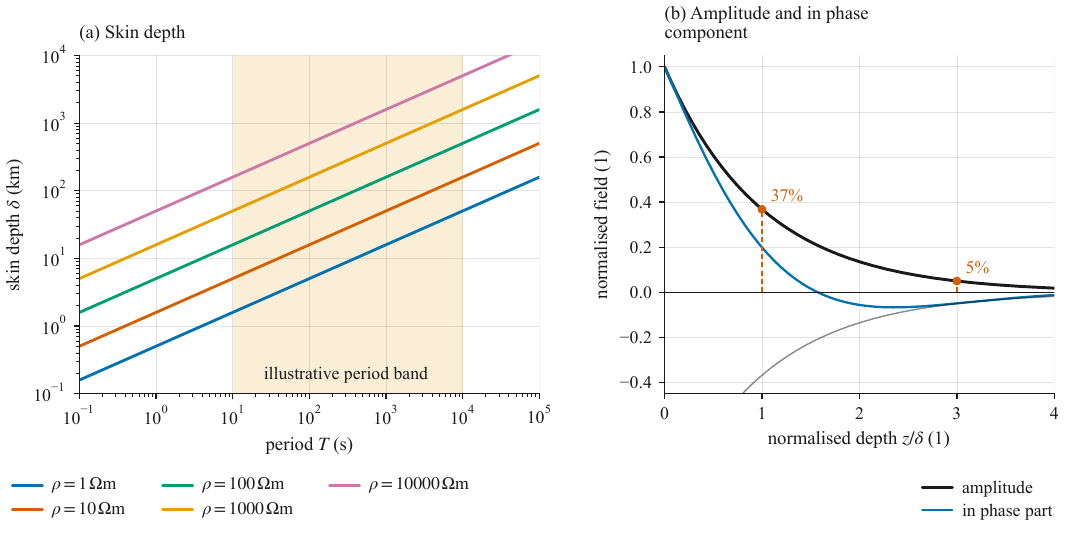}
    \caption{Attenuation of the field in a conductor. (a)~Skin depth against period for five resistivities, spanning conductive sediment to resistive shield. The shaded band marks the illustrative period range \SIrange{10}{10000}{\second} used in the MT comparison. (b)~Amplitude envelope and the component in phase with the surface field, both normalised by the surface amplitude. At one skin depth, the amplitude is reduced by $1/\ee$ and the phase lags by one radian. The component in phase with the surface field changes sign at $z=\pi\delta/2$.}
    \label{fig:skin}
\end{figure}

\begin{proposalcolumns}

\section{A uniform layer}

Within a layer of constant conductivity, differentiating the first relation in Equation~\ref{eq:coupled} and substituting the second gives the frequency domain form of the diffusion equation,
\begin{equation}
    \frac{\dd^2 E_x}{\dd z^2} = k^2 E_x, \qquad
    k = \sqrt{\ii\omega\mu\sigma},
    \label{eq:helmholtz}
\end{equation}
whose general solution in a uniform region is
\begin{equation}
    E_x(z) = E^{\downarrow}\ee^{-kz} + E^{\uparrow}\ee^{+kz}.
    \label{eq:general}
\end{equation}
The constant $k$ is the complex propagation constant in the spatial factors $\ee^{\mp kz}$, with units of inverse length. The complex electric amplitudes at $z=0$ are $E^{\downarrow}$ for the downward component and $E^{\uparrow}$ for the upward component. For $\omega>0$, the square root is chosen so that $\operatorname{Re}k>0$ and $\operatorname{Im}k>0$. The term $\ee^{-kz}$ therefore attenuates downward, while $\ee^{+kz}$ attenuates upward. These are the downward and upward components of a harmonic diffusion field. Both can occur within a finite layer. The condition at infinite depth selects the decaying component in the terminating half space.

Substitution into Faraday's law gives the magnetic field,
\begin{equation}
    H_y(z) = \frac{k}{\ii\omega\mu}\left(E^{\downarrow}\ee^{-kz} - E^{\uparrow}\ee^{+kz}\right).
    \label{eq:hfield}
\end{equation}
A single downward component therefore has the field ratio
\begin{equation}
    \eta = \frac{\ii\omega\mu}{k} = \sqrt{\frac{\ii\omega\mu}{\sigma}}
         = \sqrt{\ii\omega\mu\rho},
    \label{eq:eta}
\end{equation}
the intrinsic impedance of the medium, which is the characteristic impedance of the equivalent line.

\subsection{Skin depth}

Writing $\sqrt{\ii} = (1+\ii)/\sqrt{2}$ separates Equation~\ref{eq:helmholtz} into attenuation and phase rotation,
\begin{equation}
    k = \sqrt{\frac{\omega\mu\sigma}{2}}\,(1 + \ii),
\end{equation}
The positive real part of $k$ controls attenuation and its imaginary part controls phase. For a single downward component, $E_x(z)=E_x(0)\,\ee^{-z/\delta}\ee^{-\ii z/\delta}$, where
\begin{equation}
    \delta = \frac{1}{\operatorname{Re} k} = \sqrt{\frac{2}{\omega\mu\sigma}}
           = \sqrt{\frac{\rho T}{\pi\mu_0}}
           \approx 503\,\sqrt{\rho T}\ \si{\metre},
    \label{eq:skindepth}
\end{equation}
for $\rho$ in \si{\ohm\metre} and the period $T$ in seconds. The amplitude falls to \SI{37}{\percent} of its surface value at one skin depth and below \SI{5}{\percent} at three. Each additional skin depth introduces a phase lag of one radian. Figure~\ref{fig:skin} shows the amplitude envelope and the component in phase with the surface field.

Choosing the phase at the surface as zero gives the physical field
\begin{equation}
    E_x(z,t)=E_0\ee^{-z/\delta}\cos(\omega t-z/\delta),
    \label{eq:skin-time}
\end{equation}
where $E_0$ is the real amplitude at the surface.

At fixed permeability, Equation~\ref{eq:skindepth} gives $\delta\propto\sqrt{\rho T}$. Longer periods and higher resistivities increase the attenuation length. At $\rho=\SI{100}{\ohm\metre}$ and $T=\SI{100}{\second}$,
\begin{equation*}
    \delta \approx 503\sqrt{100\times100}
           = \SI{50.3}{\kilo\metre}.
\end{equation*}
Attenuation varies continuously with depth. Whether a layer can be resolved depends on its conductivity contrast, the periods observed, and the measurement uncertainty. Regional conductivity models describe the structure over the depth range that matters for geoelectric hazard \citep{kelbert2020}.

\end{proposalcolumns}

\begin{figure}[H]
    \centering
    \includegraphics[width=\textwidth]{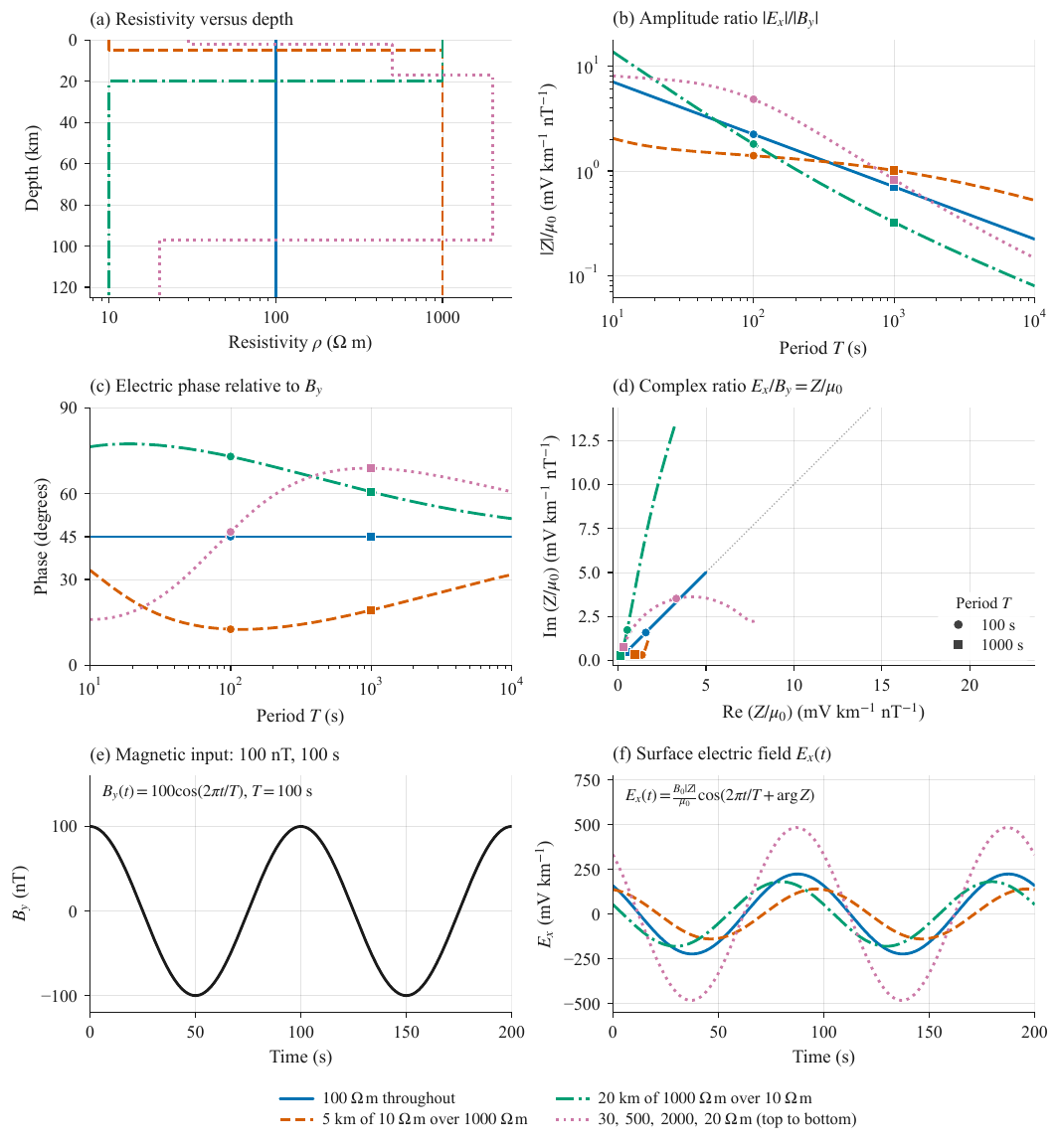}
    \caption{Magnetic input and surface electric response for four layered resistivity profiles. (a)~Resistivity versus depth. The final layer continues to infinite depth. (b,c)~Amplitude ratio $|Z|/\mu_0$ and electric phase relative to the magnetic input. (d)~The complex response, with equal real and imaginary scaling: distance from the origin gives the amplitude ratio and angle gives temporal phase. Circles mark \SI{100}{\second} and squares \SI{1000}{\second}. (e)~The common surface input $B_y(t)=B_0\cos(2\pi t/T)$, with $B_0=\SI{100}{\nano\tesla}$, $T=\SI{100}{\second}$, and $B_x=0$. (f)~Steady periodic electric fields from Equation~\ref{eq:sinusoidal-response}, evaluated at the circle markers in (b)--(d). The profile with four layers has finite thicknesses of \SI{2}{\kilo\metre}, \SI{15}{\kilo\metre}, and \SI{80}{\kilo\metre}. Table~\ref{tab:models} lists the resistivities and thicknesses.}
    \label{fig:responses}
\end{figure}

\begin{proposalcolumns}

\section{The half space}

For a uniform Earth extending to infinite depth, requiring the field to remain bounded sets the upward amplitude $E^{\uparrow}=0$ in Equation~\ref{eq:general}, because $\ee^{+kz}$ grows without bound. The field then contains only the downward component. At the surface its electric to magnetic ratio is the intrinsic impedance,
\begin{equation}
    Z_{\text{surface}} = \frac{E_x(0)}{H_y(0)} = \eta = \sqrt{\ii\omega\mu\rho}.
    \label{eq:halfspace}
\end{equation}
The equivalent line is matched at its termination because the uniform half space has no deeper interface that can return an upward component. This condition concerns the termination within the ground. Conductivity contrasts above it can produce upward components in the finite layers.

\subsection{Source amplitude and impedance}

At a fixed frequency and source geometry, linearity makes both fields proportional to the same dimensionless complex source multiplier $\alpha$. Writing the fields for unit multiplier as $\mathcal{E}_x$ and $\mathcal{H}_y$ gives
\begin{equation}
    Z=\frac{E_x}{H_y}
     =\frac{\alpha\mathcal{E}_x}{\alpha\mathcal{H}_y}
     =\frac{\mathcal{E}_x}{\mathcal{H}_y}.
    \label{eq:source-cancellation}
\end{equation}
The impedance is independent of source amplitude under the plane wave assumption and depends on the layer properties, thicknesses, and frequency. The field $H_y$ in this ratio is the total local magnetic field, including the contribution induced in the ground. An observed magnetic amplitude is needed to recover an absolute electric field \citep{cagniard1953,chave2012}.

This cancellation concerns source amplitude. A change in spatial source structure can change the response when horizontal uniformity is inadequate \citep{wait1954}.

\section{The recursion of Wait}
\label{sec:wait}

Layer $n$ has resistivity $\rho_n$, thickness $h_n$, intrinsic impedance $\eta_n$, and propagation constant $k_n$. The response of all material below it is represented by the impedance $Z_{n+1}$ at its lower boundary. A local coordinate $\zeta$ is zero at the top of the layer and equals $h_n$ at its base. Equations~\ref{eq:general} and~\ref{eq:hfield} apply with $z$ replaced by $\zeta$ and amplitudes $E_n^{\downarrow}$ and $E_n^{\uparrow}$.

\subsection{Matching the fields at the lower boundary}

The downward and upward electric amplitudes evaluated at the base are $E_{n,b}^{\downarrow}=E_n^{\downarrow}\ee^{-k_nh_n}$ and $E_{n,b}^{\uparrow}=E_n^{\uparrow}\ee^{k_nh_n}$. Continuity of the tangential fields gives
\begin{equation}
    E_b=E_{n,b}^{\downarrow}+E_{n,b}^{\uparrow},
    \qquad
    H_b=\frac{E_{n,b}^{\downarrow}-E_{n,b}^{\uparrow}}{\eta_n},
    \label{eq:interface-fields}
\end{equation}
where the subscript $b$ denotes the common field value on either side of the boundary. The material below imposes $E_b=Z_{n+1}H_b$. Substituting Equation~\ref{eq:interface-fields} gives
\begin{equation}
    \eta_n(E_{n,b}^{\downarrow}+E_{n,b}^{\uparrow})=Z_{n+1}(E_{n,b}^{\downarrow}-E_{n,b}^{\uparrow}).
    \label{eq:interface-load}
\end{equation}
Solving for the ratio of upward to downward amplitude yields
\begin{equation}
    R_n=\frac{E_{n,b}^{\uparrow}}{E_{n,b}^{\downarrow}}
       =\frac{Z_{n+1}-\eta_n}{Z_{n+1}+\eta_n}.
    \label{eq:interface-reflection}
\end{equation}
The electric field at the boundary satisfies $E_b/E_{n,b}^{\downarrow}=1+R_n$. These relations specify complex field amplitudes. Fractions of transmitted and reflected power require evaluation of the energy flux \citep{ward1988}.

\subsection{Transferring the boundary response to the surface of the layer}

Propagation across the layer relates the amplitudes at its top to those at its base:
\begin{equation}
    \frac{E_n^{\uparrow}}{E_n^{\downarrow}}=R_n\ee^{-2k_nh_n}.
    \label{eq:reflection-transfer}
\end{equation}
The factor of two accounts for the downward and upward paths through the layer. At the top, $E_x=E_n^{\downarrow}+E_n^{\uparrow}$ and $H_y=(E_n^{\downarrow}-E_n^{\uparrow})/\eta_n$, so their ratio is
\begin{equation}
    R_n = \frac{Z_{n+1} - \eta_n}{Z_{n+1} + \eta_n},
    \qquad
    Z_n = \eta_n\,\frac{1 + R_n\ee^{-2k_n h_n}}{1 - R_n\ee^{-2k_n h_n}}.
    \label{eq:reflection}
\end{equation}
Substituting for $R_n$ and using $\tanh(k_nh_n)=(1-\ee^{-2k_nh_n})/(1+\ee^{-2k_nh_n})$ gives
\begin{equation}
    \boxed{\;
    Z_n = \eta_n\,
    \frac{Z_{n+1} + \eta_n\tanh(k_n h_n)}
         {\eta_n + Z_{n+1}\tanh(k_n h_n)}\;}
    \label{eq:wait}
\end{equation}
This is the input impedance of the equivalent line section terminated by $Z_{n+1}$, as derived by \citet{wait1954}. The derivation retains the response of the entire underlying stack through that terminating impedance.

When $Z_{n+1}=\eta_n$, the reflection coefficient vanishes and $Z_n=\eta_n$. For a layer many skin depths thick, the reflected contribution decays as $\ee^{-2k_nh_n}$ and $Z_n$ approaches $\eta_n$. As the thickness tends to zero, $Z_n$ approaches $Z_{n+1}$. The exponential form uses the attenuation across the finite layer without evaluating $\ee^{+k_nh_n}$.

The recursion starts at the terminating half space with $Z_N=\eta_N$. Equation~\ref{eq:wait} is then evaluated successively for the overlying layers. The final value, $Z_1$, is the impedance at the surface.

\section{Responses derived from the impedance}
\label{sec:responses}

The \emph{Cagniard apparent resistivity} is the resistivity of the uniform half space that would produce the observed impedance modulus,
\begin{equation}
    \rho_a(\omega) = \frac{|Z(\omega)|^2}{\omega\mu_0},
    \label{eq:rhoa}
\end{equation}
where $|Z|$ is the magnitude of the complex impedance. The \emph{impedance phase}, $\arg Z$, is the phase of $E_x$ relative to $H_y$ at the same location. With the $\ee^{\ii\omega t}$ convention, positive phase means that $E_x$ leads $H_y$. A uniform half space has $\rho_a=\rho$ at every period and a phase of \SI{45}{\degree}. The layered examples in Figure~\ref{fig:responses} show how conductivity contrasts change the ratio of electric to magnetic amplitude, and the phase, with period. Apparent resistivity follows from the same impedance through Equation~\ref{eq:rhoa}.

The \emph{$C$ response} of \citet{schmucker1970} and \citet{weidelt1972} is the impedance rescaled to a length,
\begin{equation}
    C(\omega) = \frac{Z(\omega)}{\ii\omega\mu_0}.
    \label{eq:cresponse}
\end{equation}
For a uniform half space, $C=\delta(1-\ii)/2$ and $\operatorname{Re}C=\delta/2$. Figure~\ref{fig:cresponse} compares this length with the interface depths of the illustrative model. Resolving an interface requires an assessment of response sensitivity and measurement uncertainty.

\end{proposalcolumns}

\begin{table}[H]
    \centering
    \caption{Illustrative conductivity models. Resistivities are listed from the surface downward, with the last value representing the terminating half space. Thicknesses refer to the finite layers above it.}
    \label{tab:models}
    \small
    \begin{tabular}{@{}p{0.40\textwidth}ll@{}}
\toprule
Model & Resistivity (\si{\ohm\metre}) & Thickness (\si{\kilo\metre}) \\
\midrule
Uniform half space & \numlist{100} & None \\
Conductive cover over resistive basement & \numlist{10;1000} & \numlist{5} \\
Resistive shield over conductive mantle & \numlist{1000;10} & \numlist{20} \\
Sediment, crust, resistive lower crust, conductive mantle & \numlist{30;500;2000;20} & \numlist{2;15;80} \\
\bottomrule
\end{tabular}

\end{table}

\begin{figure}[H]
    \centering
    \includegraphics[width=\textwidth]{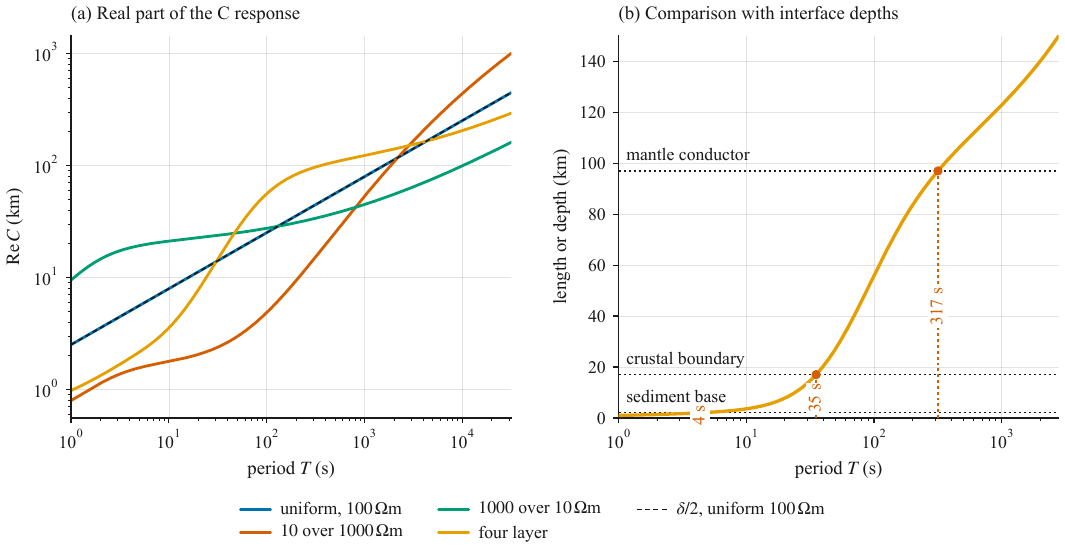}
    \caption{Characteristic induction length from the $C$ response in Equation~\ref{eq:cresponse}. (a)~$\operatorname{Re} C$ against period for the four illustrative conductivity models. For the uniform \SI{100}{\ohm\metre} model, the curve coincides with the dashed reference $\delta/2$. The layered models show how conductivity contrasts change this length. (b)~The response of the model with four layers, compared with its interface depths. The labelled periods, where $\operatorname{Re}C$ equals an interface depth, are illustrative markers of response length. The equality does not show that the interface can be resolved, which depends on response sensitivity and measurement uncertainty.}
    \label{fig:cresponse}
\end{figure}

\begin{proposalcolumns}

\subsection{Amplitude and phase of the surface electric response}
\label{sec:frequency-response}

For a magnetic input expressed as flux density, substituting $H_y=B_y/\mu_0$ into $E_x=ZH_y$ gives
\begin{equation}
    E_x=\frac{Z}{\mu_0}B_y,
    \qquad \frac{E_x}{B_y}=\frac{Z}{\mu_0}.
    \label{eq:magnetic-response}
\end{equation}
The frequency response $Z/\mu_0$ relates the amplitudes of the electric field and the magnetic flux density at the same surface location. Its magnitude and argument give the amplitude ratio and phase difference.

Writing $Z=|Z|\ee^{\ii\phi}$, where $\phi=\arg Z$, and applying a sinusoidal magnetic field of amplitude $B_0$ gives the steady periodic fields
\begin{align}
    B_y(t)&=B_0\cos(\omega t),\\
    E_x(t)&=\frac{B_0|Z|}{\mu_0}\cos(\omega t+\phi).
    \label{eq:sinusoidal-response}
\end{align}
The electric field oscillates at the input frequency $\omega$ with amplitude $B_0|Z|/\mu_0$ and phase shift $\phi$. In the complex plane, $Z/\mu_0$ has radius $|Z|/\mu_0$ and angle $\phi$. Figure~\ref{fig:responses} plots these quantities and the electric time series for the four conductivity profiles in Table~\ref{tab:models}.

For a uniform half space, Equation~\ref{eq:halfspace} gives
\begin{equation}
    \frac{Z}{\mu_0}=\sqrt{\frac{\omega\rho}{\mu_0}}\ee^{\ii\pi/4}.
    \label{eq:halfspace-response}
\end{equation}
At $\rho=\SI{100}{\ohm\metre}$ and $T=\SI{100}{\second}$, the amplitude ratio is $\halfSpaceGain\,(\si{\milli\volt\per\kilo\metre})/\si{\nano\tesla}$. A magnetic amplitude of \SI{100}{\nano\tesla} therefore produces a surface electric amplitude of \SI{\halfSpaceField}{\milli\volt\per\kilo\metre}, with an electric phase lead of \SI{45}{\degree}. The frequency domain multiplication extends to magnetic records containing many frequencies, as described in Section~\ref{sec:time-domain} \citep{kelbert2017}.

\end{proposalcolumns}

\begin{figure}[H]
    \centering
    \includegraphics[width=\textwidth]{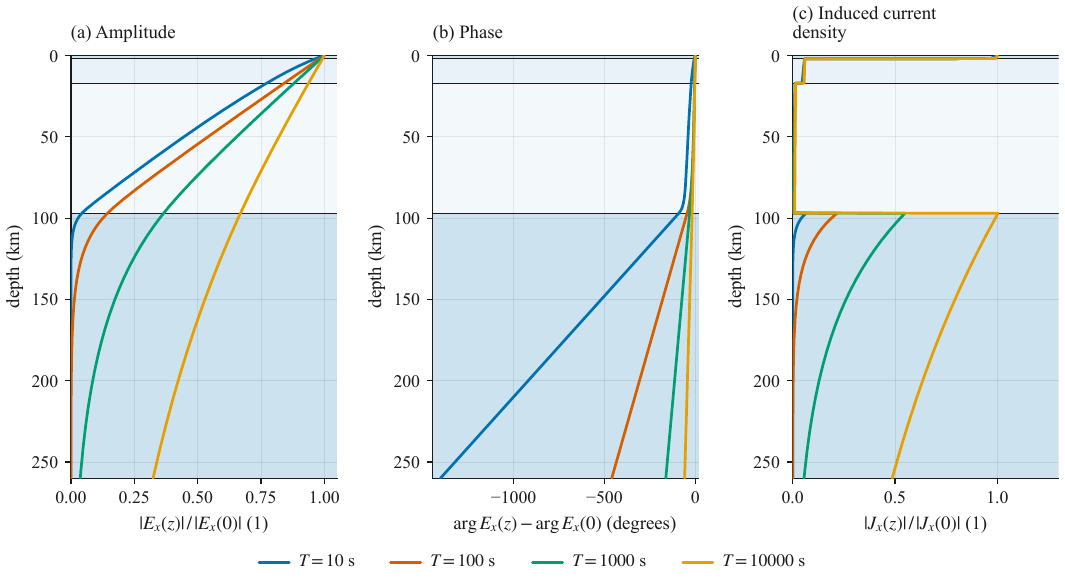}
    \caption{Downward continuation of the field through the model with four layers at four periods, from Equation~\ref{eq:continuation}. Shading marks layer resistivity, with darker tones more conductive. (a)~Amplitude relative to the surface. (b)~Electric field phase relative to the surface, followed continuously with depth rather than wrapped at $\pm\SI{180}{\degree}$. Negative values indicate a lag. (c)~Induced current density relative to its surface value. The tangential electric field is continuous at each interface. The current density $J_x=\sigma E_x$ changes abruptly when conductivity changes, including at the top of the mantle conductor.}
    \label{fig:profile}
\end{figure}

\begin{proposalcolumns}

\subsection{The MT impedance tensor and the layered limit}
\label{sec:tensor}

The MT impedance tensor describes the linear relationship between the horizontal electric and magnetic fields at a site in the frequency domain. Under plane wave forcing, a general conductivity structure requires both horizontal magnetic components to describe each electric component \citep{chave2012},
\begin{equation}
    \begin{pmatrix} E_x \\ E_y \end{pmatrix}
    =
    \begin{pmatrix} Z_{xx} & Z_{xy} \\ Z_{yx} & Z_{yy} \end{pmatrix}
    \begin{pmatrix} H_x \\ H_y \end{pmatrix}.
    \label{eq:tensor}
\end{equation}
Each coefficient $Z_{ij}(\omega)$ relates magnetic component $j$ to electric component $i$ in orthogonal horizontal coordinates. The tensor uses the $E/H$ convention, with coefficients in \si{\ohm}. The corresponding $E/B$ response is $\mathbf{Z}/\mu_0$.

For isotropic horizontal layers under the assumed plane wave forcing, $Z_{xx}=Z_{yy}=0$ and $Z_{xy}=-Z_{yx}=Z$. The tensor is therefore specified by a single complex impedance that is invariant under rotation of the horizontal axes \citep{ward1988,simpson2005}. Differences between measured tensors and this form provide diagnostics of the layered approximation, subject to the uncertainties of the measurements \citep{chave2012}.

Equation~\ref{eq:cresponse} defines the MT $C$ response using horizontal electric and magnetic fields \citep{weidelt1972}. In geomagnetic depth sounding, a related response, denoted $C_{\mathrm{GDS}}$ here, is defined using the vertical magnetic field and the horizontal divergence of the horizontal magnetic field,

\begin{equation}
    C_{\mathrm{GDS}}(\omega) = \frac{H_z}
    {\dfrac{\partial H_x}{\partial x} + \dfrac{\partial H_y}{\partial y}},
    \label{eq:gds}
\end{equation}
with $z$ positive downward \citep{schmucker1970}. Under the horizontally uniform source assumed in this derivation, both the numerator and denominator vanish. Application of Equation~\ref{eq:gds} therefore requires treatment of horizontal variation in the source \citep{weaver1994}.

\section{Fields inside the Earth}

The impedance recursion also determines the fields within the Earth. Once $Z_n$ is known at the top of every layer, the field at depth $\zeta$ below the top of layer $n$ follows from Equation~\ref{eq:general} with the reflection coefficient referred to the base of that layer,
\begin{equation}
    \frac{E_x(\zeta)}{E_x(0_n)} =
    \frac{\ee^{-k_n\zeta} + R_n\,\ee^{-k_n(2h_n-\zeta)}}
         {1 + R_n\,\ee^{-2k_n h_n}} ,
    \label{eq:continuation}
\end{equation}
Here $E_x(0_n)$ is the electric field at the top of layer $n$, where $\zeta=0$. It follows from the surface field by evaluating Equation~\ref{eq:continuation} at the base of each overlying layer in turn. The corresponding magnetic field, normalised by the same electric amplitude at the top, is
\begin{equation}
    \frac{H_y(\zeta)}{E_x(0_n)} =
    \frac{\ee^{-k_n\zeta}-R_n\ee^{-k_n(2h_n-\zeta)}}
         {\eta_n\left(1+R_n\ee^{-2k_nh_n}\right)}.
    \label{eq:magnetic-continuation}
\end{equation}
Neither exponent has a positive real part for $0\le\zeta\le h_n$, so no growing exponential is evaluated within the layer.

Figure~\ref{fig:profile} shows the fields within the illustrative model. The electric field and the magnetic field are continuous at every interface, but the induced current density $J_x = \sigma E_x$ is not, because conductivity jumps. The amplitude and phase profiles show how the attenuation scale varies with period in a layered medium. Short periods attenuate over shorter distances within each layer. The response of a deeper interface also depends on attenuation through the overlying layers and on the conductivity contrast. The skin depth is a scale for attenuation, and resolving an interface requires consideration of the sensitivity and measurement uncertainty.

\end{proposalcolumns}

\begin{figure}[H]
    \centering
    \includegraphics[width=\textwidth]{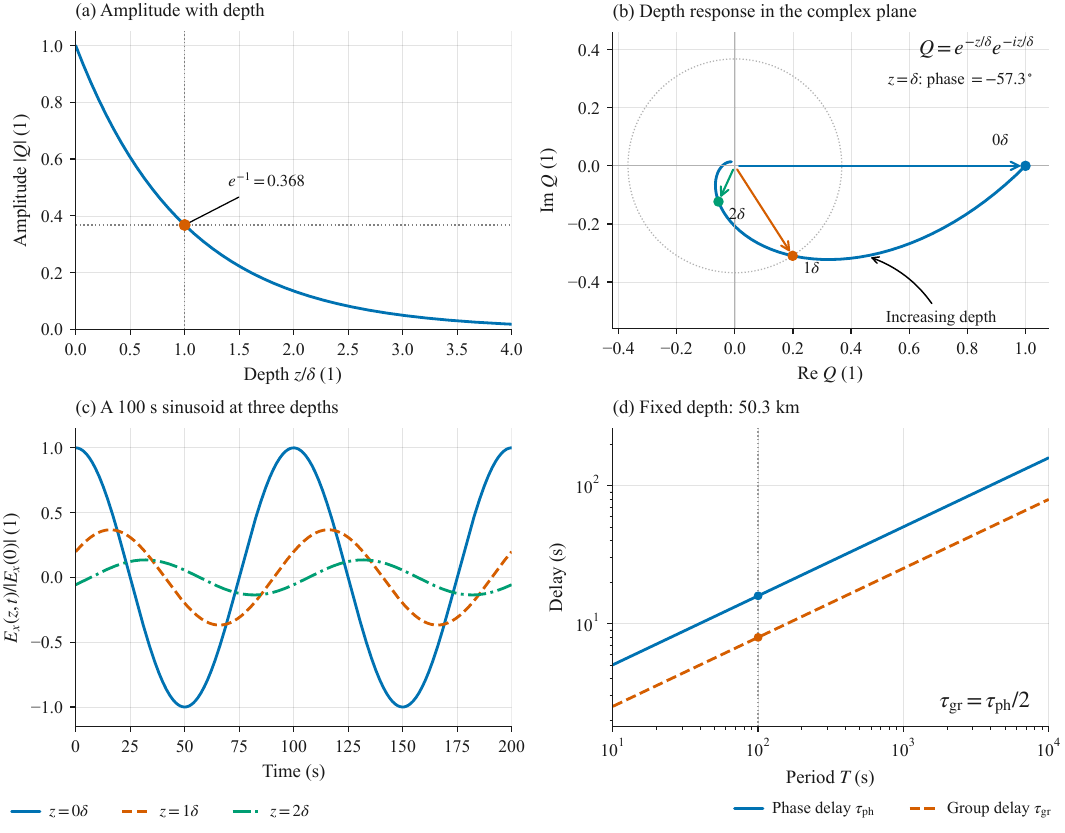}
    \caption{Attenuation, phase rotation, and delay in a uniform \SI{100}{\ohm\metre} half space. (a)~The amplitude of $Q=E_x(z)/E_x(0)$ decreases as $\ee^{-z/\delta}$. (b)~The complex response rotates clockwise and contracts with increasing depth. The arrows end at the responses for $z=0$, $\delta$, and $2\delta$. The dashed circle has radius $\ee^{-1}$. (c)~Real electric fields at those depths for a \SI{100}{\second} sinusoid, divided by the surface amplitude. (d)~Phase and group delays from Equation~\ref{eq:diffusion-delays}, evaluated at a fixed physical depth of \SI{\exampleSkinDepthKm}{\kilo\metre} as period varies. The marked \SI{100}{\second} values are \SI{\examplePhaseDelay}{\second} and \SI{\exampleGroupDelay}{\second}, respectively. The phase used to calculate delay is continuous with frequency.}
    \label{fig:diffusion-delay}
\end{figure}

\begin{proposalcolumns}

\subsection{Phase rotation and delay with depth}
\label{sec:depth-delay}

The half space provides an exact example connecting attenuation, complex phase, and the real time series. Dividing the downward solution by its surface value defines a dimensionless depth response,
\begin{equation}
    Q(z,\omega)=\frac{E_x(z,\omega)}{E_x(0,\omega)}
       =\ee^{-z/\delta}\ee^{-\ii z/\delta}.
    \label{eq:depth-response}
\end{equation}
At depth $z$, $Q$ has magnitude $\ee^{-z/\delta}$ and phase $-z/\delta$. In Figure~\ref{fig:diffusion-delay}b, increasing $z$ moves $Q$ clockwise toward the origin. The arrows end at $Q$ for $z=0$, $\delta$, and $2\delta$. Panel~(c) shows the corresponding time series from Equation~\ref{eq:skin-time}.

Let $\phi_Q=-z/\delta$ denote the continuous phase of $Q$, followed without wrapping at $\pm\pi$. Phase delay and group delay are defined by \citep{smith2007filters}
\begin{equation}
    \tau_{\mathrm{ph}}=-\frac{\phi_Q}{\omega},
    \qquad \tau_{\mathrm{gr}}=-\left.\frac{\partial\phi_Q}{\partial\omega}\right|_z.
    \label{eq:delay-definitions}
\end{equation}
At fixed physical depth and resistivity, $1/\delta=\sqrt{\omega\mu_0/(2\rho)}$. Substituting this relation and differentiating gives
\begin{align}
    \phi_Q&=-z\sqrt{\frac{\omega\mu_0}{2\rho}},\\
    \tau_{\mathrm{ph}}&=\frac{z}{\omega\delta},\qquad
    \tau_{\mathrm{gr}}=\frac{z}{2\omega\delta}=\frac{\tau_{\mathrm{ph}}}{2}.
    \label{eq:diffusion-delays}
\end{align}
\end{proposalcolumns}

\begin{figure}[H]
    \centering
    \includegraphics[width=\textwidth]{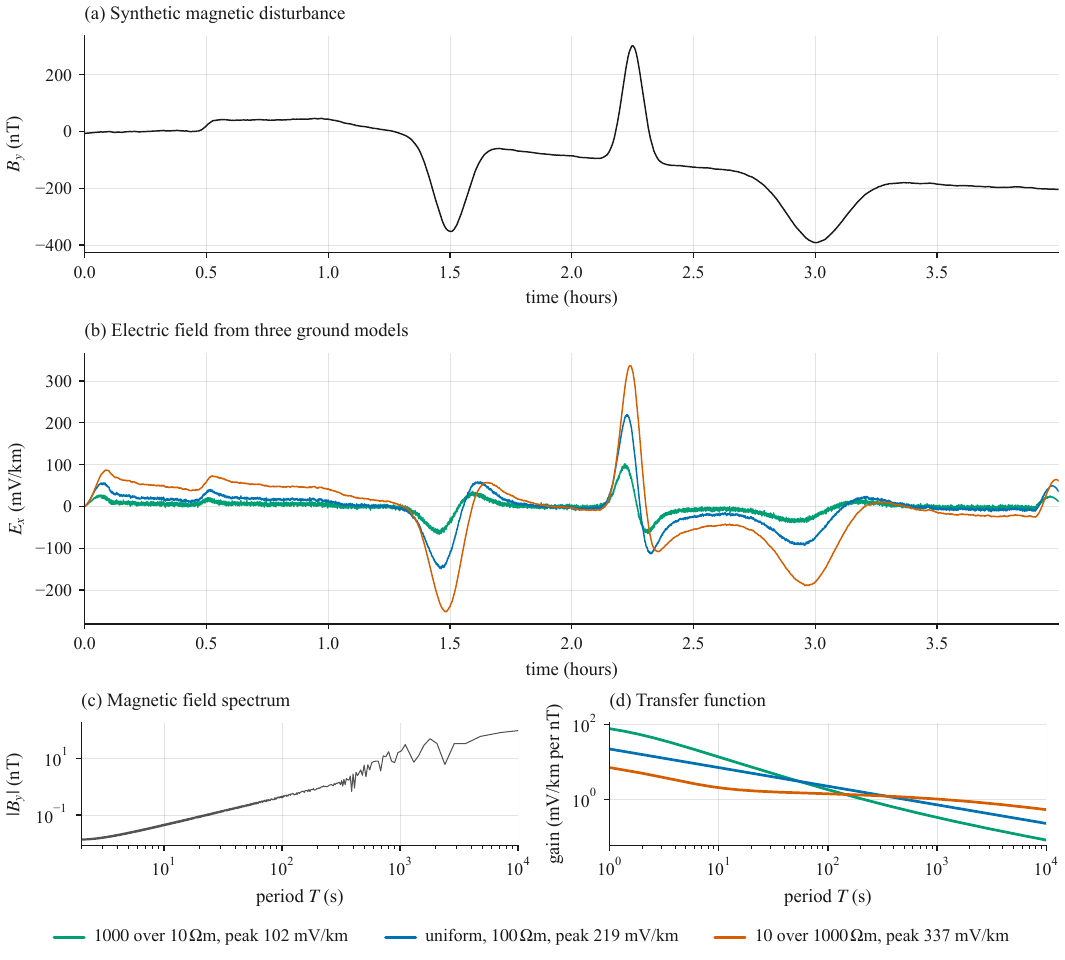}
    \caption{Geoelectric responses calculated with Equation~\ref{eq:timedomain}. (a)~A synthetic horizontal magnetic disturbance containing a sudden commencement, a main phase depression, three substorm pulses, and coloured background noise. (b)~The geoelectric field for three conductivity models subjected to the same magnetic disturbance. Because the impedance rises with frequency, a short pulse in the magnetic field produces a larger electric field than a slow excursion of the same size. (c)~Amplitude spectrum of the magnetic disturbance. (d)~Electric field amplitude per unit magnetic field amplitude, calculated from $|Z|/\mu_0$ and expressed in the plotted units.}
    \label{fig:time}
\end{figure}

\begin{proposalcolumns}

Figure~\ref{fig:diffusion-delay}d holds the depth at \SI{\exampleSkinDepthKm}{\kilo\metre}, one skin depth for \SI{100}{\ohm\metre} at \SI{100}{\second}, while varying period. At that period the phase delay is \SI{\examplePhaseDelay}{\second} and the group delay is \SI{\exampleGroupDelay}{\second}. Differentiating at fixed $z/\delta$ would instead change the physical observation depth with frequency.

The interpretation as an envelope delay requires a sufficiently narrow frequency band in which phase is approximately linear. Amplitude variation over that band can also distort the envelope \citep{smith2007filters}. The delays follow from the diffusion solution and do not define a wavefront arrival time or an interface depth. The depth lag compares $E_x$ at two depths, whereas the positive surface phase in Figure~\ref{fig:responses} compares $E_x$ with $B_y$ at the same location.

The same definition applied to the surface response gives the group delay $-\mathrm{d}\arg Z/\mathrm{d}\omega$ of the impedance. A uniform half space has constant phase, so its group delay is zero, although its amplitude still varies as $\sqrt{\omega}$ and reshapes a disturbance. Layering makes the phase vary with frequency and so introduces a group delay. A storm spans a broad band and cannot be assigned one delay. The waveform comparisons below carry the phase through the complex impedance but do not estimate a timing offset separately.

\section{The geoelectric field in the time domain}
\label{sec:time-domain}

The quantity needed for a hazard calculation is the electric field as a function of time. Because the layered Earth is linear and time invariant, the surface impedance acts as a transfer function and the calculation is a multiplication in the frequency domain,
\begin{equation}
    E_x(t) = \mathcal{F}^{-1}\!\left\{
        Z(\omega)\,\frac{B_y(\omega)}{\mu_0}
    \right\},
    \label{eq:timedomain}
\end{equation}
where $\mathcal{F}$ denotes the Fourier transform and $\mathcal{F}^{-1}$ its inverse. The zero frequency value is $Z(0)=0$, the long period limit of the layered response. For a layered model, $Z$ is evaluated directly at each frequency of the discrete transform. The magnetic record has its mean removed, is multiplied by a cosine taper over the first and last \SI{\taperEachPercent}{\percent} of its length, and is padded with zeros to \padWord{} its length. The transform is taken over nonnegative frequencies, the negative frequencies being implied by the conjugate symmetry of a real series, and the inverse is truncated to the original length. Figure~\ref{fig:time} uses this route.

The frequency dependence of $Z$ determines the amplitude and phase of each component of the electric field. For a uniform half space, Equation~\ref{eq:halfspace} gives $|Z|\propto\sqrt{\omega}$. Figure~\ref{fig:time} illustrates the response to a synthetic disturbance and compares the electric fields for different conductivity models.

For measured tensors, we calculate the electric field using Equation~\ref{eq:tensor}, with the tensor interpolation and the padding of the Fourier method of the bezpy package \citep{bezpy,lucas2020}. A tensor known only at its reported periods is interpolated to the transform frequencies by a smoothing cubic spline in $\log_{10}T$, fitted separately to the real and imaginary parts of each entry with weights equal to the reciprocal square root of its reported variance. The smoothing condition limits the weighted sum of squared residuals to the number of periods. Outside the band of reported periods, each entry continues in proportion to the square root of frequency from its value at the band edge, with the edge phase held. The magnetic record is padded with zeros to a power of two between two and four times its length, without a taper. Every magnetic input to a measured tensor in this paper is restricted to the band of that tensor twice, before padding and again on the padded grid immediately before multiplication by the tensor (Section~\ref{sec:ground}). The second restriction is needed because a record band limited before padding has part of its spectrum outside the band once padded. For the synthetic disturbance, the continuation of the tensor would otherwise change the reference field by a median of \SI{\procLeakMedian}{\percent} and by up to \SI{\procLeakMax}{\percent}. With it, the continuation multiplies no signal. The $E/H$ convention used here requires $\mathbf{H}=\mathbf{B}/\mu_0$. Published responses may instead relate $\mathbf{E}$ directly to $\mathbf{B}$, so their units must be converted before substitution. \citet{kelbert2017} compare Fourier and discrete impulse response methods for measured tensors and validate both against electric fields recorded during storms in Japan. The examples here use the Fourier method.

The resulting electric field provides the forcing for a network calculation. Computing GIC then requires the resistances and topology of the grounded conductors \citep{lehtinen1985}.

\end{proposalcolumns}

\begin{figure}[H]
    \centering
    \includegraphics[width=\textwidth]{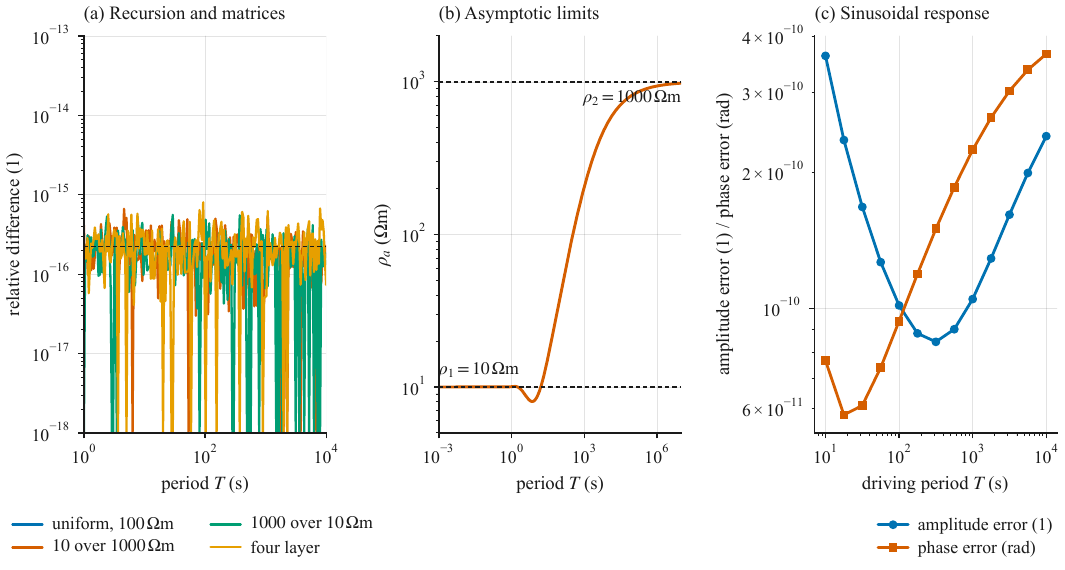}
    \caption{Three numerical checks from Table~\ref{tab:verification}, evaluated across period. (a)~Relative difference between surface impedances calculated by Wait's recursion and by layer matrices for the four models. Differences are comparable to numerical rounding. The dashed line marks double precision machine epsilon, $2^{-52}$. (b)~Apparent resistivity for a \SI{5}{\kilo\metre} layer of \SI{10}{\ohm\metre} over a \SI{1000}{\ohm\metre} half space. The short and long period limits approach the respective material resistivities. (c)~Errors in the electric response to a sinusoidal magnetic field: relative amplitude error and absolute phase error in radians. The reference values are $|Z|B_0/\mu_0$ and $\arg Z$. The calculation in the time domain includes tapering and the finite record length.}
    \label{fig:verification}
\end{figure}

\begin{proposalcolumns}

\section{Numerical verification}
\label{sec:verification}

We verify the numerical calculations against analytic half space solutions, limiting layer configurations, and an independent formulation using layer matrices. We also compare the impedance and the resulting electric field with those from an independent implementation, the bezpy package \citep{bezpy,lucas2020} at version \bezpyVersion{} (repository commit \texttt{\bezpyCommit}), for the four models of Table~\ref{tab:models} at \num{\verifyPeriodCount} periods from \SIrange{\verifyPeriodMin}{\verifyPeriodMax}{\second}. bezpy returns impedance in (\si{\milli\volt\per\kilo\metre})/\si{\nano\tesla}, which is converted to \si{\ohm} by the factor $10^{3}\mu_0$. The two implementations share no code beyond the numerical libraries they call. For the comparison in the time domain, the taper and mean removal of this implementation are disabled and its padding is matched to that of bezpy, so the comparison tests the transfer function alone. Sinusoidal forcing tests the amplitude and phase of the electric response in the time domain. Table~\ref{tab:verification} reports the deviations and tolerances for the \nChecks{} checks. Figure~\ref{fig:verification} shows selected comparisons across period. The tests with finite layer thickness, finite sampling distance, and finite record length include the corresponding approximation errors.

\end{proposalcolumns}

\begin{proposalcolumns}

To isolate the effect of interpolating impedance between measured periods, we evaluate a known layered response in two ways: directly at the Fourier frequencies, and by interpolation from \num{\floorPeriodCount} periods between \SI{10}{\second} and \SI{10000}{\second} using the same procedure as for measured MT tensors. Both responses are applied to the same magnetic time series. The relative differences in peak electric field are \SI{\floorSynthetic}{\percent} for the broadband synthetic disturbance and \SI{\floorStorm}{\percent} for a synthetic disturbance sampled once per minute. These values are the processing discrepancy under the control conditions of one model, one period grid, and two magnetic inputs. They are not a lower bound at every site, so Section~\ref{sec:ground} repeats the control at each site on its own period grid.

\end{proposalcolumns}

\begin{table}[H]
    \centering
    \caption{Verification of the implementation. Every entry compares a computed result against a value obtained independently: in closed form, by a second numerical route through layer matrices, or from an independent implementation. Deviations are relative unless the reference value is an angle, in which case they are absolute. A check is within tolerance when its deviation is finite and no greater than the stated tolerance. These checks assess numerical agreement for the specified configurations.}
    \label{tab:verification}
    \small
    \renewcommand{\arraystretch}{1.18}
    \begin{tabular}{@{}llrrl@{}}
\toprule
Check & Reference & Tolerance & Deviation & Within tolerance \\
\midrule
Half space apparent resistivity & $\rho_a=\rho$ & $10^{-14}$ & $4.1 \times 10^{-16}$ & yes \\
Half space impedance phase & $45^\circ$ & $10^{-12}$ & $0$ & yes \\
Half space C response & $C=\delta(1-i)/2$ & $10^{-14}$ & $2.7 \times 10^{-16}$ & yes \\
Recursion against layer matrices & identical & $10^{-12}$ & $7.0 \times 10^{-16}$ & yes \\
Layer of vanishing thickness & half space below & $10^{-6}$ & $9.1 \times 10^{-7}$ & yes \\
Uniform layer split in two & unsplit half space & $10^{-13}$ & $4.0 \times 10^{-16}$ & yes \\
Layer of 11 skin depths & $\eta_1+O(e^{-2h/\delta})$ & $10^{-8}$ & $2.8 \times 10^{-10}$ & yes \\
Short period limit & $\rho_a\to\rho_1$ & $10^{-9}$ & $7.1 \times 10^{-16}$ & yes \\
Long period limit & $\rho_a\to\rho_N$ & $10^{-2}$ & $2.0 \times 10^{-3}$ & yes \\
Half space depth profile & $e^{-(1+i)z/\delta}$ & $10^{-13}$ & $2.3 \times 10^{-16}$ & yes \\
Surface ratio from the profile & recursion & $10^{-13}$ & $2.2 \times 10^{-16}$ & yes \\
Tangential fields at an interface & continuous & $10^{-6}$ & $10^{-7}$ & yes \\
Driven sinusoid amplitude & $|Z|B_0/\mu_0$ & $10^{-7}$ & $3.6 \times 10^{-10}$ & yes \\
Driven sinusoid phase & $\arg Z$ & $10^{-7}$ & $3.7 \times 10^{-10}$ & yes \\
Impedance comparison & Independent recursion & $10^{-12}$ & $1.1 \times 10^{-15}$ & yes \\
Time series comparison & Independent convolution & $10^{-12}$ & $2.8 \times 10^{-15}$ & yes \\
Field conversion, broadband & identical electric fields & $10^{-2}$ & $6.3 \times 10^{-5}$ & yes \\
Field conversion, 1 min sampling & identical electric fields & $10^{-2}$ & $2.0 \times 10^{-3}$ & yes \\
\bottomrule
\end{tabular}

\end{table}

\begin{proposalcolumns}

\section{Worked example: measured MT responses}
\label{sec:ground}

In this example we quantify, at all \num{\nSites} EarthScope MT sites in the contiguous United States, the discrepancy between the electric field calculated from a measured tensor and from the layered model fitted to it. \citet{kelbert2026usmtarray} describe the survey. The transfer functions are distributed through the EarthScope repository \citep{kelbert_em_2011} and are the reference responses for every fit and comparison below.

\subsection{Field measurements and tensor estimation}
\label{sec:mtmeasurements}

An MT site records the natural variation of the horizontal electric and magnetic fields together. Orthogonal electrode pairs give $E_x$ and $E_y$, and a magnetometer gives the magnetic components. The records are cut into segments and transformed to the frequency domain, and regression across segments estimates the tensor and its uncertainty in each band \citep{kelbert2026usmtarray}. Robust estimators reduce the weight of contaminated segments \citep{egbert1986}, and a simultaneous magnetic reference at another station reduces bias from noise the two stations do not share \citep{gamble1979}. The example uses these published tensors and their reported uncertainties.

\subsection{Data selection and fitting}

Each of the \num{\nSites} sites in the site list is matched by its identifier to exactly one EMTF XML file among \num{\nArchiveFiles} files in \nArchiveBundles{} bundles downloaded from the EarthScope repository on \emtfBundleDate{}. No threshold is applied to the quality rating recorded in each file. Every file declares the $\ee^{+\ii\omega t}$ time dependence used here, an impedance unit of (\si{\milli\volt\per\kilo\metre})/\si{\nano\tesla}, and measurement axes aligned with geographic north, so no rotation is applied. The impedance is converted to \si{\ohm} by multiplying by $10^{3}\mu_0$, with $\mu_0$ in \si{\henry\per\metre}, and the variance by the square of that factor. The analysis uses periods from \SI{10}{\second} to \SI{10000}{\second}. Within this band a period estimate is retained when every tensor entry and its variance are finite and the variance is positive, and when the square root of the complex variance of the impedance projection $s=(Z_{xy}-Z_{yx})/2$ is at most \SI{\errorScreenPercent}{\percent} of $|s|$. A site needs at least \num{\minimumPeriods} retained periods. Every site meets that requirement, with \numrange{\minRetainedPeriods}{\maxRetainedPeriods} retained periods each. Table~\ref{tab:selection} gives the count of sites and period estimates after each stage.

Each file reports a variance for every tensor entry. It is the variance of the complex estimate, the mean squared modulus of its error, so its real and imaginary parts each have variance $\operatorname{VAR}(Z_{ij})/2$ \citep{egbert_zfiles,kelbert_emtffcu}. The variance assigned to $s$ is $\operatorname{VAR}(s)=[\operatorname{VAR}(Z_{xy})+\operatorname{VAR}(Z_{yx})]/4$, which omits the covariance between $Z_{xy}$ and $Z_{yx}$. Appendix~\ref{sec:covariance} verifies the convention against the covariance matrices stored in each file and bounds the omitted term.

The fitted response is $s$, which is unchanged by rotation of the horizontal coordinate axes. For an isotropic layered Earth, $s$ equals its scalar impedance. Rotational invariance does not make $s$ the response of a layered Earth: a measured $s$ need not be matched by any profile of positive conductivities, and the fit reports how closely it can be. Layered profiles are fitted to its apparent resistivity and phase by regularised least squares: the objective combines agreement with the data and a penalty on differences in log resistivity between adjacent layers. The model has \num{\fitLayerCount} layers, including a terminal half space. Its finite interfaces are fixed and logarithmically spaced from \SI{\fitShallowInterfaceM}{\metre} to \SI{\fitDeepInterfaceKm}{\kilo\metre}. Log resistivity is the fitted parameter. The profile is one of many that fit, because the inverse problem has no unique solution \citep{simpson2005}.

Residuals in $\log_{10}\rho_a$ and phase are divided by their standard errors, propagated from $\operatorname{VAR}(s)/2$, with minimum uncertainties of \num{0.02} in $\log_{10}\rho_a$ and \SI{0.5}{\degree} in phase. These minimum values limit the influence of period estimates with very small reported errors, and they set the uncertainty at \SI{\covGroundRhoFloor}{\percent} of retained periods for apparent resistivity and \SI{\covGroundPhaseFloor}{\percent} for phase. The root mean square of these normalised residuals measures the fit. A value of unity indicates residuals comparable to the uncertainties used in fitting. The regularisation weight is decreased over a prescribed sequence until this misfit reaches unity or less. If that target is unattainable, the profile with the lowest misfit among the tested weights is retained and its residual is reported. Table~\ref{tab:inversions} lists the settings. Of the \num{\nSites} fits, \num{\nFitTarget} reach the target misfit, \num{\nFitAboveTwo} exceed a misfit of \num{2}, and the largest misfit is \num{\fitWorstRms}. Fits are retained regardless of misfit. The electric field comparisons apply the period coverage requirements specified below. Each fitted profile is then compared with both off diagonal components and with the electric field calculated from the measured tensor.

\end{proposalcolumns}

\begin{table}[H]
    \centering
    \caption{Sites and period estimates remaining after each stage of selection, over the \num{\nSites} sites of the site list. Period estimates are summed over sites. The last three rows count sites only. The two storm rows differ in the lower period limit of the comparison.}
    \label{tab:selection}
    \small
    \renewcommand{\arraystretch}{1.18}
    \begin{tabular}{@{}lrr@{}}
\toprule
Stage & Sites & Period estimates \\
\midrule
Sites in the site list, one EMTF file each & 1616 & 50266 \\
Periods from \SIrange{10}{10000}{\second} & 1616 & 42016 \\
Finite tensor entries and finite positive variances & 1616 & 42012 \\
Complex standard error of $s$ at most \SI{25}{\percent} of $|s|$ & 1616 & 41901 \\
At least 8 retained periods & 1616 & 41901 \\
Layered fit at or below the target misfit & 1150 &  \\
Recorded storm, periods from \SI{150}{\second} & 1614 &  \\
Recorded storm, periods from \SI{300}{\second} & 1611 &  \\
\bottomrule
\end{tabular}

\end{table}

\begin{figure}[H]
    \centering
    \includegraphics[width=\textwidth]{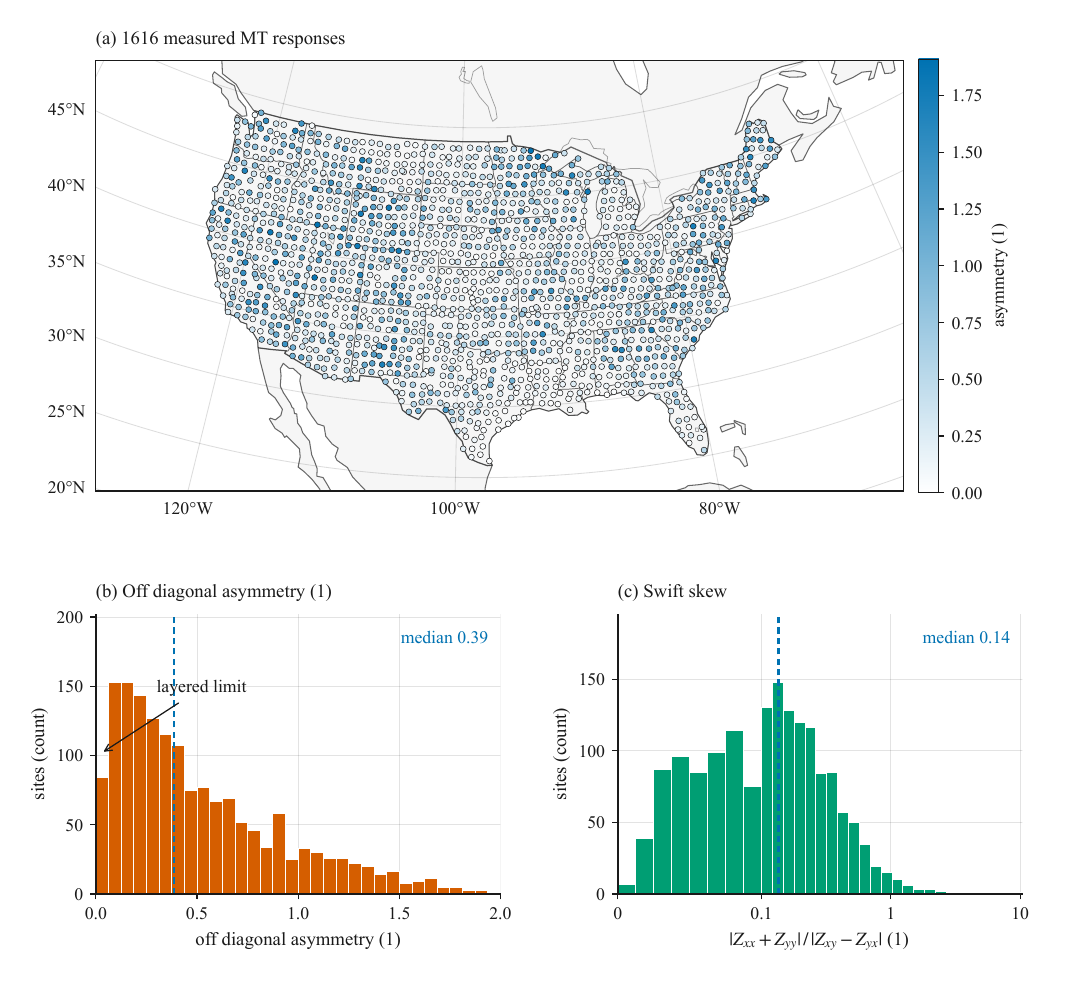}
    \caption{Off diagonal asymmetry and Swift skew at \num{\nSites} EarthScope MT sites. (a)~Sites coloured by the normalised difference between the two off diagonal impedance magnitudes, $\bigl||Z_{xy}|-|Z_{yx}|\bigr|$ divided by their mean, taken as a median over the usable band. A layered Earth gives zero everywhere. (b)~The distribution of that quantity. (c)~Swift skew, $|Z_{xx}+Z_{yy}|/|Z_{xy}-Z_{yx}|$, summarised as the median over usable periods at each site. This statistic is zero for a structure that is ideally one or two dimensional. Its horizontal axis is linear below \num{0.1} and logarithmic above it, retaining all sites. Its interpretation depends on uncertainty and distortion \citep{chave2014invariants}. }
    \label{fig:emtfsites}
\end{figure}

\begin{proposalcolumns}

\subsection{Departure from the layered response}

Figure~\ref{fig:emtfsites} summarises two diagnostics of the measured tensors before fitting. The off diagonal magnitude asymmetry is
\begin{equation}
    a(\omega)=\frac{\bigl||Z_{xy}|-|Z_{yx}|\bigr|}
    {\left(|Z_{xy}|+|Z_{yx}|\right)/2}.
    \label{eq:asymmetry}
\end{equation}
A layered Earth gives $a=0$ because $Z_{yx}=-Z_{xy}$. This statistic compares magnitudes in the measurement coordinates. Phase differences and diagonal entries require separate examination. Each site's value is the median over its usable periods. Across sites the median asymmetry is \num{\medianAsymmetry} and the upper quartile is \num{\upperAsymmetry}. Fewer than half the sites, \num{\nLowAsymmetry} of \num{\nSites}, have asymmetry below \num{0.2}. The map shows the geographical distribution of these departures.

Swift skew, $|Z_{xx}+Z_{yy}|/|Z_{xy}-Z_{yx}|$, vanishes for a structure that is ideally one or two dimensional. Its interpretation requires measurement uncertainty and local distortion to be considered \citep{chave2014invariants}. Figure~\ref{fig:emtfsites}c shows the median skew over usable periods at each site.

\end{proposalcolumns}

\begin{figure}[H]
    \centering
    \includegraphics[width=\textwidth]{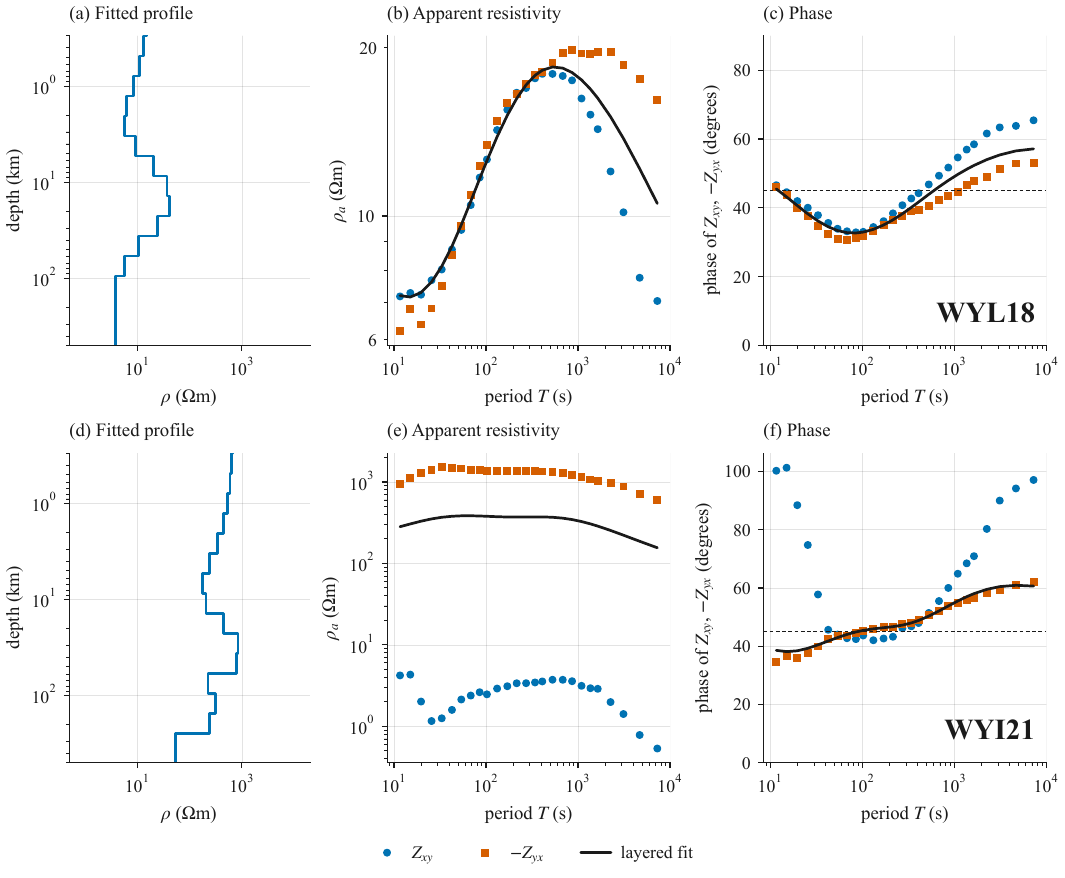}
    \caption{Layered profiles fitted to measured MT tensors at EarthScope sites \fitLowSite{} (upper row) and \fitHighSite{} (lower row). The sites illustrate contrasting off diagonal magnitude asymmetry. Site locations and their coordinate reference system are given in Section~\ref{sec:ground}. (a)~and~(d) the fitted profiles. (b)~and~(e) the measured apparent resistivity of each off diagonal entry, compared with the same fitted layered impedance. (c)~and~(f) the phases of $Z_{xy}$ and $-Z_{yx}$. Reversing the sign of $Z_{yx}$ accounts for the relation $Z_{yx}=-Z_{xy}$ in a layered Earth and permits direct comparison of the phases. The fitted curve represents one conductivity profile obtained by inversion. Additional constraints are needed to infer a unique conductivity structure.}
    \label{fig:emtffit}
\end{figure}

\begin{proposalcolumns}

Figure~\ref{fig:emtffit} compares measured MT tensors and fitted layered responses at two EarthScope survey sites in Wyoming. Site \fitLowSite{} (\fitLowName{}) is at latitude \ang{\fitLowLatitude} and longitude \ang{\fitLowLongitude}. Site \fitHighSite{} (\fitHighName{}) is at latitude \ang{\fitHighLatitude} and longitude \ang{\fitHighLongitude}. These positions are geographic coordinates in the World Geodetic System 1984 (WGS84), as reported in the site descriptions \citep{kelbert_em_2011}. The sites illustrate low and high off diagonal magnitude asymmetry, respectively. Each fitted layered impedance is evaluated against both measured off diagonal components. A layered model requires equal off diagonal magnitudes and equal phases after reversing the sign of $Z_{yx}$.

\end{proposalcolumns}

\begin{figure}[H]
    \centering
    \includegraphics[width=\textwidth]{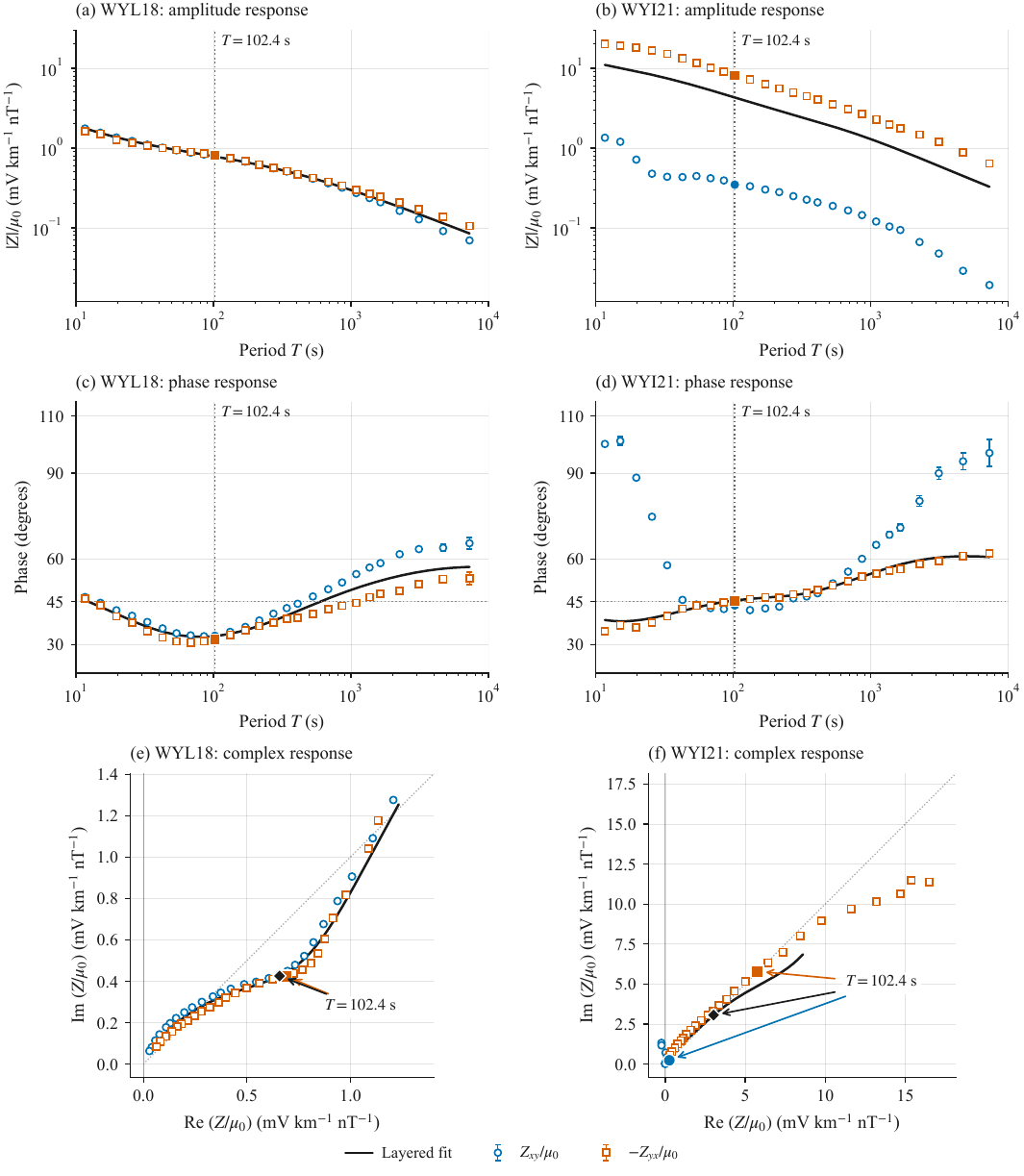}
    \caption{Frequency responses estimated from field measurements at WYL18 (left) and WYI21 (right), with coordinates given in Section~\ref{sec:ground}. (a,b)~Ratio of electric to magnetic amplitude. (c,d)~Phase. (e,f)~Complex response. Blue circles show $Z_{xy}/\mu_0$ and orange squares show $-Z_{yx}/\mu_0$ at all \num{\measuredPeriodCount} retained periods. Black curves give the layered fits to $s=(Z_{xy}-Z_{yx})/2$. Error bars show the standard errors defined in the text. Dotted vertical lines mark $T=\SI{\measuredReferencePeriod}{\second}$. At that period, the measured responses are a filled blue circle and orange square, and the fitted response is a black diamond. Arrows point to these three values in (e,f). Amplitude and phase limits are shared across sites. Limits in the complex plane differ between sites, with equal real and imaginary scaling within each panel. The two sites have contrasting off diagonal asymmetry. Data from the EarthScope EMTF repository \citep{kelbert_em_2011}, at sites of the USArray transportable array \citep{schultz2006usarray}.}
    \label{fig:measured-response}
\end{figure}

\begin{proposalcolumns}

Figure~\ref{fig:measured-response} expresses the same measured tensors as ratios of electric to magnetic amplitude, phases, and complex responses. Both sites retain \num{\measuredPeriodCount} period estimates, spanning \SIrange{\measuredPeriodMin}{\measuredPeriodMax}{\second}. The observations are plotted at their reported periods. At $T=\SI{\measuredReferencePeriod}{\second}$, the filled blue circle and orange square are measured responses, and the black diamond is the layered fit.

For an off diagonal component, $u_{ij}=\sqrt{\operatorname{VAR}(Z_{ij})/2}$ denotes the standard error of each part of the complex estimate in the $E/H$ convention. The amplitude and approximate phase standard errors are $u_{ij}/\mu_0$ and $u_{ij}/|Z_{ij}|$ radians, respectively. Joint confidence regions require the full error covariance. The transfer functions are from the USArray transportable array \citep{schultz2006usarray}.

\end{proposalcolumns}

\begin{proposalcolumns}

\subsection{Discrepancies in the geoelectric field}

Table~\ref{tab:emtfstudy} summarises the discrepancies between electric fields calculated from the measured tensors and from the fitted layered models. The field from the measured tensor is the reference. At each site the same magnetic input is applied to both ground responses by the method of Section~\ref{sec:time-domain}. The input is restricted to the retained band of the site by a raised cosine window in $\log_{10}T$ whose tapers extend \num{\bandEdgeDecades} decades inside each band limit, after removal of its mean, and the same window is applied again to its padded spectrum (Section~\ref{sec:time-domain}). Two inputs are used. The synthetic disturbance has two independent horizontal components of $2^{\syntheticExponent}$ samples at \SI{\syntheticIntervalS}{\second}, prescribed at each site, so it requires no spatial interpolation of magnetic observations. The recorded storm of 10 and 11 May 2024 is reconstructed at each site from the observatory network by the method of Section~\ref{sec:source} and sampled once per minute. The comparison under the storm therefore uses periods of \SI{150}{\second} and longer, above the \SI{120}{\second} Nyquist period of that sampling, and requires the longest retained period of a site to be at least four times that limit. For this comparison, \refusalClause. It therefore uses \num{\nStormSites} sites. All \num{\nSites} sites remain in the tensor diagnostics, the layered fits, and the comparison under the synthetic disturbance.

Three metrics compare the two fields. The peak error is the absolute difference between the maxima of the horizontal magnitude $(E_x^2+E_y^2)^{1/2}$, divided by the reference maximum. The error in horizontal magnitude is the root mean square difference between the two magnitudes over the record, divided by the root mean square reference magnitude. Neither depends on direction, and a field reversed at every instant has no error in either. The vector error,
\begin{equation}
    \varepsilon_{\mathrm{vec}}=\left[\frac{\sum_t\left|\mathbf{E}_{\mathrm{pred}}(t)-\mathbf{E}_{\mathrm{ref}}(t)\right|^2}{\sum_t\left|\mathbf{E}_{\mathrm{ref}}(t)\right|^2}\right]^{1/2},
    \label{eq:vector-error}
\end{equation}
sums over the samples of both horizontal components, where $\mathbf{E}_{\mathrm{pred}}$ is the field from the fitted model and $\mathbf{E}_{\mathrm{ref}}$ the reference. It responds to rotation and reversal of the field, which change the voltage induced along a line, and it is never smaller than the error in horizontal magnitude.

The discrepancy has three sources, which three intermediate calculations separate under identical processing. The projected response is the antisymmetric projection $\mathbf{Z}_{\mathrm{a}}=(\mathbf{Z}-\mathbf{Z}^{\mathsf{T}})/2$ of the interpolated measured tensor, with diagonal entries zero and off diagonal entries $s$ and $-s$. Its field differs from the reference only through this dimensional reduction. The projection is taken after interpolation, which keeps it exactly antisymmetric. Projecting before interpolation would let the separate smoothing of the two off diagonal entries, whose reported variances differ, change the field by a median of \SI{\procAntisymMedian}{\percent} and by up to \SI{\procAntisymMax}{\percent}. For the sampled fit, a tensor with zero diagonal entries and off diagonal entries $Z_{\mathrm{fit}}$ and $-Z_{\mathrm{fit}}$, where $Z_{\mathrm{fit}}$ is the fitted layered impedance, is constructed at the retained periods. Its entries are interpolated with the reported variances of the corresponding measured entries, after which the antisymmetric projection is applied. Its difference from the projected field is due to the inversion misfit. The directly evaluated layered response is computed at every transform frequency, so its difference from the sampled fit is due to interpolation between periods. The successive field differences sum to the total field difference. Their normalised norms, each relative to the reference of Equation~\ref{eq:vector-error}, are reported separately because cancellation and reinforcement affect the norm of the sum.

Under the synthetic disturbance the median vector error is \SI{\syntheticVectorMedian}{\percent} and the ninetieth percentile is \SI{\syntheticVectorNinety}{\percent}. The median peak error is \SI{\syntheticPeakMedian}{\percent} and the median error in horizontal magnitude is \SI{\syntheticMagnitudeMedian}{\percent}. Under the recorded storm the corresponding medians are \SI{\stormVectorMedian}{\percent}, \SI{\stormPeakMedian}{\percent}, and \SI{\stormMagnitudeMedian}{\percent}, so the two inputs give similar median peak errors. The dimensional reduction alone gives median vector errors of \SI{\syntheticReductionMedian}{\percent} and \SI{\stormReductionMedian}{\percent} for the two inputs, the inversion misfit \SI{\syntheticMisfitMedian}{\percent} and \SI{\stormMisfitMedian}{\percent}, and processing \SI{\syntheticProcessingMedian}{\percent} and \SI{\stormProcessingMedian}{\percent}. Tensor reduction produces the largest median vector discrepancy. The median discrepancies associated with the layered fit and numerical processing are substantially smaller. The processing term is the control of Section~\ref{sec:verification} repeated at each site on its own period grid. Its ninetieth percentile is \SI{\syntheticProcessingNinety}{\percent} under the synthetic disturbance and \SI{\stormProcessingNinety}{\percent} under the storm. The median peak error under the recorded storm exceeds the control processing discrepancy of \SI{\floorStorm}{\percent}.

The lower period limit of the storm comparison is a choice. Raising it to \SI{300}{\second} requires a longest retained period of at least four times that limit, \SI{1200}{\second}. This criterion retains \num{\nLateStormSites} sites, with a median vector error of \SI{\lateStormVectorMedian}{\percent} and a median peak error of \SI{\lateStormPeakMedian}{\percent}.

\end{proposalcolumns}

\begin{table}[H]
    \centering
    \caption{Percentiles across MT sites of tensor diagnostics, fit misfits, and discrepancies in the electric field. All entries are dimensionless. Off diagonal magnitude asymmetry is defined in Equation~\ref{eq:asymmetry}. Misfits to $Z_{xy}$ and $Z_{yx}$ are median relative differences over period. The weighted fit misfit is the root mean square of apparent resistivity and phase residuals divided by the uncertainties used in fitting, including the stated minimum uncertainties. A value of unity indicates residuals comparable to those uncertainties. Field discrepancies are relative to the field from the measured tensor. The vector error is Equation~\ref{eq:vector-error}, and the indented rows below it give the dimensional reduction, the inversion misfit, and processing, each normalised by the same reference.}
    \label{tab:emtfstudy}
    \small
    \renewcommand{\arraystretch}{1.18}
    \begin{tabular}{@{}lrrrr@{}}
\toprule
Quantity & 25th & Median & 75th & 90th \\
\midrule
\multicolumn{5}{@{}l}{\emph{Measured tensors and layered fits, 1616 sites}} \\
\quad off diagonal magnitude asymmetry & 0.19 & 0.39 & 0.73 & 1.14 \\
\quad Swift skew & 0.07 & 0.14 & 0.25 & 0.45 \\
\quad weighted fit misfit & 0.86 & 0.94 & 1.08 & 2.19 \\
\quad relative misfit to $Z_{xy}$ & 0.12 & 0.22 & 0.38 & 0.76 \\
\quad relative misfit to $Z_{yx}$ & 0.13 & 0.23 & 0.39 & 0.83 \\
\addlinespace
\multicolumn{5}{@{}l}{\emph{Synthetic disturbance, 1616 sites}} \\
\quad peak error & 0.099 & 0.197 & 0.336 & 0.494 \\
\quad error in horizontal magnitude & 0.176 & 0.276 & 0.402 & 0.548 \\
\quad vector error, fitted model against measured tensor & 0.294 & 0.425 & 0.588 & 0.944 \\
\quad \quad projection against measured tensor & 0.289 & 0.421 & 0.583 & 0.944 \\
\quad \quad sampled fit against projection & 0.014 & 0.021 & 0.035 & 0.061 \\
\quad \quad direct evaluation against sampled fit & 0.008 & 0.013 & 0.022 & 0.033 \\
\addlinespace
\multicolumn{5}{@{}l}{\emph{Recorded storm, periods from \SI{150}{\second}, 1614 sites}} \\
\quad peak error & 0.095 & 0.197 & 0.318 & 0.422 \\
\quad error in horizontal magnitude & 0.211 & 0.303 & 0.407 & 0.480 \\
\quad vector error, fitted model against measured tensor & 0.322 & 0.461 & 0.604 & 0.732 \\
\quad \quad projection against measured tensor & 0.321 & 0.459 & 0.604 & 0.729 \\
\quad \quad sampled fit against projection & 0.013 & 0.019 & 0.029 & 0.052 \\
\quad \quad direct evaluation against sampled fit & 0.008 & 0.013 & 0.019 & 0.028 \\
\addlinespace
\multicolumn{5}{@{}l}{\emph{Recorded storm, periods from \SI{300}{\second}, 1611 sites}} \\
\quad peak error & 0.110 & 0.204 & 0.331 & 0.425 \\
\quad vector error, fitted model against measured tensor & 0.327 & 0.465 & 0.606 & 0.739 \\
\bottomrule
\end{tabular}

\end{table}

\begin{figure}[H]
    \centering
    \includegraphics[width=\textwidth]{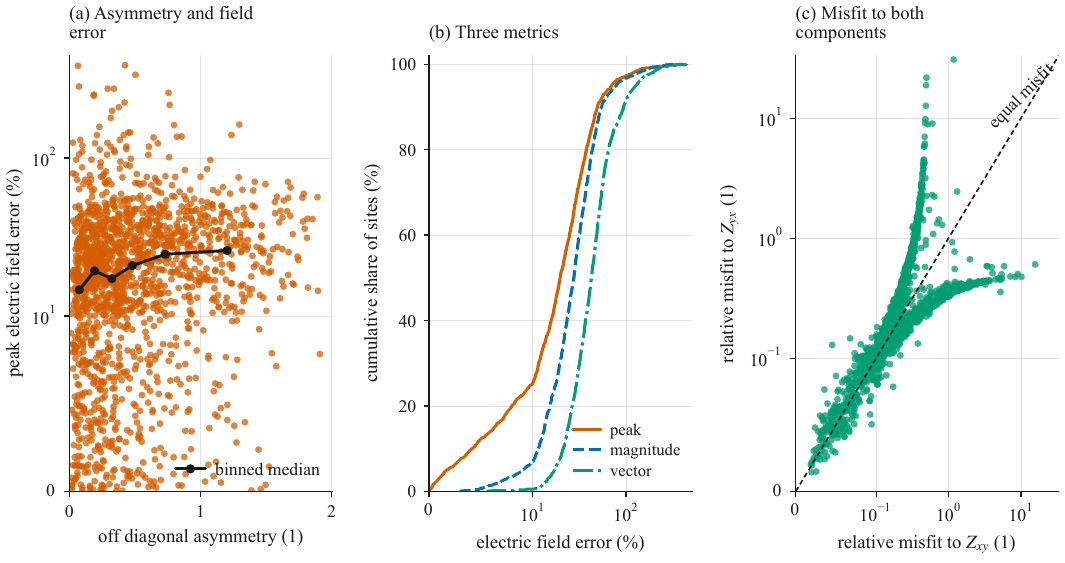}
    \caption{Effect of replacing the measured tensor with a layered response. (a)~Relative error in the peak horizontal electric field against off diagonal asymmetry, with the median within each bin. (b)~Cumulative distributions over all \num{\nSites} sites of the peak error, the error in horizontal magnitude, and the vector error of Equation~\ref{eq:vector-error}. (c)~Relative misfit of the fitted response to each off diagonal component. All sites are retained. Error axes use a linear interval near zero and logarithmic scaling beyond it. The electric fields are calculated using the same synthetic magnetic disturbance for both conductivity responses at each site. Figure~\ref{fig:source} places the same comparison under the recorded storm beside the magnetic reconstruction.}
    \label{fig:emtfconsequence}
\end{figure}

\begin{figure}[H]
    \centering
    \includegraphics[width=\textwidth]{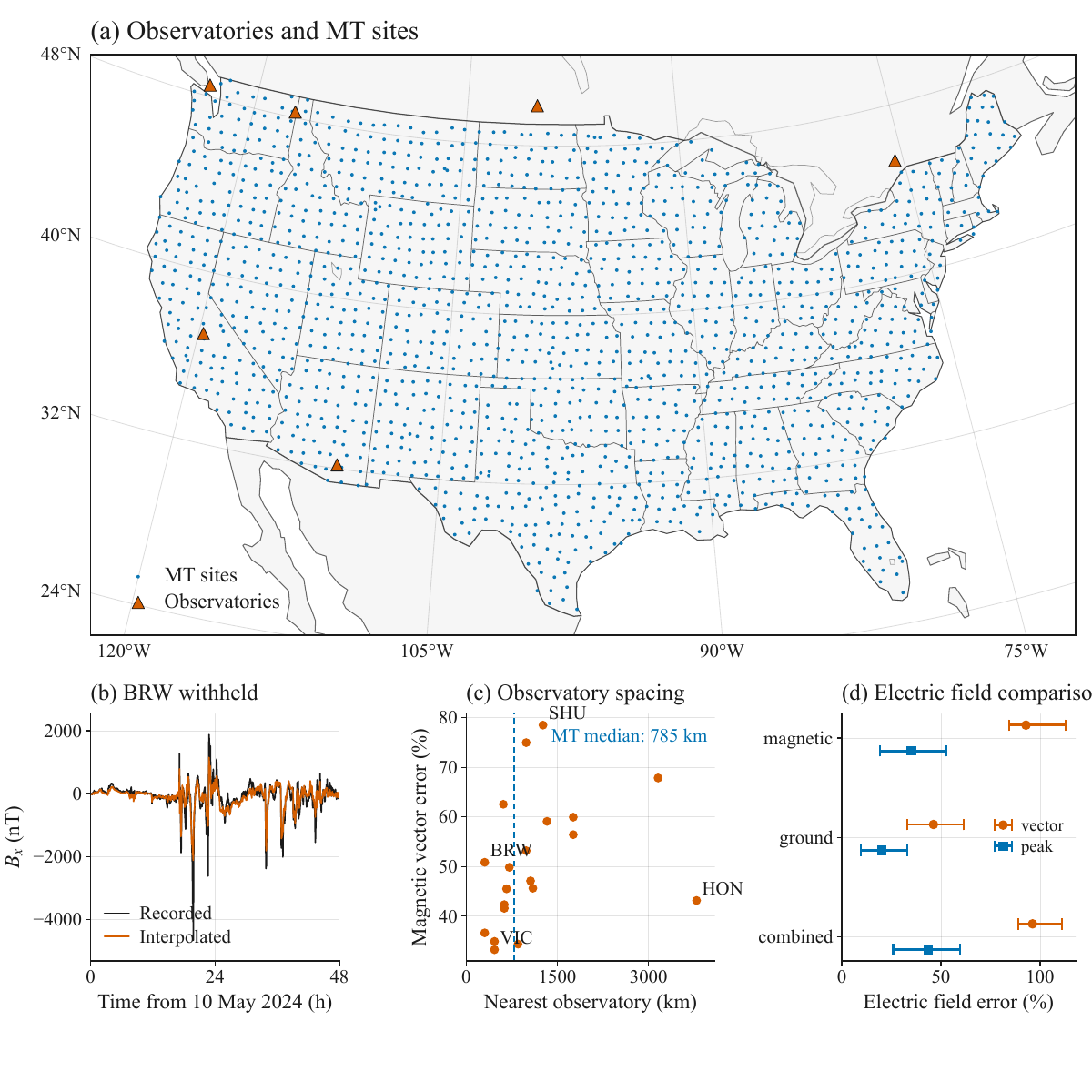}
    \caption{Magnetic field reconstruction during the storm of May 2024. (a)~MT sites and magnetic observatories over the contiguous United States. Interpolation uses the full observatory network, including stations outside the map. (b)~The northward field at the station with the largest horizontal disturbance, withheld and predicted from the remaining network. (c)~Magnetic vector error over the full record against distance to the nearest remaining observatory, with selected stations labelled and the median distance from an MT site to its nearest observatory marked. (d)~Median and interquartile range of the vector error (circles) and peak error (squares) in the electric field over all pairs of measured tensor and withheld observatory, for the magnetic reconstruction, the layered ground representation, and their combination (Table~\ref{tab:budget}). Every term is relative to the field from the recorded magnetic field and the measured tensor. Median distances to the nearest observatory are \SI{\observatoryKm}{\kilo\metre} for the withheld observatories and \SI{\siteKm}{\kilo\metre} for the MT sites.}
    \label{fig:source}
\end{figure}

\begin{proposalcolumns}

Figure~\ref{fig:emtfconsequence} relates the field discrepancies to off diagonal asymmetry, compares the three metrics, and compares the misfit to each impedance component. The component misfits measure the disagreement with each observed impedance, and the field discrepancies measure the resulting difference in the calculated electric field.

\section{Worked example: magnetic interpolation}
\label{sec:source}

For the recorded storm, geoelectric field estimation requires the magnetic disturbance at each MT site. Figure~\ref{fig:source}a shows that the observatories and MT sites occupy different locations. The magnetic field at each MT site is therefore reconstructed from surrounding observations. This reconstruction introduces an additional source of uncertainty. \citet{bonner2017} estimate local magnetic fields from remote observatories using transfer functions between stations, then apply measured MT tensors to obtain electric fields. We use a spatial current representation for the magnetic interpolation.

\citet{wilkerson2026gic} assembled magnetic and GIC records from the event and compared measured currents with model estimates. We evaluate the magnetic reconstruction by withholding individual observatory records. This tests the reconstruction of the recorded field. It does not test the plane wave approximation, which would require a comparison at the scale of the source or with a finite source.

\subsection{Reconstructing the magnetic field}

The magnetic interpolation uses spherical elementary current systems (SECS), a set of current patterns whose magnetic fields can be calculated at both observatories and target locations \citep{amm1999}, as implemented in version \pysecsVersion{} of the pysecs package \citep{pysecs}. The patterns have zero divergence and represent currents that close within a spherical sheet \SI{\secsHeightKm}{\kilo\metre} above the ground. The current systems are located at the nodes of a grid of \num{\secsLatitudes} latitudes from \ang{\secsLatMin} to \ang{\secsLatMax}~N and \num{\secsLongitudes} longitudes from \ang{\secsLonWest} to \ang{\secsLonEast}~W, \num{\secsSystems} systems in all. Their amplitudes are fitted separately at each minute to the horizontal components recorded at \num{\nObservatories} observatories through the storm of 10 and 11 May 2024, with equal weights, and the fitted amplitudes then determine the field at other locations. There are more unknown current amplitudes than observed field components. The fit therefore uses a singular value decomposition that discards singular values below \num{\secsEpsilon} of the largest, which suppresses combinations of amplitudes that the observations constrain poorly. The vertical component is excluded, because at ground level it carries a large contribution induced within the Earth that a sheet current above the ground does not represent. Internal induction also contributes to the measured horizontal field. The fitted sheet is an equivalent magnetic representation and does not independently separate external and internal contributions.

The records are IAGA-2002 files of one minute values from the United States Geological Survey and the Geological Survey of Canada. Of the \num{\nObservatories} stations, \obsDefinitive{} provide definitive data, \obsQuasiDefinitive{} quasi-definitive data, \obsProvisional{} provisional data, and \num{\obsVariation} variation data. The median over the two days is removed from each component at each station, which removes any constant offset whatever the data type. Gaps of up to \num{\gapSamples} samples are bridged linearly, and a station with more than \SI{\gapPercent}{\percent} of its samples missing, or with a longer gap, is not used.

Each station is withheld in turn and its magnetic record predicted from the remaining observatories. The grid, the height, and the cutoff are fixed in advance. The cutoff is the default of the pysecs package, and none of the three was adjusted using the withheld records. Stations outside the domain of the current sheet contribute to the field reconstruction but are excluded as validation targets, leaving \num{\nObservatoriesTested} withheld stations. The magnetic vector error is Equation~\ref{eq:vector-error} applied to the two horizontal magnetic components over the full record. The peak error compares the maxima of the horizontal magnitude. The electric comparison of Section~\ref{sec:ground} uses only periods of \SI{150}{\second} and longer, so the vector error is also computed after the same restriction, from \SI{150}{\second} to \SI{10000}{\second}. Its median is \SI{\sourceVectorMedian}{\percent} for the full record and larger in that band, \SI{\sourceBandVectorMedian}{\percent}. Figure~\ref{fig:source}b shows one such prediction at the Barrow observatory (station code \withheldCode{}), where the recorded horizontal field peaks at \SI{\withheldPeak}{\nano\tesla}. The interpolated peak of \SI{\withheldPredictedPeak}{\nano\tesla} falls short of the recorded value.

\end{proposalcolumns}

\begin{proposalcolumns}

\subsection{Station results}

Table~\ref{tab:sourcestudy} reports magnetic reconstruction errors at the withheld observatories. Figure~\ref{fig:source}c shows the vector error at individual stations. For the quarter of withheld stations with the shortest distances to a remaining observatory, the largest vector error is \SI{\closeVectorWorst}{\percent}. The shortest distance in this group is \SI{\closeKm}{\kilo\metre}. At station \farCode{}, the most isolated station at \SI{\farKm}{\kilo\metre}, it is \SI{\farVector}{\percent} for this storm. Distance alone therefore does not order the errors, which also depend on the spatial structure of the disturbance.

\end{proposalcolumns}

\begin{table}[H]
    \centering
    \caption{Magnetic reconstruction at the withheld observatories during the storm of 10 and 11 May 2024, with percentiles across stations. Errors are dimensionless relative values, so \num{0.2} corresponds to \SI{20}{\percent}. The vector error is Equation~\ref{eq:vector-error} applied to the horizontal magnetic field. The last row gives, for comparison of isolation, the distance from each MT site in the storm comparison to its nearest observatory.}
    \label{tab:sourcestudy}
    \small
    \renewcommand{\arraystretch}{1.18}
    \begin{tabular}{@{}lrrr@{}}
\toprule
& 25th & Median & 75th \\
\midrule
\multicolumn{4}{@{}l}{\emph{Magnetic field at 20 withheld observatories}} \\
\quad peak error & 0.15 & 0.24 & 0.38 \\
\quad vector error, full record & 0.42 & 0.48 & 0.59 \\
\quad vector error, periods from \SI{150}{\second} & 0.68 & 0.80 & 1.02 \\
\quad distance to nearest remaining observatory (\si{\kilo\metre}) & 623 & 920 & 1280 \\
\addlinespace
\multicolumn{4}{@{}l}{\emph{1614 MT sites in the recorded storm comparison}} \\
\quad distance to nearest observatory (\si{\kilo\metre}) & 486 & 785 & 1177 \\
\bottomrule
\end{tabular}

\end{table}

\begin{proposalcolumns}

\subsection{Electric field comparison of magnetic reconstruction and layered representation}

The magnetic errors are in the magnetic field and the representation discrepancies of Section~\ref{sec:ground} are in the electric field, so the two sets of percentages cannot be compared directly. To place them on one scale, we pass the recorded and the reconstructed magnetic fields at each withheld observatory through each of the \num{\budgetTensors} measured tensors of the storm comparison, and through the layered model fitted to each tensor. The reference is the electric field from the recorded magnetic field and the measured tensor. The magnetic term uses the reconstructed field with the measured tensor, the representation term uses the recorded field with the layered model, and the combined calculation uses the reconstructed field with the layered model. Each magnetic input is restricted to the band of the tensor from \SI{150}{\second}, as in Section~\ref{sec:ground}, and every term uses the vector error of Equation~\ref{eq:vector-error}. The tensors were measured elsewhere, so each pair combines a record and a ground response from different places. The \num{\budgetPairs} observatory--tensor pairs reuse every record and every tensor, and they compare the two terms under common ground responses.

Table~\ref{tab:budget} gives the result. The median vector error is \SI{\budgetMagneticMedian}{\percent} for the magnetic term, \SI{\budgetGroundMedian}{\percent} for the representation term, and \SI{\budgetChainMedian}{\percent} for the combined calculation. The median peak errors are \SI{\budgetMagneticPeakMedian}{\percent}, \SI{\budgetGroundPeakMedian}{\percent}, and \SI{\budgetChainPeakMedian}{\percent}. The magnetic term is the larger in \SI{\budgetExceedPercent}{\percent} of pairs, and the combined calculation is closer to the magnetic term than to the representation term. The magnetic term is measured at withheld observatories a median \SI{\observatoryKm}{\kilo\metre} from the nearest remaining observatory, whereas the MT sites lie a median \SI{\siteKm}{\kilo\metre} from the nearest observatory. The magnetic term is therefore measured under greater isolation than a typical MT site, by a factor of \num{\isolationRatio} in median distance.

\end{proposalcolumns}

\begin{table}[H]
    \centering
    \caption{Electric field comparison of the magnetic reconstruction and the layered representation, over \num{\budgetPairs} observatory--tensor pairs formed from \num{\budgetTensors} measured tensors and \num{\budgetObservatories} withheld observatories, for the storm of 10 and 11 May 2024. Every electric field is compared with the field from the recorded magnetic field and the measured tensor, over the band of that tensor from \SI{150}{\second}. Errors are dimensionless relative values. The withheld observatories lie a median \SI{\observatoryKm}{\kilo\metre} from the nearest remaining observatory, and the MT sites a median \SI{\siteKm}{\kilo\metre} from the nearest observatory.}
    \label{tab:budget}
    \small
    \renewcommand{\arraystretch}{1.18}
    \begin{tabular}{@{}lllrrrrrr@{}}
\toprule
& & & \multicolumn{3}{c}{Vector error} & \multicolumn{3}{c}{Peak error} \\
\cmidrule(lr){4-6}\cmidrule(l){7-9}
Term & Magnetic field & Ground response & 25th & Median & 75th & 25th & Median & 75th \\
\midrule
magnetic reconstruction & reconstructed & measured tensor & 0.84 & 0.93 & 1.13 & 0.19 & 0.35 & 0.53 \\
conductivity representation & recorded & layered fit & 0.33 & 0.46 & 0.61 & 0.10 & 0.20 & 0.33 \\
both combined & reconstructed & layered fit & 0.89 & 0.96 & 1.11 & 0.26 & 0.43 & 0.60 \\
\bottomrule
\end{tabular}

\end{table}

\begin{figure}[H]
    \centering
    \includegraphics[width=\textwidth]{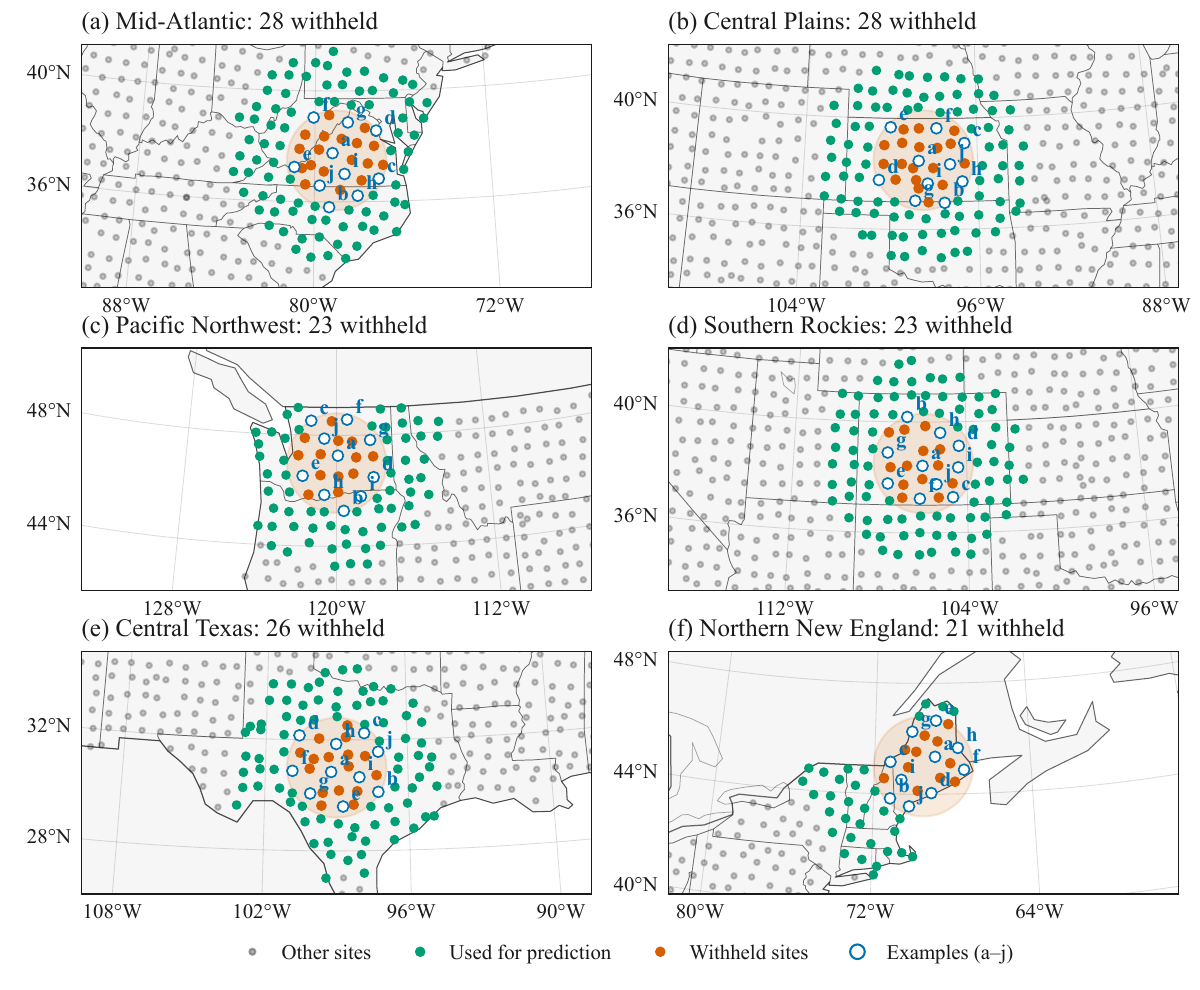}
    \caption{Regional tests of conductivity interpolation. Orange circles bound gaps of \SI{\krigeRadius}{\kilo\metre} radius, and orange points mark all withheld sites. Green points are the union of retained sites used for predictions within each gap. Each target uses its nearest \num{\krigeNeighbours} eligible sites. Grey points show other observations. Blue outlines and letters identify the examples in Appendix~\ref{sec:krigingdetails}, selected using coordinates without reference to prediction errors. All maps have equal frames and the same projected width and height.}
    \label{fig:kriginggaps}
\end{figure}

\begin{proposalcolumns}

The peak and vector metrics describe the reconstructed fields. Their consequences for equipment exposure require a network calculation and a model of the equipment response.

\section{Conductivity between MT sites}
\label{sec:kriging}

At a location without MT observations, conductivity must be estimated from information elsewhere. One approach fits layered conductivity profiles at measured sites and interpolates their logarithms in space. Kriging estimates an unobserved quantity by a linear combination of retained estimates, with weights chosen to minimise prediction variance subject to unbiasedness under the assumed mean model \citep{wackernagel2003}. A Gaussian process formulation gives the corresponding predictive mean and covariance \citep{rasmussen2006}. Propagating uncertainty from both inversion and spatial interpolation has precedent in airborne electromagnetic resistivity modelling \citep{pryet2011}. We apply that approach below to the \num{\krigeSites} MT sites and test the resulting impedances against withheld observations.

All \num{\krigeSites} sites contribute to the evaluation. At each site a profile with \num{\krigeLayers} layers is fitted to the impedance projection $s$ at the retained periods of that site. We call this the local fit. The projection retains the part of the measured tensor that a layered Earth can represent. The local fit differs from the fit of Section~\ref{sec:ground} in its data, parameter, error model, and penalty (Table~\ref{tab:inversions}). The first inversion was used to evaluate fit misfit and electric field discrepancies. The second supplies local modes and covariance estimates for spatial prediction under a common prior specification, with the same prior weights at every site. Independent Gaussian errors are assumed for the real and imaginary parts of $s$, with standard deviations specified by the reported variances and the stated uncertainty floor, which avoids the linearisation needed for apparent resistivity and phase. Every spatial prediction below is compared with the local fit from this same procedure. A prediction target is a withheld site at which the impedance is predicted, and the retained sites are the sites not withheld, whose local fits supply the data for kriging. Let $\mathbf{r}_p$ denote the position of retained site $p$ and $\mathbf{r}_*$ the prediction location. The profile vector $\mathbf{m}_p$ contains one dimensionless log conductivity per layer,
\begin{equation}
    m_{pn}=\log_{10}\!\left[\frac{\sigma_n(\mathbf{r}_p)}{\sigma_{\mathrm{ref}}}\right],
    \qquad \sigma_{\mathrm{ref}}=\SI{1}{\siemens\per\metre}.
    \label{eq:log-conductivity}
\end{equation}
The profile has $N$ layers, indexed by $n=1,\ldots,N$. A hat denotes an estimated profile. For $M$ retained sites, ordinary multivariate kriging has the form
\begin{equation}
    \widehat{\mathbf{m}}_*=\sum_{p=1}^{M}\mathbf{W}_p\widehat{\mathbf{m}}_p,
    \qquad \sum_{p=1}^{M}\mathbf{W}_p=\mathsf{I}_N.
    \label{eq:kriging-weights}
\end{equation}
Each $N\times N$ weight matrix $\mathbf{W}_p$ can combine information across depths. The identity constraint preserves any spatially constant mean profile \citep{wackernagel2003}. The weights depend on site separations, covariance between depths, and uncertainty in the retained inversions. This example uses the nearest \num{\krigeNeighbours} retained sites and treats each local fit as an observation of the latent profile at its site, with an error whose covariance is the local posterior covariance. Appendix~\ref{sec:krigingdetails} gives this observation model, the covariance model, and the equations used to calculate the weights and prediction uncertainty.

The calculation draws \num{\krigeDraws} profiles from the resulting Gaussian prediction for $\mathbf{m}_*$, converts each component to conductivity through $\sigma_n=\sigma_{\mathrm{ref}}10^{m_{*n}}$, and evaluates its layered impedance. The displayed central profile uses $\sigma_{\mathrm{ref}}10^{\widehat m_{*n}}$. This is the marginal median conductivity under the Gaussian approximation in log conductivity, rather than its arithmetic mean. The resulting impedance likewise need not be the mean of the simulated responses. We call the draws at one target its conditional ensemble. Draws are generated separately for each target. They retain the dependence between depths within a target but not the dependence between targets, so they are not joint realisations of conductivity across a region. A line integral or a network calculation that spans several targets would need draws from the joint prediction covariance of those targets.

For validation we first withhold each site and any record within \SI{1}{\kilo\metre}, then repeat the calculation after removing all observations within \SI{150}{\kilo\metre} of the target. A further \num{\krigeRegions} experiments withhold all sites within \SI{\krigeRadius}{\kilo\metre} of fixed geographic centres, yielding \num{\krigeTargets} regional targets (Figure~\ref{fig:kriginggaps}). These gaps cross state boundaries. All remaining sites are eligible neighbours regardless of state. The withheld profiles are excluded from the estimation of the spatial covariance, correlation length, and mean.

\end{proposalcolumns}

\begin{figure}[H]
    \centering
    \includegraphics[width=\textwidth]{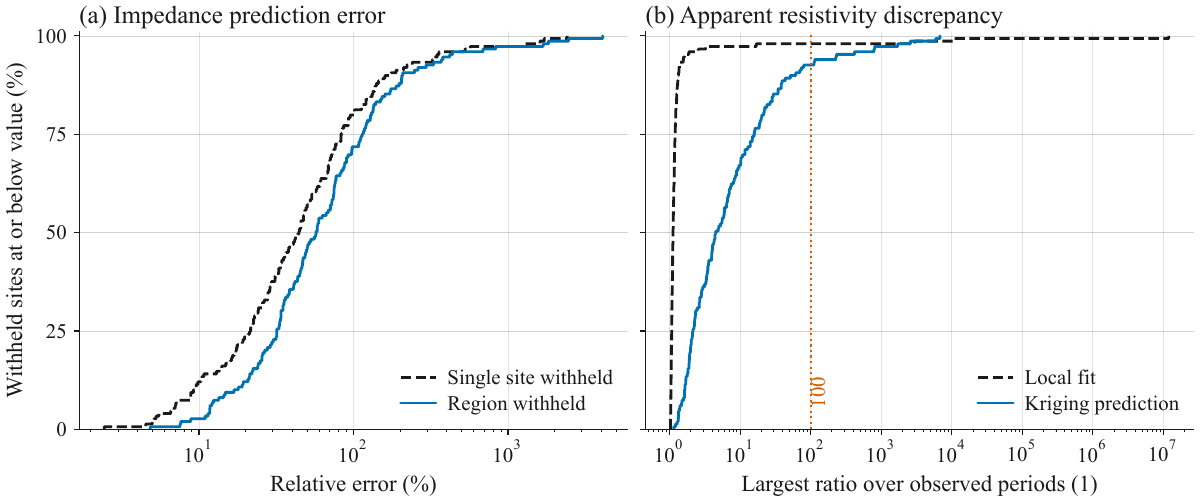}
    \caption{Response discrepancies at all \num{\krigeTargets} regional targets. (a)~Cumulative distributions of relative error between the predicted layered impedance and the measured $s$ compare the same targets under individual and regional withholding. (b)~The largest apparent resistivity discrepancy over each site's reported measurement periods is the larger of the predicted/measured ratio and its reciprocal. Both overprediction and underprediction therefore exceed unity. The dashed curve uses the local fit to the target's own observations as a diagnostic reference. The dotted line marks a factor of 100. Every target and all extreme values are retained.}
    \label{fig:krigingperformance}
\end{figure}

\begin{proposalcolumns}

The impedance prediction error compares the predicted layered impedance with the measured $s$. It is the Euclidean norm of their complex difference over the reported measurement periods divided by the norm of $s$. A separate comparison evaluates all four measured tensor entries. The predicted tensor has zero diagonal entries and off diagonal entries equal to the predicted impedance and its negative. Its error is the norm of the difference across all entries and periods divided by the corresponding norm of the measured tensor. Individual withholding gives median errors for the impedance projection and all tensor entries of \SI{\krigeSingleScalar}{\percent} and \SI{\krigeSingleTensor}{\percent}, respectively, at a median nearest retained distance of \SI{\krigeSingleDistance}{\kilo\metre}. Removing observations within \SI{150}{\kilo\metre} gives \SI{\krigeGapScalar}{\percent} and \SI{\krigeGapTensor}{\percent}, at \SI{\krigeGapDistance}{\kilo\metre}.

For the same \num{\krigeTargets} regional targets, the median impedance prediction error increases from \SI{\krigeMatchedScalar}{\percent} under individual withholding to \SI{\krigeRegionalScalar}{\percent} under regional withholding. The corresponding medians using all tensor entries are \SI{\krigeMatchedTensor}{\percent} and \SI{\krigeRegionalTensor}{\percent}. The median nearest retained distance increases from \SI{\krigeMatchedDistance}{\kilo\metre} to \SI{\krigeRegionalDistance}{\kilo\metre}. Figure~\ref{fig:krigingperformance} retains the complete distributions, including large errors that a median alone conceals.

\end{proposalcolumns}

\begin{proposalcolumns}

Two simple predictors serve as baselines under the same regional withholding: the local fit at the nearest retained site, and the mean of the local fits at the \num{\krigeNeighbours} nearest retained sites. Their median errors in $s$ are \SI{\sensNearestScalar}{\percent} and \SI{\sensMeanScalar}{\percent}, against \SI{\sensReferenceScalar}{\percent} for kriging, the lowest of the three. When all tensor entries are evaluated, the retained mean has the lowest median error, \SI{\sensMeanTensor}{\percent}, against \SI{\sensReferenceTensor}{\percent} for kriging and \SI{\sensNearestTensor}{\percent} for the nearest profile. The retained mean also has fewer targets with an apparent resistivity factor of 100 or more, \num{\sensMeanHundred}, than kriging with \num{\sensReferenceHundred}, and the nearest profile has more, \num{\sensNearestHundred}. The relative performance of the predictors therefore depends on the response quantity evaluated. Appendix~\ref{sec:krigingsensitivity} repeats the regional predictions under \num{\sensSettings} alternative settings of the covariance model and the local prior. The median error in $s$ ranges from \SI{\sensErrorLow}{\percent} to \SI{\sensErrorHigh}{\percent} across them, and the mean coverage defined below from \SI{\sensCoverageLow}{\percent} to \SI{\sensCoverageHigh}{\percent}.

At \num{\krigeHundredSites} of the \num{\krigeTargets} regional targets, predicted apparent resistivity differs from that of the measured $s$ by a factor of 100 or more at one or more reported measurement periods, \num{\krigeHundredPeriods} of the \num{\krigePeriods} target periods in all. At a fixed period, apparent resistivity is proportional to impedance amplitude squared, so this threshold corresponds to an amplitude discrepancy of at least a factor of ten. These are multiplicative discrepancies in amplitude, distinct from the normalised complex errors, which include phase. At \num{\extremeBandSites} of these \num{\extremeSites} targets the factor reaches 100 over a band of three or more consecutive periods, and at \num{\extremeIsolatedSites} at a single period. Of the \num{\krigeHundredPeriods} periods, \num{\extremeOver} are overpredictions and \num{\extremeUnder} underpredictions. The local fit to the target's own data already differs from the measurement by a factor of 100 or more at \num{\extremeLocalAlso} of these periods, all at \extremeLocalSites{}. At the other extreme targets the local fit agrees with the measurement, and the median relative standard error of the measurement at the affected periods is at most \SI{\extremeMeasurementMax}{\percent}, so these discrepancies arise in the interpolation rather than in poorly determined data. Table~\ref{tab:extremes} lists every case. They concern the predicted response. Multiple conductivity profiles are consistent with one measured response \citep{grandis1999}, so they do not measure error in conductivity.

For each target, the response interval at each period is the central \SI{95}{\percent} range of the real part of $s$ over the conditional ensemble, between its \num{2.5} and \num{97.5} percentiles, and separately that of the imaginary part. Coverage counts the real and imaginary parts of the measured $s$ that fall inside their intervals. Pooled over all targets, \num{\krigeCoverageCovered} of \num{\krigeCoverageComponents} components fall inside, or \SI{\krigeCoveragePooled}{\percent}. The mean over targets, which weights each target equally, is \SI{\krigeCoverageSiteMean}{\percent}, and the means for the \num{\krigeRegions} regions range from \SI{\krigeCoverageLow}{\percent} to \SI{\krigeCoverageHigh}{\percent}. Every component falls inside at \num{\krigeCoverageFullSites} targets and none at \num{\krigeCoverageEmptySites}. Targets within a region share retained sites and spatial structure and are not independent, so we report the regional values rather than a sampling interval for the pooled rate. The intervals describe the response without measurement noise under the stated model, and they are compared with noisy measurements, so coverage below the nominal \SI{95}{\percent} is a descriptive diagnostic and not a test of calibration. The intervals for apparent resistivity and phase in the appendix figures are percentiles of those quantities over the same ensemble.

\end{proposalcolumns}

\begin{table}[H]
    \centering
    \caption{The two layered inversions. Both fit $s=(Z_{xy}-Z_{yx})/2$ at the retained periods of each site, with $\operatorname{VAR}(s)=[\operatorname{VAR}(Z_{xy})+\operatorname{VAR}(Z_{yx})]/4$. The fit to measured tensors is compared with the full tensor in Section~\ref{sec:ground}. The fit supplied to kriging provides the local fits and covariances of Section~\ref{sec:kriging}.}
    \label{tab:inversions}
    \small
    \renewcommand{\arraystretch}{1.18}
    \begin{tabular}{@{}>{\raggedright\arraybackslash}p{0.15\textwidth}*{2}{>{\raggedright\arraybackslash}p{0.39\textwidth}}@{}}
\toprule
& Fit to measured tensors & Fit supplied to kriging \\
\midrule
Data & $\log_{10}\rho_a$ and phase of $s$ & real and imaginary parts of $s$ \\
Parameter & $\log_{10}\rho$ of each layer, $\rho$ in \si{\ohm\metre} & $\log_{10}\sigma$ of each layer, $\sigma$ in \si{\siemens\per\metre} \\
Layers & 16, interfaces log spaced from \SI{300}{\metre} to \SI{247.6}{\kilo\metre} & 10, interfaces at \numlist{0.5;2;5;10;20;40;80;160;320}~\si{\kilo\metre} \\
Standard error & propagated from $\operatorname{VAR}(s)/2$ for each real and imaginary part, at least \num{0.02} in $\log_{10}\rho_a$ and \SI{0.5}{\degree} in phase & $\sqrt{\operatorname{VAR}(s)/2}$ for each part, at least \SI{5}{\percent} of $|s|$ \\
Penalty & first differences of $\log_{10}\rho$ & Gaussian prior: adjacent differences with standard deviation \num{0.75}, departures from $-2$ with standard deviation \num{2} \\
Penalty weight & largest of 13 weights from $10^{1.5}$ to $10^{-1.5}$ that reaches a misfit of 1, otherwise the lowest misfit & fixed, the same at every site \\
Bounds & $-1\le\log_{10}\rho\le5$ & none \\
Optimiser & trust region reflective least squares & trust region reflective least squares from two starting profiles \\
Result & one profile & profile at the posterior mode and the inverse Gauss--Newton covariance \\
\bottomrule
\end{tabular}

\end{table}

\begin{proposalcolumns}

\section{Discussion}

The transmission line analogy connects the propagation and attenuation within each layer to the impedance at the surface. That impedance also determines apparent resistivity, phase, and the conversion of a magnetic record to an electric field. 

The worked examples relate the layered solution to measured responses that can depart from its assumptions. Across the \num{\nSites} sites, \num{\nHighAsymmetry} have off diagonal asymmetry of at least \num{0.2}. This diagnostic measures departure from the impedance form imposed by the layered approximation. Interpretation of the measured tensor also requires consideration of measurement error and the dimensionality of the conductivity structure \citep{chave2012}.

The plane wave approximation assumes that horizontal variation in the source is small over the relevant induction scales. Treatments that retain finite source geometry are needed when this approximation is inadequate \citep{boteler1998}. Lateral conductivity contrasts, including coastlines, require a representation of the Earth beyond horizontal layering \citep{kelbert2020}. The source geometry and conductivity representation should therefore be assessed for the region and periods of interest.

\subsection{Uncertainty in geoelectric estimation}
\label{sec:uncertainty}

Geoelectric estimation combines uncertainty in magnetic observations and their spatial reconstruction with uncertainty in the Earth response. Where conductivity is inferred and interpolated, additional assumptions enter through inversion, regularisation, and spatial covariance. Misspecification of source geometry or conductivity structure adds error of a different kind, which none of the covariances in this paper represents. A covariance describes spread conditional on a statistical model, and a residual measures disagreement with one observation.

The tensors used as references are themselves estimates from finite, noisy records (Section~\ref{sec:mtmeasurements}). Their uncertainty should be propagated through subsequent calculations, including inversion and field estimation. A small residual against a measured tensor does not mean the tensor is known exactly.

Inversion adds uncertainty because different profiles explain the same responses, to a degree set by the measurement errors, the period range, and the prior \citep{grandis1999}. In the conductivity example the Gaussian approximation is conditional on fixed layer interfaces, one fitted mode, and the layered representation of conductivity.

Interpolation adds uncertainty about how conductivity varies between sites. Carrying the inversion covariance into kriging accounts for the imprecision of the neighbouring profiles, conditional on the chosen covariance model \citep{pryet2011}. It does not average over other covariance models or unknown geological boundaries. The withholding experiments of Section~\ref{sec:kriging} test the predictions where observations exist, under the specified gaps.

Error in the reconstructed magnetic field must be propagated through the tensor of Equation~\ref{eq:tensor}, because each electric component depends on both magnetic components. The common electric field comparison of Section~\ref{sec:source} does this for one storm, with the measured tensors as common ground responses. A joint ensemble of magnetic records and Earth responses would extend it to uncertainty, retaining the specified dependence between components, periods, depths, and locations. Draws of conductivity already carry their uncertainty into the impedance, so adding a separate term for it would count it twice. The conductivity ensemble has not been combined with the magnetic reconstruction, and its errors cannot be added to those of Table~\ref{tab:budget}.

A predicted response and a future noisy observation have different uncertainties. Calibrating intervals for the second requires an observation model for the target measurement, which the conductivity example does not include.

\subsection{Implications for infrastructure estimates}

Network and equipment models provide the subsequent steps in infrastructure risk assessment. The voltage driving each line is the integral of the electric field along its route. \citet{kelbertlucas2020} derive line responses that depend on frequency from MT impedances that vary in space, with an approximation for the magnetic variation along each line. A discrepancy in the local electric field therefore needs to be propagated through the line integral and the network before it can be interpreted as an error in GIC. Because the voltage depends on the direction of the field along the route, the vector error of Equation~\ref{eq:vector-error} is the relevant measure for this step, not a measure of magnitude. The conductivity draws of Section~\ref{sec:kriging} are separate at each target, so such a calculation along a line would first need draws that are joint across targets. The induced currents depend on conductor topology and resistances \citep{lehtinen1985,viljanen1994}. \citet{bor2026components} construct a regional GIC model from substation components identified in imagery. Unknown transformer configurations and grounding resistances introduce further uncertainty. The coupled model of \citet{oughton2026coupled} places these inputs within an assessment of transformer vulnerability and economic consequences.

\section{Conclusions}

This paper reviews MT theory for a layered Earth and its use in geoelectric estimation. Maxwell's equations, the diffusion approximation, and interface conditions lead to the classical impedance recursion. The transmission line representation connects this recursion to attenuation with depth and the response to a magnetic time series.

The worked evaluations measure three steps of a geoelectric estimate. Replacing measured tensors by fitted layered models changes the electric field by a median vector error of \SI{\syntheticVectorMedian}{\percent} at \num{\nSites} sites under a synthetic disturbance and \SI{\stormVectorMedian}{\percent} at \num{\nStormSites} sites under the storm of May 2024. Tensor reduction produces the largest median vector discrepancy, and the median discrepancies associated with the layered fit and numerical processing are substantially smaller. Measures of magnitude alone understate this discrepancy, with medians of \SI{\syntheticMagnitudeMedian}{\percent} and \SI{\stormMagnitudeMedian}{\percent}. Reconstructing the magnetic field at \num{\nObservatoriesTested} withheld observatories gives a median vector error of \SI{\sourceVectorMedian}{\percent}. Propagated through the same measured tensors, it gives a median electric vector error of \SI{\budgetMagneticMedian}{\percent}, against \SI{\budgetGroundMedian}{\percent} for the layered representation, at observatories more isolated than typical MT sites. Kriging conductivity into withheld regions predicts $s$ with a median relative error of \SI{\krigeRegionalScalar}{\percent} at \num{\krigeTargets} targets, modestly lower than the nearest retained profile or the mean of retained profiles. Over all tensor entries the mean of retained profiles has the lowest median error. At \num{\krigeHundredSites} targets the predicted apparent resistivity differs by a factor of 100 or more, in most of them over a band of periods. The central \SI{95}{\percent} response intervals contain \SI{\krigeCoveragePooled}{\percent} of the measured components.

For the May 2024 storm, the electric vector error from magnetic reconstruction exceeded that from the layered representation in \SI{\budgetExceedPercent}{\percent} of the tested observatory--tensor pairs. The vector comparison retains differences in direction as well as amplitude, which are relevant to calculating induced voltages along transmission line routes.

\subsection{Limitations}

The theoretical treatment assumes horizontal layering and plane wave forcing. The evaluations use \num{\nSites} MT sites and one storm recorded at \num{\nObservatories} observatories, and their geographic coverage and retained periods limit the conclusions. The common electric field comparison combines observatory records with tensors measured elsewhere, and its magnetic term is measured under greater isolation than a typical MT site. Different conductivity profiles can produce similar surface impedances, and some measured tensors are poorly represented by a layered response. The variance assigned to $s$ omits the covariance between $Z_{xy}$ and $Z_{yx}$, and the reference fields interpolate each tensor with spline weights that bezpy takes from the reported complex variance. The conductivity interpolation uses a local Gaussian approximation to the inverse problem, fixed layer interfaces, a stationary spatial covariance, and draws that are independent between targets. Its \num{\krigeRegions} regional gaps are separate experiments with spatially dependent targets, not an exhaustive national validation. Propagating these discrepancies into line voltages and currents in power networks is left to future work.

\section{Data availability}

The magnetotelluric transfer functions are available from the EarthScope repository at \url{https://ds.iris.edu/spud/emtf} \citep{kelbert_em_2011}. The one minute geomagnetic records for 10 and 11 May 2024 are available from the United States Geological Survey at \url{https://geomag.usgs.gov} and through INTERMAGNET at \url{https://intermagnet.org}. The code, the processed results, and a list of the transfer function files used are available at \url{https://github.com/denniesbor/cagniard}.

\section{Acknowledgements}

The author thanks Robert Weigel for discussions of electromagnetic induction and its transmission line representation.

\section{Use of generative artificial intelligence}

I used generative artificial intelligence tools, Claude (Anthropic) and Codex (OpenAI), to edit the text, write and test the code for the calculations and figures, and draw the schematic in Figure~\ref{fig:geometry}. I directed this work, reviewed all output from these tools, and take full responsibility for the content of this paper.

\end{proposalcolumns}

\clearpage
\section{Symbols, units, and conventions}
\label{sec:notation}

Depth $z$ increases downward, and harmonic fields have time dependence $\ee^{+\ii\omega t}$. Equations use SI units.

\begingroup
\small
\setlength{\tabcolsep}{5pt}
\renewcommand{\arraystretch}{1.12}
\begin{longtable}{@{}p{0.25\textwidth}p{0.17\textwidth}p{\dimexpr0.58\textwidth-4\tabcolsep\relax}@{}}
\caption{Symbols and SI units.}\label{tab:symbols}\\
\toprule
Symbol & SI unit & Meaning and convention \\
\midrule
\endfirsthead
\multicolumn{3}{l}{\tablename~\thetable{} (continued)}\\
\toprule
Symbol & SI unit & Meaning and convention \\
\midrule
\endhead
\midrule
\multicolumn{3}{r}{Continued on next page}\\
\endfoot
\bottomrule
\endlastfoot
\multicolumn{3}{@{}l}{\textbf{Fields and material properties}}\\*
$\mathbf{E},\ E_x,E_y,E_z$ & $\mathrm{V\,m^{-1}}$ & Electric field and its components. $E_x,E_y$ are horizontal electric components, and $E_z$ is vertical. Plotted unit: mV/km. \\
$\mathbf{B},\ B_x,B_y,B_z$ & $\mathrm{T}$ & Magnetic flux density and its components. $\mathbf{B}=\mu_0\mathbf{H}$. Plotted unit: nT. \\
$\mathbf{H},\ H_x,H_y,H_z$ & $\mathrm{A\,m^{-1}}$ & Magnetic field strength and its components. Total local magnetic field, including the contribution induced in the ground. \\
$\mathbf{J},\ J_x$ & $\mathrm{A\,m^{-2}}$ & Conduction current density. $\mathbf{J}=\sigma\mathbf{E}$. \\
$\mathbf{J}_{\mathrm{conduction}},\ \mathbf{J}_{\mathrm{displacement}}$ & $\mathrm{A\,m^{-2}}$ & Conduction and displacement current densities. $\mathbf{J}_{\mathrm{conduction}}=\sigma\mathbf{E}$ and $\mathbf{J}_{\mathrm{displacement}}=\partial\mathbf{D}/\partial t$. \\
$\mathbf{D}$ & $\mathrm{C\,m^{-2}}$ & Electric displacement field. $\mathbf{D}=\epsilon\mathbf{E}$. \\
$\sigma,\ \sigma_n$ & $\mathrm{S\,m^{-1}}$ & Electrical conductivity. Positive, isotropic and constant within each layer in this derivation. \\
$\rho,\ \rho_n$ & $\Omega\,\mathrm{m}$ & Electrical resistivity. $\rho_n=1/\sigma_n$ in layer $n$. \\
$\epsilon$ & $\mathrm{F\,m^{-1}}$ & Electric permittivity. Displacement current is neglected where $\omega\epsilon/\sigma\ll1$. \\
$\mu,\ \mu_0$ & $\mathrm{H\,m^{-1}}$ & Magnetic permeability and free-space permeability. The ground is taken to have $\mu=\mu_0$. \\
\multicolumn{3}{@{}l}{\textbf{Coordinates, periods and layer amplitudes}}\\*
$x,y,z$ & $\mathrm{m}$ & Cartesian coordinates. $x$ and $y$ are horizontal. $z=0$ at the surface and increases downward. \\
$\zeta,\ h_n$ & $\mathrm{m}$ & Depth within layer and finite layer thickness. $\zeta=0$ at the top of layer $n$ and $\zeta=h_n$ at its base. \\
$n,\ N$ & $1$ & Layer index and total number of layers. Layer $1$ is uppermost. Layer $N$ is the terminating half space. \\
$t,\ T$ & $\mathrm{s}$ & Time and oscillation period. $\omega=2\pi/T$. \\
$\omega$ & $\mathrm{rad\,s^{-1}}$ & Angular frequency. $\omega=2\pi/T$. \\
$k,\ k_n$ & $\mathrm{m^{-1}}$ & Complex propagation constant. $k^2=\mathrm{i}\omega\mu\sigma$, with $\operatorname{Re}k>0$. The downward component varies as $\exp(-kz)$. \\
$\delta$ & $\mathrm{m}$ & Skin depth. $\delta=\sqrt{2/(\omega\mu\sigma)}$. A downward component decreases in amplitude by $\mathrm{e}^{-1}$ over this depth. \\
$E_n^{\downarrow},\ E_n^{\uparrow}$ & $\mathrm{V\,m^{-1}}$ & Downward and upward electric amplitudes at the layer top. $E_n^{\downarrow}\exp(-k_n\zeta)$ and $E_n^{\uparrow}\exp(+k_n\zeta)$ are the two electric components at depth $\zeta$ within layer $n$. \\
$E_{n,b}^{\downarrow},\ E_{n,b}^{\uparrow}$ & $\mathrm{V\,m^{-1}}$ & Downward and upward electric amplitudes at the layer base. $E_{n,b}^{\downarrow}=E_n^{\downarrow}\exp(-k_nh_n)$ and $E_{n,b}^{\uparrow}=E_n^{\uparrow}\exp(+k_nh_n)$. \\
$E_b,\ H_b$ & $\mathrm{V\,m^{-1}},\ \mathrm{A\,m^{-1}}$ & Total tangential fields at an interface. Continuous across the interface under the stated boundary conditions. \\
$E_0,\ B_0$ & $\mathrm{V\,m^{-1}},\ \mathrm{T}$ & Real surface electric or magnetic sinusoidal amplitude. $E_x(0,t)=E_0\cos(\omega t)$ in the depth example. $B_y(t)=B_0\cos(\omega t)$ in the example of the surface response. \\
$\alpha$ & $1$ & Complex source multiplier. Dimensionless scaling of a reference forcing at fixed spatial geometry. \\
$\mathcal{E}_x,\ \mathcal{H}_y$ & $\mathrm{V\,m^{-1}},\ \mathrm{A\,m^{-1}}$ & Fields for unit source multiplier. $E_x=\alpha\mathcal{E}_x$, $H_y=\alpha\mathcal{H}_y$. \\
\multicolumn{3}{@{}l}{\textbf{Impedance, frequency response and delay}}\\*
$Z,\ Z_n$ & $\Omega$ & Scalar MT impedance and impedance at the top of a layer. $Z=E_x/H_y$. The surface impedance is $Z_1$. \\
$\eta,\ \eta_n$ & $\Omega$ & Intrinsic impedance of a uniform medium. $\eta=\sqrt{\mathrm{i}\omega\mu\rho}$. It is also the characteristic impedance of the equivalent line. \\
$\mathbf{Z},\ Z_{ij}$ & $\Omega$ & MT impedance tensor and its entries. $E_x=Z_{xx}H_x+Z_{xy}H_y$ and $E_y=Z_{yx}H_x+Z_{yy}H_y$. \\
$i,j$ & $1$ & Horizontal tensor component indices. $i,j\in\{x,y\}$. The imaginary unit is upright $\mathrm{i}$. \\
$Z/\mu_0,\ Z_{ij}/\mu_0$ & $(\mathrm{V/m})/\mathrm{T}$ & Electric response per unit magnetic flux density. Plotted units: (mV/km)/nT. \\
$R_n$ & $1$ & Electric amplitude reflection coefficient at a layer base. $R_n=E_{n,b}^{\uparrow}/E_{n,b}^{\downarrow}$. \\
$\rho_a$ & $\Omega\,\mathrm{m}$ & Apparent resistivity. $\rho_a=|Z|^2/(\omega\mu_0)$. \\
$\phi$ & $\mathrm{rad}$ & Surface impedance phase. $\phi=\arg Z$. For $\phi>0$, $E_x$ leads $H_y$. Plotted unit: degrees. \\
$C$ & $\mathrm{m}$ & MT induction response of Schmucker and Weidelt. $C=Z/(\mathrm{i}\omega\mu_0)$. Plotted unit: km. \\
$C_{\mathrm{GDS}}$ & $\mathrm{m}$ & Response of geomagnetic depth sounding. Defined from vertical magnetic field and horizontal divergence. Requires horizontal source variation. \\
$Q$ & $1$ & Electric field at depth divided by its surface value. $Q=E_x(z)/E_x(0)$. In a half space, $Q=\exp[-(1+\mathrm{i})z/\delta]$. \\
$\phi_Q$ & $\mathrm{rad}$ & Continuous phase of the depth response. $\phi_Q=-z/\delta$ in a half space. The phase is continuous, including complete rotations. \\
$\tau_{\mathrm{ph}},\ \tau_{\mathrm{gr}}$ & $\mathrm{s}$ & Phase delay and group delay of the depth response. $\tau_{\mathrm{ph}}=-\phi_Q/\omega$, $\tau_{\mathrm{gr}}=-\partial\phi_Q/\partial\omega$ at fixed physical depth. \\
\multicolumn{3}{@{}l}{\textbf{Circuit analogy, data diagnostics and operators}}\\*
$V,\ I$ & $\mathrm{V},\ \mathrm{A}$ & Voltage and current on the equivalent transmission line. The differential equations correspond to those for $E_x,H_y$. Each field equation carries one more factor of inverse length. \\
$Z',\ Y'$ & $\Omega\,\mathrm{m^{-1}},\ \mathrm{S\,m^{-1}}$ & Series impedance and shunt admittance per unit length. $Z'=\mathrm{i}\omega\mu$ and $Y'=\sigma$. With displacement current, $Y'=\sigma+\mathrm{i}\omega\epsilon$. \\
$a$ & $1$ & Off diagonal magnitude asymmetry. Absolute difference between the two off diagonal magnitudes divided by their mean. Site values are medians over usable periods. \\
$s$ & $\Omega$ & Impedance projection. $s=(Z_{xy}-Z_{yx})/2$, unchanged by rotation of the horizontal axes. It equals $Z$ for a layered Earth. \\
$\mathbf{Z}_{\mathrm{a}}$ & $\Omega$ & Antisymmetric projection of the impedance tensor. $\mathbf{Z}_{\mathrm{a}}=(\mathbf{Z}-\mathbf{Z}^{\mathsf{T}})/2$, with diagonal entries zero and off diagonal entries $s$ and $-s$. \\
$Z_{\mathrm{fit}}$ & $\Omega$ & Impedance of the layered profile fitted to $s$. Evaluated at the retained periods of a site to form the sampled fit. \\
$\varepsilon_{\mathrm{vec}}$ & $1$ & Vector error of a predicted horizontal field. Root of the summed squared vector difference over the summed squared reference, Equation~\ref{eq:vector-error}. Plotted in per cent. \\
$u_{ij}$ & $\Omega$ & Standard error of each part of an impedance entry. $u_{ij}=\sqrt{\operatorname{VAR}(Z_{ij})/2}$ after conversion to the $E/H$ convention. Used for the amplitude and approximate phase error bars. \\
$\operatorname{VAR}(Z_{ij}),\ \operatorname{VAR}(s)$ & $\Omega^2$ & Reported variance of an impedance entry and variance assigned to the projection. Variance of the complex estimate, the mean squared modulus of its error. The real and imaginary parts each carry half. $\operatorname{VAR}(s)=[\operatorname{VAR}(Z_{xy})+\operatorname{VAR}(Z_{yx})]/4$. \\
$\mathcal{F},\ \mathcal{F}^{-1}$ & $\text{operator}$ & Fourier transform and inverse transform. The inverse reconstructs time dependence with $\exp(+\mathrm{i}\omega t)$. \\
$\mathrm{i}$ & $1$ & Imaginary unit. $\mathrm{i}^2=-1$. \\
$\mathrm{e}$ & $1$ & Base of the natural exponential. $\mathrm{e}=2.71828\ldots$. \\
\multicolumn{3}{@{}l}{\textbf{Conductivity inference and spatial prediction}}\\*
$\mathbf{r}_p,\ \mathbf{r}_*$ & $\mathrm{m}$ & Site positions. $p$ denotes a retained site and $*$ an unobserved target, both on the same sphere. \\
$p,q,\ M$ & $1$ & Retained site indices and count. $p,q=1,\ldots,M$. Layers use $n=1,\ldots,N$. \\
$\sigma_{\mathrm{ref}}$ & $\mathrm{S\,m^{-1}}$ & Conductivity reference. $\sigma_{\mathrm{ref}}=1\,\mathrm{S\,m^{-1}}$ makes the logarithm dimensionless. \\
$m_{pn},\ \mathbf{m}_p$ & $1$ & Log conductivity and profile vector. $m_{pn}=\log_{10}[\sigma_n(\mathbf{r}_p)/\sigma_{\mathrm{ref}}]$. Hats denote estimates. The subscript $\mathrm{ret}$ denotes stacked retained profiles. \\
$\mathbf{e}_p,\ \mathcal{N}$ & $1$ & Error of a local fit as an observation of the latent profile, and the Gaussian distribution. $\mathbf{e}_p\sim\mathcal{N}(\mathbf{0},\boldsymbol{\Sigma}_{\mathrm{inv},p})$ in Equation~\ref{eq:local-observation}. \\
$\mathbf{W}_p$ & $1$ & Kriging weight matrix. An $N\times N$ matrix for site $p$, with $\sum_{p=1}^{M}\mathbf{W}_p=\mathsf{I}_N$. \\
$\mathsf{I}_N,\ \mathsf{U}$ & $1$ & Identity and mean design matrices. $\mathsf{I}_N$ is the $N\times N$ identity. $\mathsf{U}$ stacks $M$ copies vertically. Superscript $\mathsf{T}$ denotes transpose. \\
$\boldsymbol{\beta}$ & $1$ & Spatial mean of log conductivity. One unknown constant per layer within each target neighbourhood. A hat denotes its estimate. \\
$d_{pq},\ d_{*p},\ \ell$ & $\mathrm{m}$ & Site separations and correlation length. Chord distances between sites, or between target and site. Distances and $\ell$ use the same unit. \\
$\psi_\ell(d)$ & $1$ & Spatial correlation. Mat\'ern correlation with smoothness $3/2$. \\
$\nu_{\mathrm{nug}}$ & $1$ & Relative process nugget. $\nu_{\mathrm{nug}}=0.1$ multiplies $\boldsymbol{\Sigma}_{\mathrm{depth}}$ to represent unresolved spatial variability. \\
$\boldsymbol{\Sigma}_{\mathrm{inv},p},\ \boldsymbol{\Sigma}_{\mathrm{depth}}$ & $1$ & Inversion and depth covariances. $N\times N$ matrices for the local inversion at site $p$ and the spatial process across layers. \\
$\boldsymbol{\Sigma}_{\mathrm{ret}},\ \boldsymbol{\Sigma}_{*\mathrm{ret}}$ & $1$ & Retained and target cross covariances. The $MN\times MN$ retained covariance includes inversion uncertainty. The target cross covariance is $N\times MN$. \\
$\boldsymbol{\Sigma}_{\mathrm{prior}},\ \boldsymbol{\Sigma}_{\mathrm{pred}},\ \boldsymbol{\Sigma}_{\beta}$ & $1$ & Prior, prediction and mean covariances. $N\times N$ matrices. Prediction uncertainty includes mean estimation and the process nugget, but no target measurement noise. \\
$\mathsf{A}_*$ & $1$ & Mean correction matrix. $\mathsf{A}_*=\mathsf{I}_N-\boldsymbol{\Sigma}_{*\mathrm{ret}}\boldsymbol{\Sigma}_{\mathrm{ret}}^{-1}\mathsf{U}$. \\
\end{longtable}
\endgroup

\proposalbibliography{mt_transmission_line}

\clearpage
\appendix
\section{Reported impedance variance}
\label{sec:covariance}

The variance reported with each impedance entry in an EMTF file is the variance of the complex estimate. It is the product of the diagonal element of the residual covariance for the output electric channel and the diagonal element of the inverse signal covariance for the input magnetic channel, and the real and imaginary parts each carry half of it \citep{egbert_zfiles,kelbert_emtffcu}. Each file stores both matrices at every period. For all \num{\nSites} files the stored variances equal these products to a relative difference of \num{\covVarianceMismatch}, which confirms the convention for the files used. Both inversions therefore take $\operatorname{VAR}(s)/2$ as the variance of each part of $s$, and treat the two parts as independent, as for a circular complex error.

The covariance between two entries is the product of the corresponding off diagonal elements of the same two matrices \citep{egbert_zfiles}. The sources consulted do not state which factor is complex conjugated, so we bound the omitted term rather than evaluate it. When both stored matrices are valid covariance matrices, the magnitude of the term omitted from $4\operatorname{VAR}(s)$ is at most twice the product of the moduli of the residual covariance between $E_x$ and $E_y$ and of the inverse signal covariance between $H_y$ and $H_x$. For any valid joint covariance of $Z_{xy}$ and $Z_{yx}$ with the reported variances, it is also at most $2[\operatorname{VAR}(Z_{xy})\operatorname{VAR}(Z_{yx})]^{1/2}$, by the Cauchy--Schwarz inequality. At \num{\covInvalidPeriods} periods, at the \covInvalidSiteCount{} sites \covInvalidSites{}, the stored residual or inverse signal covariance violates the Cauchy--Schwarz inequality and so is not a valid covariance matrix. Its off diagonal elements then justify no bound, so at these periods the omitted term is bounded by the Cauchy--Schwarz bound from the reported variances alone. At every other period the smaller of the two bounds is used. Periods with invalid stored matrices are retained in both inversions and weighted by their reported variances, as at every other period. The selection of periods does not use the stored matrices, and neither inversion uses the cross covariance, so the inversions are unaffected. Relative to $\operatorname{VAR}(Z_{xy})+\operatorname{VAR}(Z_{yx})$, the bound has a median of \SI{\covCrossMedian}{\percent} and a ninetieth percentile of \SI{\covCrossNinety}{\percent} over the \num{\nPeriodsRetained} retained periods. It exceeds \SI{10}{\percent} at \num{\covCrossAboveTen} periods at \num{\covCrossSitesAboveTen} sites and reaches \SI{\covCrossMax}{\percent} at the most. Covariance between periods is not reported and is taken as zero.

The weights of both inversions are set mostly by their minimum uncertainties rather than by the reported variance. The \SI{5}{\percent} floor of the local fit sets the uncertainty at \SI{\covKrigingFloor}{\percent} of retained periods. The floors of the fit in Section~\ref{sec:ground} set it at \SI{\covGroundRhoFloor}{\percent} of periods for apparent resistivity and \SI{\covGroundPhaseFloor}{\percent} for phase.

\section{Conductivity interpolation details}
\label{sec:krigingdetails}

The conductivity experiment uses interfaces at depths of \num{0.5}, \num{2}, \num{5}, \num{10}, \num{20}, \num{40}, \num{80}, \num{160}, and \SI{320}{\kilo\metre}, followed by a half space. Each local inversion fits the real and imaginary parts of $s$ independently, using the larger of $\sqrt{\operatorname{VAR}(s)/2}$ and \SI{5}{\percent} of $|s|$ as the assumed standard deviation of each part (Appendix~\ref{sec:covariance}). A Gaussian prior centres log conductivity on $-2$, with conductivity expressed in \si{\siemens\per\metre}. Gaussian penalties scale adjacent differences by \num{0.75} and deviations from the reference by \num{2}. The covariance uses the inverse Gauss--Newton curvature at the fitted mode. This local approximation does not explore the full distribution considered in Bayesian sampling of layered MT models \citep{grandis1999}.

\subsection{Spatial covariance and prediction}

The spatial calculation treats each local fit as an observation of a latent log conductivity profile,
\begin{equation}
    \widehat{\mathbf{m}}_p=\mathbf{m}_p+\mathbf{e}_p,
    \qquad \mathbf{e}_p\sim\mathcal{N}(\mathbf{0},\boldsymbol{\Sigma}_{\mathrm{inv},p}),
    \label{eq:local-observation}
\end{equation}
where $\mathbf{m}_p$ is the latent profile at site $p$ and $\boldsymbol{\Sigma}_{\mathrm{inv},p}$ the covariance of the local fit. Errors from different local inversions are assumed independent, while covariance between depths within each profile is retained. This is an approximation in two stages, not a joint inversion of all MT observations. The local mode is drawn toward the prior mean and toward a smooth profile, and $\boldsymbol{\Sigma}_{\mathrm{inv},p}$ is a posterior covariance that already contains the prior, so Equation~\ref{eq:local-observation} treats a shrunken estimate as an unbiased observation. The data narrow the prior substantially: across layers, the median ratio of posterior to prior variance ranges from \num{\priorRatioLow} to \num{\priorRatioHigh}. The prior still dominates at some sites in the shallowest layer and in the half space, where the ratio exceeds one half at \num{\priorTopSites} and \num{\priorBottomSites} sites, respectively. Write $\boldsymbol{\Sigma}_{\mathrm{depth}}$ for the covariance between layers in the spatial process. It and $\boldsymbol{\Sigma}_{\mathrm{inv},p}$ both describe dimensionless log conductivity. The spatial correlation is the Mat\'ern form with smoothness $3/2$ \citep[][Equation~4.17]{rasmussen2006},
\begin{equation}
    \psi_\ell(d)=\left(1+\frac{\sqrt{3}d}{\ell}\right)\exp\!\left(-\frac{\sqrt{3}d}{\ell}\right),
    \label{eq:kriging-correlation}
\end{equation}
where $d$ is chord distance and $\ell$ is correlation length, expressed in the same unit. Neighbour selection and gap membership use distances along great circles on the same sphere. The relative process nugget $\nu_{\mathrm{nug}}=0.1$ represents unresolved spatial variability, separate from inversion uncertainty.

Stack the $M$ retained profile estimates into the vector $\widehat{\mathbf{m}}_{\mathrm{ret}}$ of $MN$ components. Its covariance has $N\times N$ blocks
\begin{equation}
    [\boldsymbol{\Sigma}_{\mathrm{ret}}]_{pq}=
    \begin{cases}
      (1+\nu_{\mathrm{nug}})\boldsymbol{\Sigma}_{\mathrm{depth}}+\boldsymbol{\Sigma}_{\mathrm{inv},p}, & p=q,\\
      \psi_\ell(d_{pq})\boldsymbol{\Sigma}_{\mathrm{depth}}, & p\ne q.
    \end{cases}
    \label{eq:kriging-retained-covariance}
\end{equation}
Here $d_{pq}$ is the distance between retained sites $p$ and $q$. For an unobserved target, the $p$th block of its covariance with the retained profiles and its prior covariance are
\begin{equation}
    [\boldsymbol{\Sigma}_{*\mathrm{ret}}]_p=\psi_\ell(d_{*p})\boldsymbol{\Sigma}_{\mathrm{depth}},
    \qquad \boldsymbol{\Sigma}_{\mathrm{prior}}=(1+\nu_{\mathrm{nug}})\boldsymbol{\Sigma}_{\mathrm{depth}}.
    \label{eq:kriging-target-covariance}
\end{equation}
The process nugget is independent between distinct sites, so it contributes to the target variance but not to the target covariance with retained sites.

Let $\boldsymbol{\beta}$ contain one unknown spatial mean per layer and let $\mathsf{U}$ be the $MN\times N$ matrix formed by stacking $M$ copies of $\mathsf{I}_N$. Generalised least squares gives
\begin{equation}
    \boldsymbol{\Sigma}_{\beta}=(\mathsf{U}^{\mathsf{T}}\boldsymbol{\Sigma}_{\mathrm{ret}}^{-1}\mathsf{U})^{-1},
    \qquad \widehat{\boldsymbol{\beta}}=\boldsymbol{\Sigma}_{\beta}\mathsf{U}^{\mathsf{T}}\boldsymbol{\Sigma}_{\mathrm{ret}}^{-1}\widehat{\mathbf{m}}_{\mathrm{ret}}.
    \label{eq:kriging-mean}
\end{equation}
The prediction adds the spatially correlated residual to this estimated mean. Its covariance includes uncertainty in estimating that mean \citep[][Section~2.7]{rasmussen2006},
\begin{align}
    \widehat{\mathbf{m}}_*&=\widehat{\boldsymbol{\beta}}+\boldsymbol{\Sigma}_{*\mathrm{ret}}\boldsymbol{\Sigma}_{\mathrm{ret}}^{-1}(\widehat{\mathbf{m}}_{\mathrm{ret}}-\mathsf{U}\widehat{\boldsymbol{\beta}}),
    \label{eq:kriging-prediction}\\
    \mathsf{A}_*&=\mathsf{I}_N-\boldsymbol{\Sigma}_{*\mathrm{ret}}\boldsymbol{\Sigma}_{\mathrm{ret}}^{-1}\mathsf{U},\nonumber\\
    \boldsymbol{\Sigma}_{\mathrm{pred}}&=\boldsymbol{\Sigma}_{\mathrm{prior}}-\boldsymbol{\Sigma}_{*\mathrm{ret}}\boldsymbol{\Sigma}_{\mathrm{ret}}^{-1}\boldsymbol{\Sigma}_{*\mathrm{ret}}^{\mathsf{T}}+\mathsf{A}_*\boldsymbol{\Sigma}_{\beta}\mathsf{A}_*^{\mathsf{T}}.
    \label{eq:kriging-prediction-covariance}
\end{align}
Substitution of Equation~\ref{eq:kriging-mean} into Equation~\ref{eq:kriging-prediction} gives the linear weights in Equation~\ref{eq:kriging-weights}. Multiplication by $\mathsf{U}$ verifies their sum constraint. The last term in Equation~\ref{eq:kriging-prediction-covariance} accounts for the unknown spatial mean. Dropping it would treat that mean as known. The covariance includes unresolved target variability but excludes noise from a future target measurement.

For each neighbourhood, $\boldsymbol{\Sigma}_{\mathrm{depth}}$ averages the sample covariance of retained log conductivity profiles and a smooth covariance in depth, with equal weights. The sample covariance of fitted profiles also contains inversion variance, which Equation~\ref{eq:kriging-retained-covariance} counts again through $\boldsymbol{\Sigma}_{\mathrm{inv},p}$. The mean inversion variance is between \SI{\priorShareLow}{\percent} and \SI{\priorShareHigh}{\percent} of the variance of fitted profiles between sites in each layer. The smooth term correlates layers through the exponential of the negative absolute difference in $\ln(1+z/\SI{1}{\kilo\metre})$, evaluated at layer midpoints and at \SI{480}{\kilo\metre} for the half space. Its standard deviations are the larger of the sample standard deviation and \num{0.2}. A diagonal variance of $0.05^2$ is added. With this depth covariance held fixed, the correlation length is selected from \num{50}, \num{100}, \num{200}, and \SI{400}{\kilo\metre} by maximising restricted likelihood over retained profiles. Neither target observations nor target fits are used in these estimates. The reported uncertainty conditions on these covariance choices. Draws retain dependence between depths but are sampled separately at each target, so they are not joint realisations across a region.

\subsection{Sensitivity and baselines}
\label{sec:krigingsensitivity}

The mixture weight, the floor of \num{0.2}, the diagonal variance, the nugget, the candidate correlation lengths, and the representative depth of the half space are fixed choices of the method. Only the correlation length is estimated, by restricted likelihood over retained profiles. Table~\ref{tab:krigingsensitivity} varies one choice at a time, with the regional withholding and target exclusion unchanged. Two further rows remove the mean inversion covariance from the sample covariance before mixing, which addresses the double counting described above, and omit the inversion covariance entirely. Two rows refit every local profile with a different prior adjacent standard deviation. The reference row reproduces the stored predictive means to within $10^{-8}$ in log conductivity. It uses a different random seed from the main calculation, so its mean coverage, \SI{\sensReferenceCoverage}{\percent}, differs from the \SI{\krigeCoverageSiteMean}{\percent} of Section~\ref{sec:kriging} by sampling alone. Removing the inversion variance from the sample covariance gives a median error in $s$ of \SI{\sensDeconvolvedScalar}{\percent}. Across all settings the median error changes by a few percentage points, and the mean coverage by considerably more.

\begin{table}[H]
    \centering
    \caption{Regional withholding under alternative settings of the spatial model and the local prior, and two simple baselines, over the same \num{\krigeTargets} targets. Median errors are the normalised complex errors of $s$ and of all tensor entries. The Targets column gives the number of sites with an apparent resistivity factor of 100 or more at one or more periods. Coverage is the share of real and imaginary parts of the measured $s$ inside the central \SI{95}{\percent} response intervals, as a mean over targets and as the range of the six regional means. The baselines have no intervals.}
    \label{tab:krigingsensitivity}
    \small
    \renewcommand{\arraystretch}{1.12}
    \begin{tabular}{@{}lrrrrr@{}}
\toprule
& \multicolumn{2}{c}{Median error (\si{\percent})} & Targets & \multicolumn{2}{c}{Coverage (\si{\percent})} \\
\cmidrule(lr){2-3}\cmidrule(l){5-6}
Setting & $s$ & All entries & factor $\ge100$ & Mean & Regions \\
\midrule
Settings used in the main text & 57.3 & 75.7 & 11 & 82.5 & 63.7--98.4 \\
Process nugget \num{0.05} & 58.2 & 73.2 & 8 & 84.9 & 71.9--97.0 \\
Process nugget \num{0.2} & 55.3 & 74.1 & 9 & 79.7 & 65.6--98.1 \\
Sample covariance between depths only & 56.6 & 72.5 & 7 & 91.1 & 86.5--98.7 \\
Smooth covariance between depths only & 57.5 & 74.0 & 11 & 82.0 & 64.6--98.8 \\
Standard deviation floor \num{0.1} & 57.3 & 75.7 & 11 & 82.5 & 63.7--98.4 \\
Standard deviation floor \num{0.4} & 56.9 & 75.7 & 11 & 82.2 & 63.7--96.5 \\
Correlation length fixed at \SI{100}{\kilo\metre} & 55.2 & 72.7 & 12 & 85.9 & 77.5--98.5 \\
Correlation length fixed at \SI{400}{\kilo\metre} & 57.8 & 72.8 & 9 & 65.2 & 50.5--81.8 \\
Mean inversion covariance removed from sample covariance & 53.0 & 72.9 & 9 & 83.9 & 64.5--98.4 \\
Inversion covariance omitted & 57.2 & 72.9 & 11 & 82.8 & 62.6--97.9 \\
Prior adjacent standard deviation \num{0.5} & 57.0 & 73.2 & 11 & 81.6 & 63.0--98.2 \\
Prior adjacent standard deviation \num{1.5} & 53.8 & 73.8 & 9 & 84.2 & 66.9--99.4 \\
\addlinespace
Baseline: nearest retained profile & 63.5 & 76.8 & 22 &  &  \\
Baseline: mean of the retained profiles & 59.1 & 72.7 & 7 &  &  \\
\bottomrule
\end{tabular}

\end{table}

\subsection{Extreme discrepancies}

Table~\ref{tab:extremes} lists every regional target at which the predicted apparent resistivity differs from that of the measured $s$ by a factor of 100 or more at one or more periods. The longest run counts consecutive periods at or above that factor. The local fit factor is the median, over the affected periods, of the same factor for the local fit to the target's own data. The measurement error is the median, over the same periods, of the standard error of each part of $s$ relative to $|s|$.

\begin{table}[H]
    \centering
    \caption{Regional targets with an apparent resistivity factor of 100 or more. The direction states whether the prediction lies above (over) or below (under) the measured value at the affected periods. Factors are dimensionless.}
    \label{tab:extremes}
    \small
    \renewcommand{\arraystretch}{1.12}
    \begin{tabular}{@{}llrrlrrr@{}}
\toprule
& & Periods & Longest & Failing periods & Largest & Local fit & Measurement \\
Region & Site & failing & run & (\si{\second}) & factor & factor & error (\si{\percent}) \\
\midrule
Mid-Atlantic & MBB05 & 3 of 25 & 3 & \numrange{3121}{7282} (over) & \num{112} & \num{1.09} & 1.6 \\
Mid-Atlantic & VAQ55 & 25 of 25 & 25 & \numrange{12}{4681} (over) & \num{807} & \num{1.02} & 0.7 \\
Mid-Atlantic & VAS55 & 21 of 21 & 21 & \numrange{15}{4681} (over) & \num{6778} & \num{132} & 4.0 \\
Northern New England & MEB61 & 10 of 21 & 10 & \numrange{20}{171} (over) & \num{792} & \num{72.2} & 1.4 \\
Northern New England & MEB63 & 24 of 26 & 23 & \numrange{12}{7282} (over) & \num{420} & \num{1.04} & 0.6 \\
Pacific Northwest & WAB08 & 25 of 26 & 25 & \numrange{12}{4681} (under) & \num{1714} & \num{1.02} & 0.2 \\
Pacific Northwest & WAC07 & 5 of 26 & 5 & \numrange{12}{33} (under) & \num{227} & \num{1.01} & 0.2 \\
Pacific Northwest & WAC08 & 25 of 26 & 25 & \numrange{12}{4681} (under) & \num{2589} & \num{1.01} & 0.2 \\
Pacific Northwest & WAC09 & 26 of 26 & 26 & \numrange{12}{7282} (under) & \num{5955} & \num{1.01} & 0.3 \\
Southern Rockies & COP24 & 20 of 26 & 20 & \numrange{54}{7282} (under) & \num{233} & \num{1.09} & 0.3 \\
Southern Rockies & COT23 & 1 of 26 & 1 & \num{12} (over) & \num{114} & \num{1.06} & 0.5 \\
\bottomrule
\end{tabular}

\end{table}

\subsection{Regional comparisons}

The comparisons show ten sites per region. Selection uses coordinates alone, beginning nearest the mean site position and successively adding the site most distant from those selected. All withheld sites contribute to the statistics. Grey and blue bands show marginal \SI{95}{\percent} ranges for local and predicted conductivity. Response bands span the central \SI{95}{\percent} of simulated responses. The half space is displayed to \SI{500}{\kilo\metre}.

\begin{figure}[H]
    \centering
    \includegraphics[width=\textwidth]{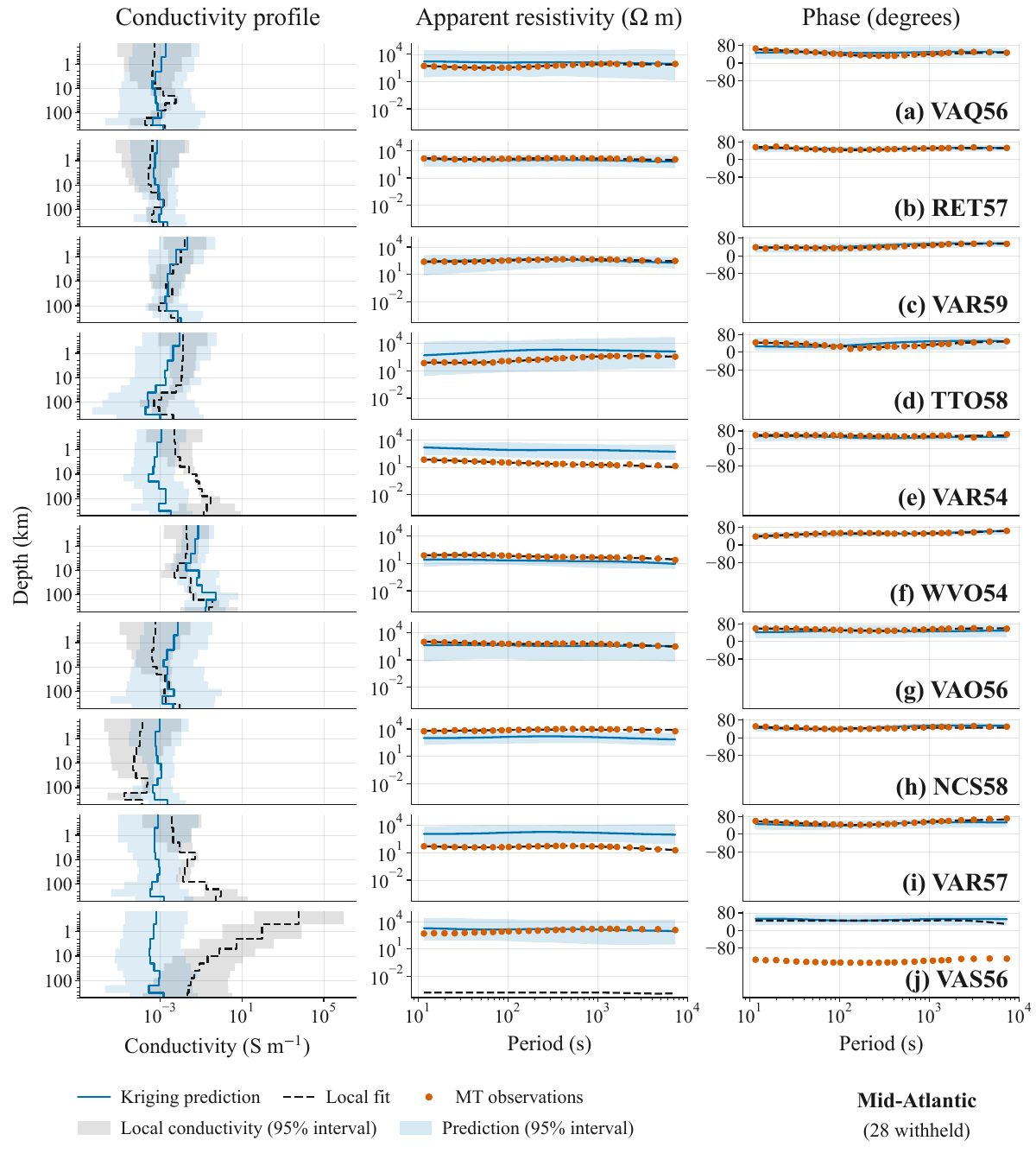}
    \caption{Mid-Atlantic gap, ten of \num{\krigeMidAtlanticSites} withheld sites.}
    \label{fig:krigingMidAtlantic}
\end{figure}

\begin{figure}[H]
    \centering
    \includegraphics[width=\textwidth]{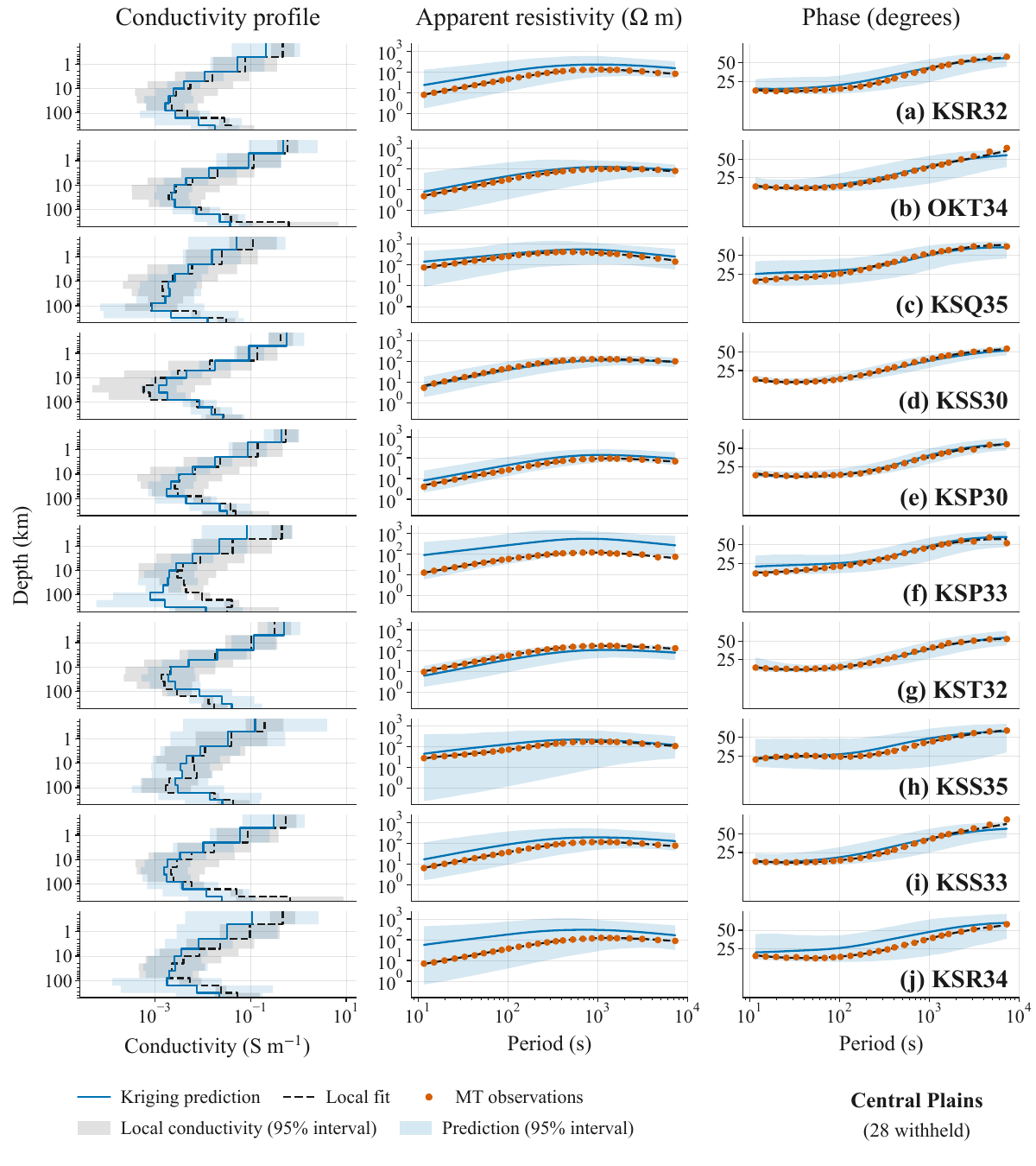}
    \caption{Central Plains gap, ten of \num{\krigeCentralPlainsSites} withheld sites.}
    \label{fig:krigingCentralPlains}
\end{figure}

\begin{figure}[H]
    \centering
    \includegraphics[width=\textwidth]{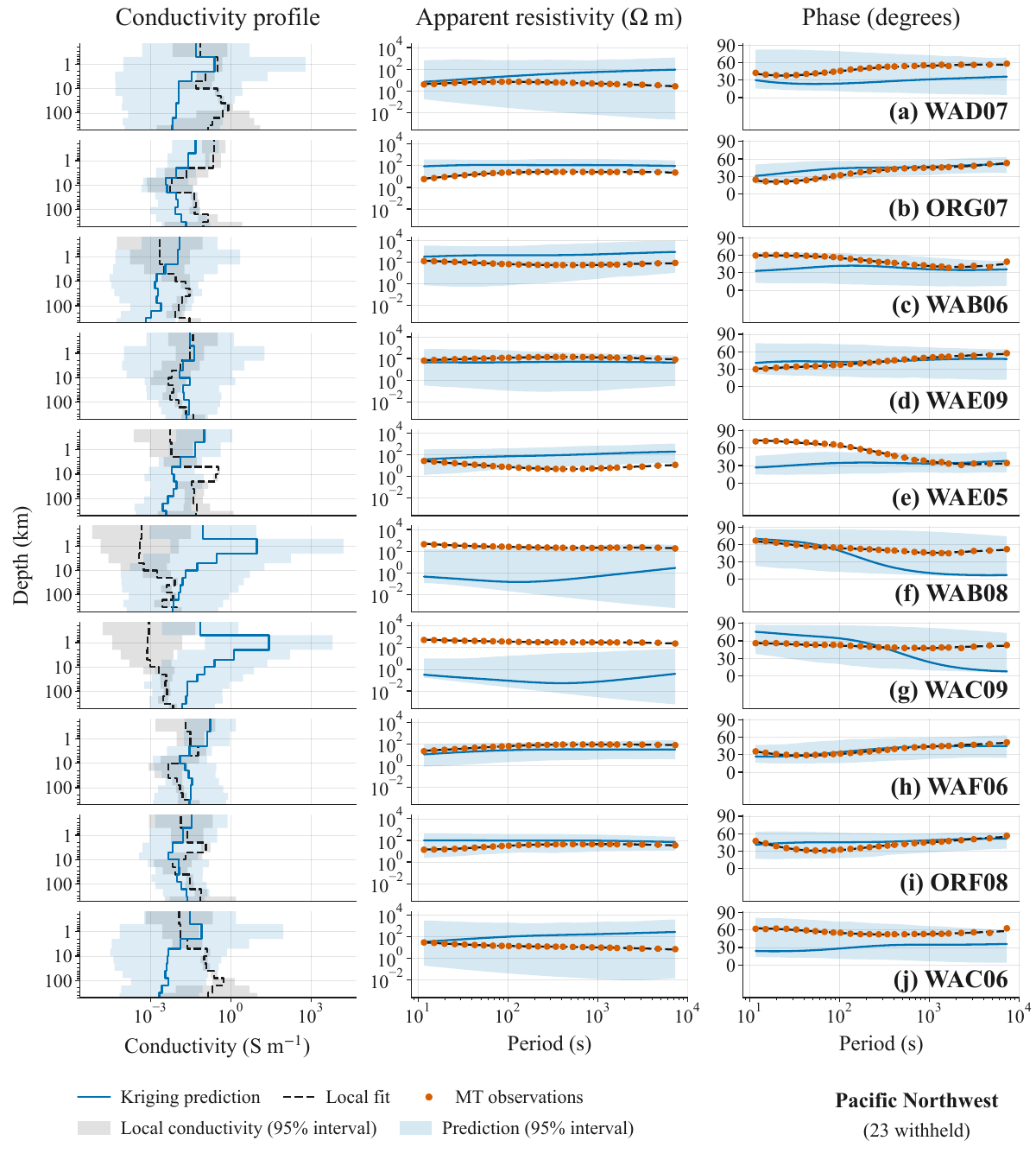}
    \caption{Pacific Northwest gap, ten of \num{\krigePacificNorthwestSites} withheld sites.}
    \label{fig:krigingPacificNorthwest}
\end{figure}

\begin{figure}[H]
    \centering
    \includegraphics[width=\textwidth]{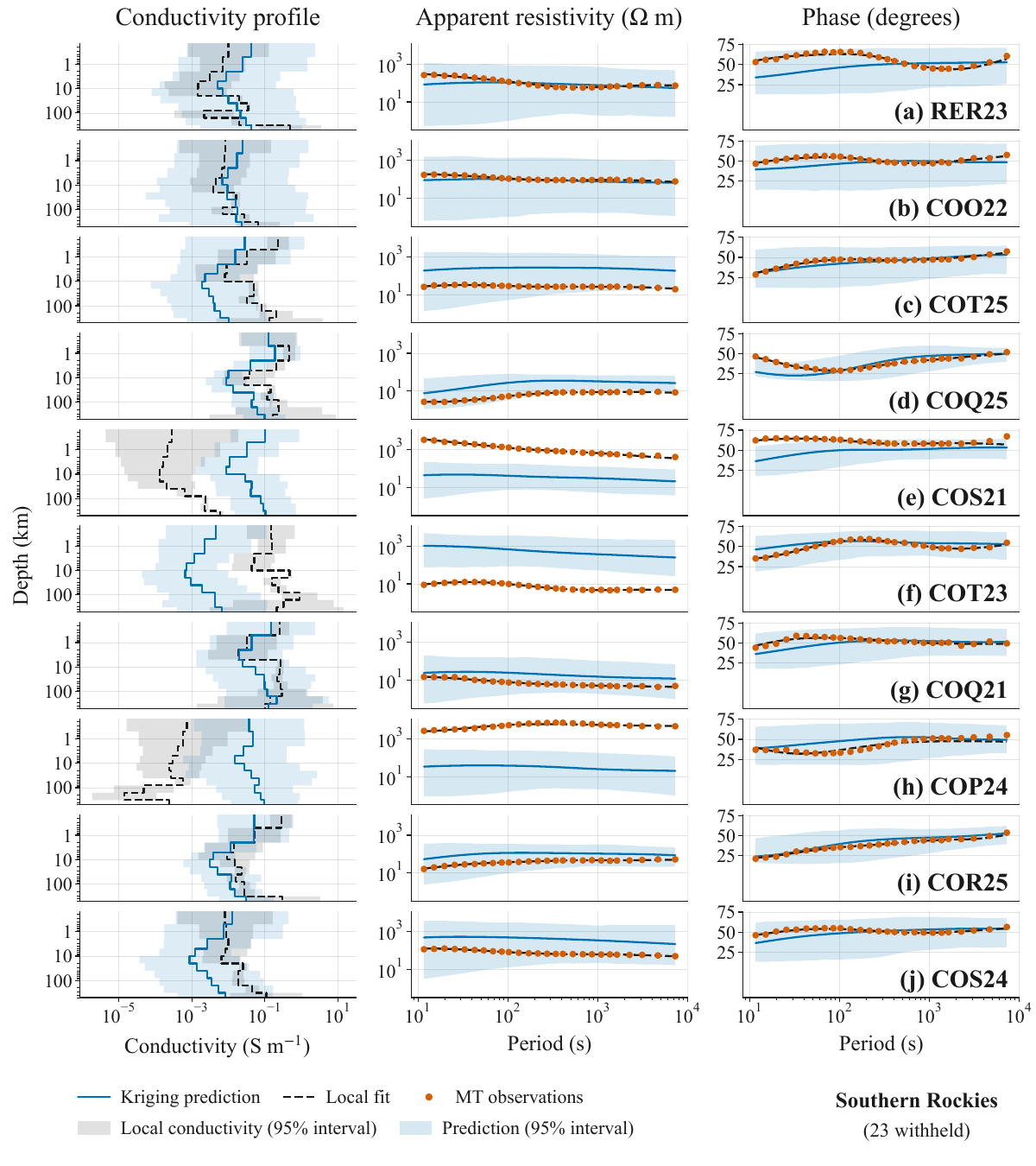}
    \caption{Southern Rockies gap, ten of \num{\krigeSouthernRockiesSites} withheld sites.}
    \label{fig:krigingSouthernRockies}
\end{figure}

\begin{figure}[H]
    \centering
    \includegraphics[width=\textwidth]{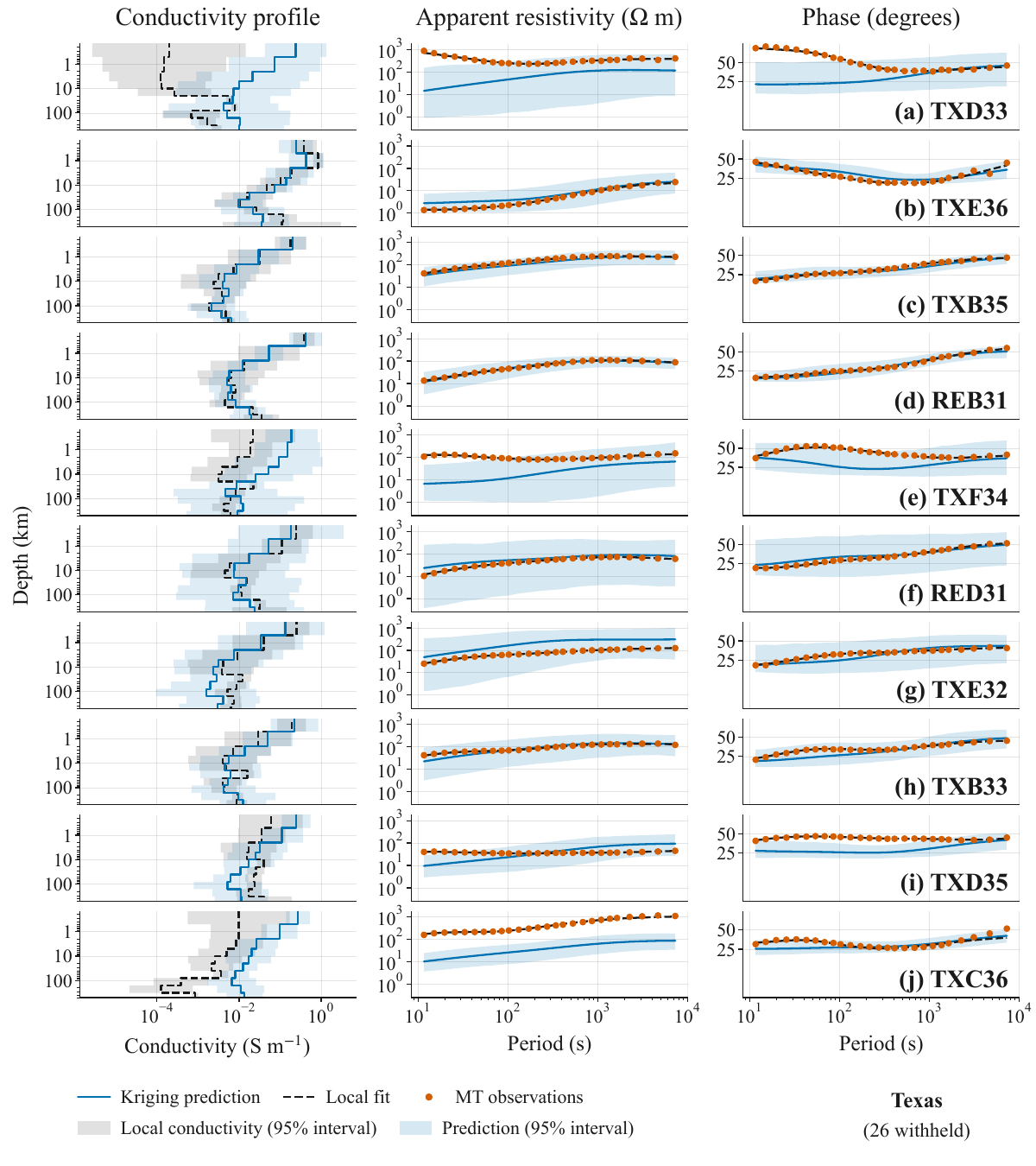}
    \caption{Central Texas gap, ten of \num{\krigeCentralTexasSites} withheld sites.}
    \label{fig:krigingCentralTexas}
\end{figure}

\begin{figure}[H]
    \centering
    \includegraphics[width=\textwidth]{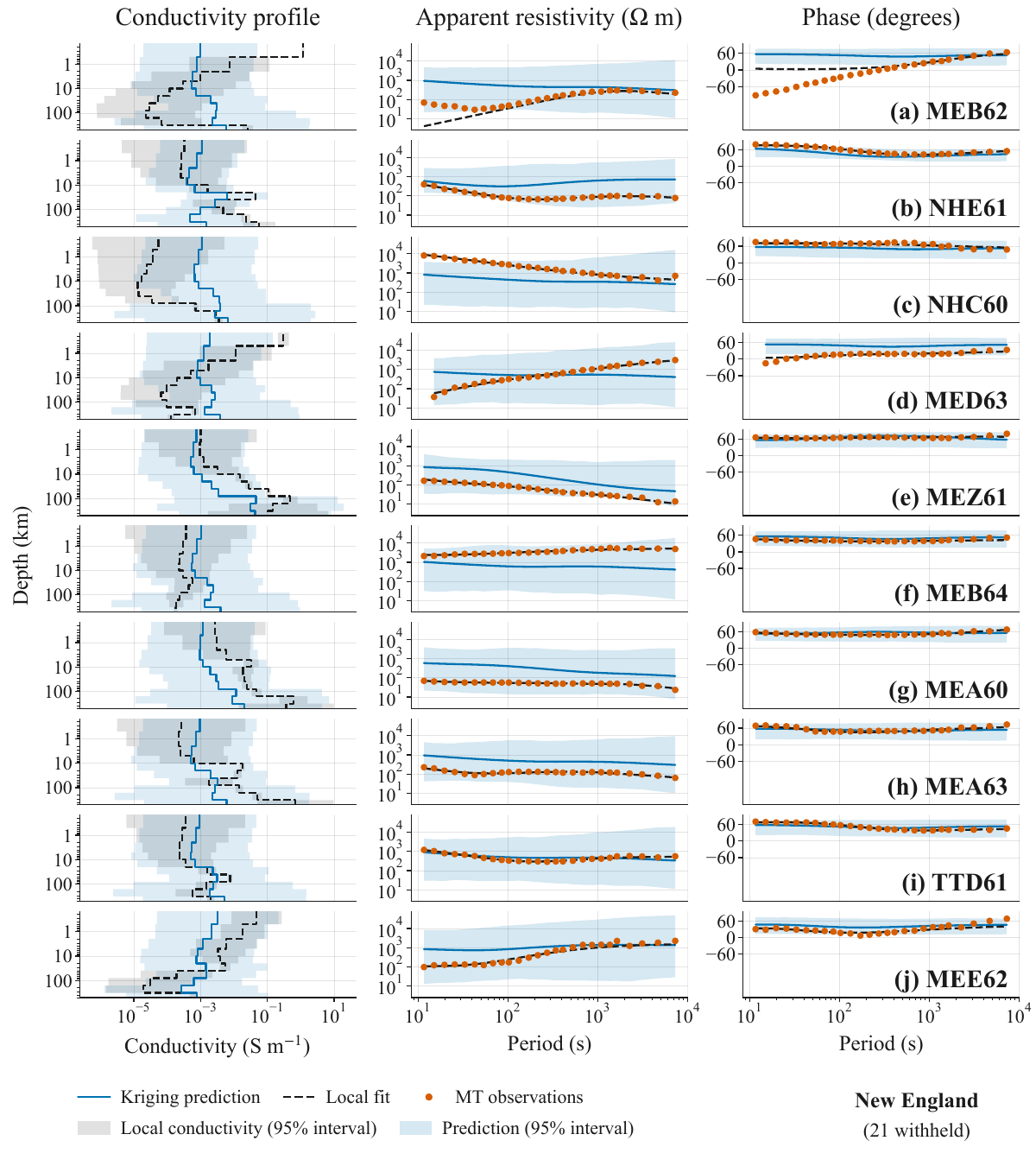}
    \caption{Northern New England gap, ten of \num{\krigeNorthernNewEnglandSites} withheld sites.}
    \label{fig:krigingNorthernNewEngland}
\end{figure}

The six gaps and their withheld sites are: Mid-Atlantic, \ang{37.5}~N \ang{79}~W, \num{\krigeMidAtlanticSites} sites, Central Plains, \ang{38.5}~N \ang{98.5}~W, \num{\krigeCentralPlainsSites} sites, Pacific Northwest, \ang{47}~N \ang{120}~W, \num{\krigePacificNorthwestSites} sites, Southern Rockies, \ang{38.5}~N \ang{106}~W, \num{\krigeSouthernRockiesSites} sites, Central Texas, \ang{31}~N \ang{99}~W, \num{\krigeCentralTexasSites} sites, Northern New England, \ang{45}~N \ang{69.5}~W, \num{\krigeNorthernNewEnglandSites} sites. Figures~\ref{fig:krigingMidAtlantic} to~\ref{fig:krigingNorthernNewEngland} show ten sites from each. In every panel, blue curves are predictions from retained sites outside the gap, black dashed curves are local fits to the withheld observations, and orange circles are the apparent resistivity and phase of the measured $(Z_{xy}-Z_{yx})/2$. Rows follow the map labels in Figure~\ref{fig:kriginggaps}.

\end{document}